%% file: main.tex
\documentclass[journal]{IEEEtran}

\usepackage{cite}
\usepackage{amsmath,amssymb,amsfonts}
\usepackage{algorithmic}
\usepackage{graphicx}
\usepackage{textcomp}
\usepackage{xcolor}
\usepackage[hidelinks]{hyperref}
\hypersetup{
  pdftitle={5G ISAC-Based UAV Detection and 3-D Tracking Using Uplink Sounding
    Reference Signals on an End-to-End O-RAN Simulation Testbed},
  pdfauthor={Arun K. Gurung, Satha K. Sathananthan, Shiva R. Pokhrel},
  pdfkeywords={ISAC, O-RAN, 5G NR, UAV detection, UL-SRS, FlexRIC, Sionna RT,
    bistatic sensing, EKF, counter-UAS}}
\usepackage{booktabs}
\usepackage{multirow}
\usepackage{array}
\usepackage{subfig}
\usepackage{bm}          % bold math
\usepackage{siunitx}     % SI units
\usepackage{placeins}    % \FloatBarrier -- keeps floats from queuing past their section
\usepackage{tikz}
\usetikzlibrary{arrows.meta,positioning,calc,fit,backgrounds,shapes.geometric}
\DeclareSIUnit{\sample}{Sa}
\DeclareSIUnit{\dBm}{dBm}
\DeclareSIUnit{\dBsm}{dBsm}

\tikzset{
  node distance = 3.6mm and 4.2mm,
  blk/.style   = {draw, rounded corners=1pt, align=center, font=\scriptsize,
                  inner sep=2pt, minimum height=5.4mm},
  phy/.style   = {blk, fill=blue!6},
  ric/.style   = {blk, fill=orange!8},
  ext/.style   = {blk, fill=black!5},
  flow/.style  = {-{Latex[length=1.6mm]}, thin},
  proto/.style = {-{Latex[length=1.6mm]}, thin, dashed, gray!70},
  lbl/.style   = {font=\scriptsize, inner sep=1pt, align=center},
  slbl/.style  = {font=\tiny, inner sep=1pt, align=center},
}


\input{campaign_numbers}
\input{campaign_bridge_numbers}
\input{primary_accuracy}

\input{detector_numbers}
\input{cfar_numbers}
\input{speed_numbers}
\input{upa_numbers}

\input{multistatic_numbers}
\input{multistatic_campaign_numbers}   % cell matrix (auto-generated by aggregate_multistatic.py)

\begin{document}

\title{5G ISAC-Based UAV Detection and 3-D Tracking Using Uplink Sounding Reference Signals
  on an End-to-End O-RAN Simulation Testbed}

%% JOURNAL-CLASS AUTHOR BLOCK. IEEEtran's [journal] option does not implement
%% \IEEEauthorblockN/\IEEEauthorblockA or \and -- it silently flattens them, which is why
%% the conference block rendered as one run-on line after the class switch. Journal style
%% is a single comma-separated author list with \IEEEmembership, and affiliations carried
%% in \thanks footnotes. Affiliation text is reproduced verbatim from the conference block;
%% nothing has been inferred from the e-mail domains.
\author{Arun~K.~Gurung,~\IEEEmembership{Senior Member,~IEEE,}
  Satha~K.~Sathananthan,
  and~Shiva~R.~Pokhrel,~\IEEEmembership{Senior Member,~IEEE}%
  \thanks{A.~K.~Gurung is in Melbourne, Australia
    (e-mail: akgurung@ieee.org).}%
  \thanks{S.~K.~Sathananthan is in Melbourne, Australia
    (e-mail: satha@nexvis.com.au).}%
  \thanks{S.~R.~Pokhrel is with Deakin University, Burwood, Australia
    (e-mail: shiva.pokhrel@deakin.edu.au).}}

%% Running heads. The left field is the journal name, left empty until the venue is
%% fixed rather than filled with a placeholder that could ship by accident.
\markboth{}{Gurung \MakeLowercase{\textit{et al.}}: An End-to-End O-RAN Simulation
  Testbed for 5G ISAC-Based UAV Detection and 3-D Tracking}

\maketitle

%% =========================================================
\begin{abstract}
  Integrated Sensing and Communication (ISAC) lets cellular infrastructure serve
  communication users and sense on the same waveform. We present an end-to-end O-RAN
  \emph{simulation} testbed for 5G ISAC targeting low-altitude UAV detection and 3-D
  tracking, built from open-source components---OpenAirInterface, FlexRIC and Sionna
  RT---in which the NR Uplink Sounding Reference Signal is repurposed as a passive
  radar waveform: a PHY-layer sensing stage inside the gNB produces detections that
  reach an Extended Kalman Filter tracking xApp over a custom E2 service model, with no
  change to the NR standard and no dedicated sensing waveform.

  A single bistatic pair leaves elevation unobservable, so the tracker needs a height
  prior; we remove it two independent ways and measure both
  --- a planar receive array supplying a
  vertical aperture, and a second transmitter supplying range diversity.
  Both live results corroborate an offline ray-traced study of the same
  estimator, which converges from a deliberately wrong initial altitude to
  \SI{\UPAoffHeightRmse}{\metre} RMSE at a consistent filter, so altitude observability
  is established both in the signal-processing chain and end to end through the live
  stack. Detection coverage is preserved under a concurrent
  \SI{10}{\mega\bit\per\second} uplink communications load.
\end{abstract}


%% =========================================================
\section{Introduction}
\label{sec:intro}

The proliferation of low-altitude unmanned aerial vehicles (UAVs)
in civil airspace has created an urgent need for scalable,
cost-effective counter unmanned-aircraft-system (C-UAS) solutions. Traditional radar
systems require dedicated spectrum, infrastructure, and
operational budgets that are prohibitive for many deployment
scenarios. Cellular networks, by contrast, already blanket
terrestrial environments and can be reused for dual-purpose
sensing, aligning with the Integrated Sensing and Communication
(ISAC) vision being studied in 3GPP SA1 and RAN study items towards
Release~19 \cite{3gpp_tr22837, 3gpp_tr38867}. The wording matters:
these are \emph{study reports}, and at the time of writing no normative 3GPP
sensing specification exists; the uplink sounding reference signal (UL-SRS) is a candidate sensing signal under study,
not a standardized one.

Open Radio Access Network (O-RAN) architecture disaggregates the
base station into the O-RAN radio unit (O-RU), distributed unit
(O-DU), central unit (O-CU), and the RAN Intelligent
Controller (RIC), exposing standardized interfaces (E2, O1, A1)
that enable third-party applications (xApps/rApps) to observe and
control the radio stack in near-real time \cite{oran_wg1_arch}.
This programmability makes O-RAN a natural host for ISAC
processing: sensing algorithms can be embedded as xApps in the
Near-RT RIC, co-located with the RAN, and fed detection results
via the E2 interface \cite{oran_wg3_e2ap}.

Despite growing theoretical interest in 5G ISAC
\cite{liu_survey_isac, zhang_dual_function}, end-to-end open
testbeds that integrate a real NR protocol stack, a standards-compliant RIC, and a ray-traced channel simulator
especially UAV surveillance remain scarce.

\textbf{Contributions.} This paper makes the following
contributions:

\begin{enumerate}
  \item \textbf{An O-RAN compliant ISAC sensing plane, end to end}: a live 5G NR
        stack, a Near-RT RIC and a ray-traced channel wired into one containerized testbed,
        with UAV echo injection and a calibrated thermal-noise link budget, so detections
        come from a real PHY driven by physically-scaled signals. Sensing runs inside the
        OAI gNB (rank-$K_c$ clutter-subspace deflation, 2-D OFDM Range-Doppler processing,
        OS-CFAR with single-peak grouping, closed-form interferometric angle of arrival) and
        every detection reaches a subscribing xApp over a prototype E2 service model
        (SM-SENS) using standard E2AP subscription/indication procedures --- with no
        modification to the NR standard and no dedicated sensing waveform.

  \item \textbf{A bistatic tracker evaluation}: an
        EKF xApp with a nonlinear range--range-rate--azimuth measurement model, Hungarian
        association and a time-driven track-maintenance pass that bridges intermittent
        detection gaps to sustain continuity, evaluated per-CPI against ray-traced ground
        truth with corrected bistatic definitions. The evaluation is deliberately
        adversarial: a height-prior ablation and single-pair observability analysis, a
        street-canyon occlusion stress test, sensing under a concurrent
        \SI{10}{\mega\bit\per\second} uplink load, and an end-to-end transport latency study
        (Section~\ref{sec:transport} and Appendix~\ref{sec:comms}).

  \item \textbf{Two routes from 2-D to 3-D tracking}: the
        single-pair height prior above is removed twice over. A two-dimensional uniform
        planar array at the gNB adds a vertical baseline, and a separable dual phase-ramp
        estimator recovers elevation alongside azimuth --- reducing exactly to the baseline
        azimuth estimator when the array is one-dimensional --- feeding a 3-D EKF that
        estimates altitude online (Section~\ref{sec:upa}). Independently, a second nrUE
        forms a second bistatic pair whose range diversity fixes altitude by geometry
        alone, on the unmodified azimuth-only array (Section~\ref{sec:multistatic}). Over
        $N=\MScampaignN$ seeded transits per configuration on one host and one build, both
        collapse the cross-seed spread of the final altitude from
        \SI{\MhfinalSDA}{\metre} to \SI{\MhfinalSDC}{\metre} --- the observability
        signature. They are not equivalent: the multistatic route is pinned \emph{and}
        unbiased (\SI{\MhfinalB}{\metre} against a \SI{\MSheightTrue}{\metre} truth,
        \SI{\MheightB}{\metre} height RMSE), while the planar array carries a residual
        \SI{\MhbiasC}{\metre} elevation bias and instead wins horizontal accuracy
        (\SI{\MposC}{\metre}).

\end{enumerate}

The remainder of this paper is organized as follows.
Section~\ref{sec:related} surveys related work.
Section~\ref{sec:sysmodel} presents the system model.
Section~\ref{sec:testbed} describes the testbed architecture.
Section~\ref{sec:sensing} details the UL-SRS sensing pipeline.
Section~\ref{sec:results} presents evaluation results.
Section~\ref{sec:discussion} discusses design insights and limitations.
Section~\ref{sec:conclusion} concludes the paper.

%% =========================================================
\section{Related Work}
\label{sec:related}

Table~\ref{tab:related} positions this testbed against the closest open
RAN digital-twin, O-RAN, and cellular-ISAC platforms along the axes a
sensing testbed must satisfy; the remainder of this section discusses
each line of work in turn.

% --- Novelty / capability comparison against prior platforms ---
\newcommand{\cmark}{\checkmark}
\begin{table*}[t]
  \centering
  \caption{Capability comparison against the closest prior platforms.
    \cmark~= provided; --~= not provided.
    \emph{Channel} column: RT~= ray-tracing simulation, sim~= link-level or
    stochastic channel simulation, HW~= real RF hardware. \emph{R/D/AoA} = range,
    Doppler (range-rate) and azimuth/bearing all estimated and evaluated.
    \emph{Coex.}~= carries concurrent live communication traffic on the same
    waveform; \emph{Open}~= open-source code release.
    $^{\ddagger}$Built on open-source components (OAI, FlexRIC, Sionna) with the
    sensing code release pending.}
  \label{tab:related}
  \footnotesize
  \begin{tabular}{@{}l l c l l c c c c c@{}}
    \toprule
    Platform                                                                           & NR stack     & RIC    & Waveform        & Channel & UAV    & R/D/AoA
                                                                                       & E2           & Coex.  & Open                                                                             \\
    \midrule
    OWDT~\cite{iye_owdt}                                                               & OAI          & \cmark & --              & RT      & --     & --      & \cmark & \cmark & --$^{\ddagger}$ \\
    Colosseum~\cite{polese_colosseum_twin}                                             & OAI/srsRAN   & \cmark & --              & HW-emu  & --     & --      & \cmark & \cmark & \cmark          \\
    Boston Twin~\cite{testolina_bostontwin}                                            & --           & --     & --              & RT      & --     & --      & --     & --     & \cmark          \\
    O-RAN ARC~\cite{villa_oran_arc}                                                    & OAI          & \cmark & --              & HW      & --     & --      & \cmark & \cmark & \cmark          \\
    Sionna RK~\cite{nvidia_sionna_rk}                                                  & OAI          & --     & --              & RT+HW   & --     & --      & --     & \cmark & \cmark          \\
    CellSense~\cite{kumar_cellsense}                                                   & 5G OFDM      & --     & passive         & sim+HW  & --     & \cmark  & --     & \cmark & --              \\
    OFDM/passive radar~\cite{nuss_5gnr_radar,falcone_dvbt_uav,zhang_passive_radar_lte} & --           & --     & OFDM/passive    & HW      & \cmark & \cmark  & --     & --     & --              \\
    \midrule
    \textbf{This work}                                                                 & \textbf{OAI} & \cmark & \textbf{UL-SRS} & RT      & \cmark & \cmark  & \cmark & \cmark & --$^{\ddagger}$ \\
    \bottomrule
  \end{tabular}
\end{table*}

\subsection{ISAC in 5G/6G Networks}

ISAC has been extensively studied as a key enabler of 6G
\cite{liu_survey_isac, zhang_6g_isac, mu_noma_isac}. Seminal
work by Liu \emph{et al.}~\cite{liu_mu_miso_beamforming}
demonstrated joint beamforming design for dual-function
radar-communication (DFRC) systems. Subsequent contributions
addressed waveform design, interference management, and
performance bounds \cite{sturm_ofdm_isac, braun_ofdm_radar}.
The 3GPP standardization community formalized ISAC in
TR~22.837~\cite{3gpp_tr22837} (sensing use cases) and
TR~38.867~\cite{3gpp_tr38867} (the Release-19 ISAC channel-model study item),
which identify UL-SRS as a \emph{candidate} sensing signal for range and
velocity estimation in the sub-6~GHz and mmWave bands.
Concurrent work by Kumar \emph{et al.}, CellSense~\cite{kumar_cellsense},
targets clutter-robust passive sensing in the sub-6~GHz cellular
band, validated via both Sionna-based OFDM link-level simulation
and a USRP (Universal Software Radio Peripheral) hardware prototype -- a validation methodology directly
complementary to this paper's Sionna RT-based simulation, with
USRP hardware validation as this testbed's own planned next step
(Section~\ref{sec:discussion}).

\subsection{Passive UAV Detection}

UAV detection leveraging existing cellular waveforms has gained
momentum as networks approach ubiquitous deployment
\cite{khawaja_uav_survey}. Zhang \emph{et
  al.}~\cite{zhang_passive_radar_lte} demonstrated passive radar
using LTE downlink signals for micro-drone detection at ranges up
to \SI{500}{\metre}. Falcone \emph{et
  al.}~\cite{falcone_dvbt_uav} exploited DVB-T signals for
multicopter detection. Drone signature extraction using Doppler
micro-Doppler signatures of rotor blades has been validated in
\cite{rahman_micro_doppler} and \cite{chen_micro_doppler_cnn}
using both radar and cellular signals. In the NR domain, an
OFDM-waveform radar has been proposed \cite{nuss_5gnr_radar}
but full-stack integration with O-RAN remains undemonstrated.

\subsection{O-RAN Testbeds and xApp Research}

OpenAirInterface (OAI) provides an open-source, 3GPP-compliant
NR stack widely used in research testbeds
\cite{kaltenberger_oai}. FlexRIC
\cite{schmidt_flexric} implements the O-RAN Near-RT RIC with E2
interface support and a programmable xApp SDK. ColO-RAN
\cite{polese_oai_flexric} demonstrated data-driven RAN control at
scale, while a broader architectural treatment of O-RAN interfaces,
algorithms, and open research challenges is given in
\cite{polese_coloran}. Data-driven O-RAN experimentation frameworks
\cite{bonati_openrangym} and machine-learning channel datasets such as
DeepMIMO~\cite{alkhateeb_deepmimo} have advanced data-driven RAN
research but do not address ISAC applications.
Sionna~\cite{hoydis_sionna} provides a GPU-accelerated link-level
simulation framework with ray-tracing (Sionna RT
\cite{hoydis_sionna_rt}) that has been used for channel
modeling; its integration with a live NR stack for ISAC purposes
is novel. NVIDIA's Sionna Research Kit~\cite{nvidia_sionna_rk}
similarly couples Sionna with a live OAI-based stack on
GPU-accelerated hardware, but targets AI-RAN neural-receiver
research on communication-side key performance indicators (KPIs), not sensing -- this work
instead uses a CPU-only, containerized OAI+Sionna RT stack
for active UAV ISAC sensing.

Separately, a growing body of work builds \emph{wireless digital
  twins} that pair a live or emulated RAN stack with ray-traced or
otherwise site-specific channel models
\cite{khan_digitaltwin_survey}. Colosseum~\cite{polese_colosseum_twin}
couples a large-scale RF emulator with a digital-twin control plane
for O-RAN experimentation, and Boston
Twin~\cite{testolina_bostontwin} provides a city-scale ray-tracing
digital twin for 6G network studies. Villa \emph{et
  al.}~\cite{villa_oran_arc} integrate OAI with an NVIDIA-accelerated
O-RAN testbed across multiple vendors. Closest to our own use of
Sionna RT, the Open Wireless Digital Twin~\cite{iye_owdt} couples
OAI and Sionna RT under FlexRIC to emulate 5G mobility and monitor
RAN KPIs (reference signal received power [RSRP], modulation and coding scheme [MCS], block error rate [BLER], throughput) via the Near-RT RIC --
architecturally the same three-layer stack (OAI + FlexRIC + Sionna
RT) as this paper, but applied to communication-side mobility
emulation rather than active sensing. Concurrently, the \emph{dApp} concept~\cite{polese_dapp} proposes
co-locating sub-\SI{10}{\milli\second} inference microservices directly inside
the O-DU alongside the physical layer (PHY), providing IQ-sample access that the E2 interface was
never designed to carry. The sensing-dApp architecture of~\cite{oran_isac_dapp}
formalizes this for monostatic ISAC: it specifies sensing dApps at the O-DU for
delay, Doppler, and angle extraction, a companion E2SM-SENS service model for
structured sensing telemetry, and the Open Fronthaul metadata required for
waveform--echo association. This work implements a bistatic sensing pipeline and
a parallel E2SM-SENS transport independently, making the sensing PHY stage
dApp-ready in principle while deferring a standards-compliant lifecycle to
hardware integration (Section~\ref{sec:future}).

\textbf{Gap.} As Table~\ref{tab:related} makes explicit, the three-layer
OAI\,+\,FlexRIC\,+\,Sionna~RT stack is not itself novel --- the Open Wireless
Digital Twin~\cite{iye_owdt}, Colosseum~\cite{polese_colosseum_twin} and Boston
Twin~\cite{testolina_bostontwin} all pair emulated or ray-traced RANs with a RIC
control plane. What every prior platform instruments is the
\emph{communication} plane --- RSRP, MCS, BLER, throughput. What this work adds
is the \emph{sensing} plane: a UL-SRS radar processing chain inside the gNB PHY,
a bistatic tracker in the Near-RT RIC delivered over a custom E2SM-SENS service
model, and an evaluation reported in sensing terms --- range, range-rate and
bearing accuracy against ray-traced ground truth, together with the geometric
limits that bound them. To our knowledge this is the first end-to-end 5G O-RAN
ISAC simulation testbed work for UAV tracking with extensive pipeline validation
and identification of key limitations towards practical deployment.

%% =========================================================
\section{System Model}
\label{sec:sysmodel}

\subsection{Coordinate System and Conventions}
\label{sec:coords}

All positions are expressed in a local east--north--up (ENU) frame:
$\hat{\mathbf{x}}$ points east, $\hat{\mathbf{y}}$ north and
$\hat{\mathbf{z}}$ up. The gNB receive array is a uniform linear array (ULA)
whose elements lie along $\hat{\mathbf{y}}$, so its boresight is
$+\hat{\mathbf{x}}$. Azimuth is measured at the \emph{array phase center} ---
i.e.\ at the antenna, at height $h_{\mathrm{gNB}}$ above the mast foot, which is
where the bearing is physically subtended --- from the
$+\hat{\mathbf{x}}$ boresight towards $+\hat{\mathbf{y}}$: a target whose
horizontal offset from the phase center is $(\Delta x, \Delta y)$ therefore has
\begin{equation}
  \theta = \mathrm{atan2}(\Delta y,\; \Delta x),
  \label{eq:az_conv}
\end{equation}
and the array's unambiguous sector is $\theta \in [-90^\circ, +90^\circ]$
(Fig.~\ref{fig:sysmodel}). Placing the array broadside to the target corridor
this way makes the inter-element phase ramp signed and monotonic as the target
crosses boresight; an endfire ($+\hat{\mathbf{y}}$-boresight) orientation would
leave the bearing sign ambiguous over the same corridor.

\subsection{Signal Model}

Consider a single 5G NR gNB equipped with $N_{\mathrm{rx}} = 4$
receive antennas operating on Band~n78 at carrier frequency
$f_c = \SI{3319.68}{\mega\hertz}$ (\SI{3.32}{\giga\hertz}) with
system bandwidth $B = \SI{40}{\mega\hertz}$ over 106 physical
resource blocks (PRBs) at 30~kHz subcarrier spacing (SCS). A UAV at 3D position $\mathbf{p}_T(t)$ and velocity
$\mathbf{v}_T(t)$ acts as a non-cooperative passive scatter
target while a registered nrUE transmits UL-SRS as part of the
normal NR uplink procedure. The received baseband signal at the
$n$-th gNB receive antenna after cyclic-prefix removal is

\begin{equation}
  y_n[k,l] = \sum_{p=1}^{P} h_{n,p}[k,l]\, s[k,l] + w_n[k,l],
  \label{eq:rx_signal}
\end{equation}

\noindent where $s[k,l]$ is the SRS orthogonal frequency-division
multiplexing (OFDM) symbol on subcarrier
$k$ and SRS occasion $l$, $h_{n,p}[k,l]$ is the channel
coefficient of the $p$-th propagation path, $P$ is the total number of paths, and
$w_n[k,l] \sim \mathcal{CN}(0,\sigma^2)$ is complex additive white
Gaussian noise (AWGN).

\emph{Path budget in this testbed.} The ray tracer returns $P = 3$ in the flat-LOS benchmark:
\begin{enumerate}
  \item the \emph{direct} UE$\to$gNB path (LOS);
  \item one \emph{static reflection} --- here the ground bounce --- retained as
        the strongest surviving non-target path; and
  \item the \emph{UAV echo}, the bistatic path of interest.
\end{enumerate}
Paths (1) and (2) are static and share the zero-Doppler subspace that
Section~\ref{sec:clutter} deflates; path~(3) is the only Doppler-bearing return,
which is precisely why a rank-$K_c = 1$ deflation suffices in this scene and why
a higher rank removes the target itself. This $P=3$ budget is compute-conscious,
but for the flat-LOS benchmark it is also \emph{physically complete}, as quantified
at the end of this subsection.

Two simplifications follow from this budget. First, only the two strongest static paths are retained, so a real
scene's weaker multipath is absent: the clutter the canceller faces is
optimistic. Second, all rays that scatter off the UAV are \emph{coherently
  summed into a single echo tap}, so the target is rendered as one point scatterer
rather than as a body with extent --- fuselage multi-bounce and rotor
micro-Doppler are not modeled. In the street-canyon scene the two static taps
persist, but the UAV echo is absent entirely whenever the target is occluded
(Appendix~\ref{sec:nlos}), so there $P$ drops to $2$ and no pipeline could detect
the target.

\emph{Sufficiency of the three-path budget in flat-LOS.} Three
checks establish that two static taps plus one echo per target is the complete
channel for this benchmark.
\emph{(i)~The taps sit well inside the cyclic prefix.} At \SI{30}{\kilo\hertz} SCS
the normal CP spans \SI{2.34}{\micro\second}, a maximum excess path of
\SI{703}{\metre}; the scene's excess bistatic path lengths of
\SIrange{40}{85}{\metre} occupy under \SI{12}{\percent} of that window, so the
rendered paths are never delay-limited and the path count could be raised freely on
this axis.
\emph{(ii)~Two static taps retain \emph{all} of the static channel here.} A
per-CPI truncation-headroom audit---comparing, at each snapshot, the strongest
\emph{discarded} non-target path against the weakest \emph{retained} one---discards
\SI{0}{\percent} of static path energy at two taps in the flat-LOS scene, because
the ray tracer returns only the LOS and ground-bounce non-target paths: two taps is
the entire static channel, not a subset. The identical audit on the street-canyon
scene discards \SI{64}{\percent} of static energy at two taps (the strongest dropped
path within \SI{1.7}{\decibel} of the weakest kept), which is why
Appendix~\ref{sec:nlos} treats the buildings result as a lower bound.
\emph{(iii)~One echo tap matches the achievable resolution.} With bistatic range
resolution $\Delta R_{\mathrm{res}} = c/B \approx \SI{8.0}{\metre}$
(Eq.~\eqref{eq:dr}), the range cell is far larger than a multirotor's
physical extent, so intra-body multi-bounce is not separable in delay and a
point-scatterer echo is the correct rendering at this bandwidth; for the
\SIrange{-20}{-10}{\dBsm} small-to-mid targets studied here the residual multi-bounce
return lies below the noise floor regardless.

The UL-SRS sequence follows the 3GPP NR specification
\cite{3gpp_ts38211}:

\begin{equation}
  s[k] = r^{(\alpha,\delta)}(k)\, e^{j\phi_{\mathrm{seq}}(k)},
  \quad k = 0, \ldots, M_{\mathrm{SRS}}-1,
  \label{eq:srs_seq}
\end{equation}

\noindent where $r^{(\alpha,\delta)}(k)$ is a Zadoff-Chu root
sequence with cyclic shift $\alpha$. The SRS is transmitted on a
comb-2 pattern (every second subcarrier), so
$M_{\mathrm{SRS}} = 624$ tones are occupied across the
$106~\text{PRBs}\times 12~\text{SC/PRB} = 1272$ subcarriers at
\SI{30}{\kilo\hertz} SCS. The effective sensing bandwidth is the
occupied span $B \approx 1272\times\SI{30}{\kilo\hertz}
  \approx\SI{38.2}{\mega\hertz}$ (taken as $B = \SI{40}{\mega\hertz}$
by the gNB PHY sensing module). A comb least-squares extractor packs the
$M_{\mathrm{SRS}}$ occupied tones gap-free before range processing; the
ISAC IFFT for range profiling is zero-padded to 4096 points.

\subsection{Bistatic Geometry and Range-Doppler Processing}

The testbed operates in a bistatic configuration:
the nrUE at position $\mathbf{p}_U$ transmits UL-SRS; the gNB
at position $\mathbf{p}_G$ receives the direct signal and the
UAV bistatic echo. The bistatic range is

\begin{equation}
  R_{\mathrm{bi}} = |\mathbf{p}_T - \mathbf{p}_G|
  + |\mathbf{p}_T - \mathbf{p}_U|,
  \label{eq:rbi}
\end{equation}

\noindent i.e.\ the \emph{total} path length of the echo, the sum of the two
legs. It is not a one-way range, and this distinction sets the range resolution
in~\eqref{eq:dr} below.

\emph{Excess path as the measured observable.} The gNB recovers the echo
delay relative to the direct path, which arrives first and is used as the
timing reference. What the sensing stage reports is therefore the \emph{excess}
bistatic range
\begin{equation}
  \Delta R_{\mathrm{bi}} = R_T + R_R - L,
  \qquad L = |\mathbf{p}_G - \mathbf{p}_U|,
  \label{eq:excess}
\end{equation}
with $L$ the (known) baseline. Since the transmitter and receiver positions are
known, the tracker restores the total path length,
$\hat{R}_{\mathrm{bi}} = \Delta\hat{R}_{\mathrm{bi}} + L$, before applying the
measurement model~\eqref{eq:h1}. The two
quantities differ by \SI{130}{\metre} in this testbed, and a range observable
is not interpretable without stating which convention it follows.

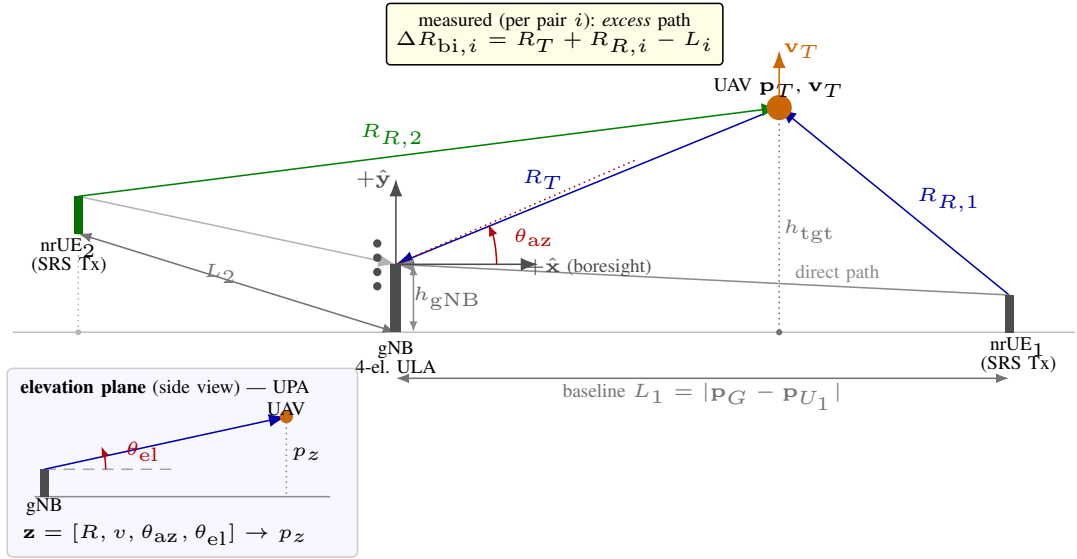
\begin{figure*}[!tbp]
  \centering
  \resizebox{0.78\textwidth}{!}{%
    \begin{tikzpicture}[scale=1.0]
      %% --- ground plane / positions -------------------------------------
      %% G  = gNB mast foot (ground);  Ga = gNB ANTENNA phase centre.
      %% Everything the gNB measures -- the echo legs, the direct paths, and the
      %% azimuth -- is referenced to Ga, NOT to the ground point: azimuth is the
      %% bearing seen at the array, which sits at h_gNB above the base.
      %% Multistatic: TWO nrUEs (nrUE_1 right, nrUE_2 left) share the ONE gNB Rx.
      \coordinate (G)  at (0,0);       % gNB mast foot
      \coordinate (Ga) at (0,0.62);    % gNB antenna (array phase center)
      \coordinate (U)  at (5.6,0);     % nrUE_1
      \coordinate (Ua) at (5.6,0.34);  % nrUE_1 antenna
      \coordinate (U2)  at (-2.9,0.9);   % nrUE_2 elevated (background plane -> pseudo-3D)
      \coordinate (U2a) at (-2.9,1.24);  % nrUE_2 antenna
      \coordinate (U2g) at (-2.9,0);     % nrUE_2 ground projection
      \coordinate (Tg) at (3.5,0);     % UAV ground projection
      \coordinate (T)  at (3.5,2.05);  % UAV

      \draw[black!25] (-3.5,0) -- (6.2,0);

      %% --- bistatic baselines (one per pair) ----------------------------
      \draw[{Latex[length=1.2mm]}-{Latex[length=1.2mm]}, black!55]
      ($(G)+(0,-0.42)$) -- node[slbl, below, pos=0.5]
        {baseline $L_1=|\mathbf{p}_G-\mathbf{p}_{U_1}|$} ($(U)+(0,-0.42)$);
      %% L_2 is a slanted baseline to the elevated (background) nrUE_2, with a
      %% dotted drop-line to its ground projection -- reads as depth / pseudo-3D.
      \draw[{Latex[length=1.2mm]}-{Latex[length=1.2mm]}, black!55]
      (G) -- node[slbl, above left=-0.6mm, pos=0.5] {$L_2$} (U2);
      \draw[densely dotted, black!30] (U2g) -- (U2);
      \fill[black!30] (U2g) circle (0.026);

      %% --- pair 1 (nrUE_1) echo legs + direct path (at ANTENNA height) ---
      \draw[flow, blue!65!black] (Ua) -- node[slbl, above right, pos=0.42]
      {$R_{R,1}$} (T);
      \draw[flow, blue!65!black] (T) -- node[slbl, above left, pos=0.55]
      {$R_T$} (Ga);
      \draw[flow, black!45] (Ua) -- node[slbl, above, pos=0.28]
      {direct path} (Ga);

      %% --- pair 2 (nrUE_2) echo leg + direct path (multistatic) ---------
      \draw[flow, green!50!black] (U2a) -- node[slbl, above left, pos=0.5]
      {$R_{R,2}$} (T);
      \draw[flow, black!30] (U2a) -- (Ga);

      %% --- gNB mast + its 4-element ULA at the top -----------------------
      \fill[black!70] (G) ++(-0.05,0) rectangle ++(0.10,0.62);
      \foreach \k in {0,1,2,3}{\fill[black!70] ($(Ga)+(-0.17,-0.20+0.13*\k)$)
          circle (0.038);}
      \draw[{Latex[length=1.0mm]}-{Latex[length=1.0mm]}, black!45]
      ($(G)+(0.16,0)$) -- node[slbl, right=-0.4mm, black!60]
        {$h_{\mathrm{gNB}}$} ($(Ga)+(0.16,0)$);
      \node[slbl, below=0.4mm of G] {gNB\\[-1pt]4-el.\ ULA};

      %% --- nrUE_1 with its own array ------------------------------------
      \fill[black!70] (U) ++(-0.04,0) rectangle ++(0.08,0.34);
      \node[slbl, below right=0pt and -3mm of U] {nrUE$_1$\\[-1pt](SRS Tx)};

      %% --- nrUE_2 (second bistatic pair) --------------------------------
      \fill[green!45!black] (U2) ++(-0.04,0) rectangle ++(0.08,0.34);
      \node[slbl, below left=0pt and -3mm of U2] {nrUE$_2$\\[-1pt](SRS Tx)};

      %% --- UAV -----------------------------------------------------------
      \fill[orange!80!black] (T) circle (0.115);
      \node[slbl, above=0.5mm of T] {UAV $\mathbf{p}_T$, $\mathbf{v}_T$};
      \draw[flow, orange!80!black] (T) -- ++(0,0.52)
      node[slbl, right=0pt] {$\mathbf{v}_T$};
      \draw[densely dotted, black!55] (T) -- node[slbl, right, pos=0.55]
      {$h_{\mathrm{tgt}}$} (Tg);
      \fill[black!55] (Tg) circle (0.028);

      %% --- azimuth: measured AT THE ANTENNA, from +x (east) toward +y ----
      \draw[flow, black!70] (Ga) -- ++(1.30,0)
      node[slbl, below right=-1mm and -1.6mm] {$+\hat{\mathbf{x}}$ (boresight)};
      \draw[flow, black!70] (Ga) -- ++(0,0.78) node[slbl, left=0pt] {$+\hat{\mathbf{y}}$};
      \draw[-{Latex[length=1.1mm]}, red!70!black]
      ($(Ga)+(0.92,0)$) arc[start angle=0, end angle=23.5, radius=0.92];
      \node[slbl, red!70!black] at (1.26,0.87) {$\theta_{\mathrm{az}}$};
      \draw[densely dotted, red!70!black] (Ga) -- ($(Ga)+(23.5:2.4)$);

      %% --- the measured observable ---------------------------------------
      \node[slbl, align=center, draw, rounded corners=1pt, fill=yellow!12,
        inner sep=2.4pt] at (1.45,2.72)
      {measured (per pair $i$): \emph{excess} path\\[-1pt]
        $\Delta R_{\mathrm{bi},i} = R_T + R_{R,i} - L_i$};

      %% --- INSET: the elevation (vertical) plane the UPA additionally resolves.
      %% The main view is the azimuth/horizontal plane (bearing theta); drawing
      %% theta_el there would collide with theta at the same vertex, so the
      %% vertical plane is shown separately here. Placed lower-left (empty space),
      %% within the main width -> does not rescale the main diagram.
      \begin{scope}[shift={(-3.55,-2.05)}]
        \draw[black!20, rounded corners=1.5pt, fill=blue!3] (0,0) rectangle (3.2,1.72);
        \node[slbl, anchor=north west] at (0.08,1.66)
        {\textbf{elevation plane} (side view) --- UPA};
        \draw[black!45] (0.25,0.55) -- (2.95,0.55);                 % horizon
        \fill[black!70] (0.30,0.55) rectangle (0.38,0.80);          % gNB (UPA)
        \node[slbl, below=-0.3mm] at (0.34,0.55) {gNB};
        \draw[densely dashed, black!35] (0.34,0.80) -- (1.5,0.80);  % horizontal ref
        \coordinate (iu) at (2.55,1.28);
        \draw[flow, blue!65!black] (0.34,0.80) -- (iu);             % LOS gNB->UAV
        \fill[orange!80!black] (iu) circle (0.06);
        \node[slbl, above=-0.3mm] at (iu) {UAV};
        \draw[-{Latex[length=1mm]}, red!70!black] (0.90,0.80)
        arc[start angle=0, end angle=22, radius=0.56];            % elevation angle
        \node[slbl, red!70!black] at (1.24,0.965) {$\theta_{\mathrm{el}}$};
        \draw[densely dotted, black!45] (iu) -- (2.55,0.55);        % altitude
        \node[slbl] at (2.74,0.94) {$p_z$};
        \node[slbl, anchor=south west] at (0.10,0.07)
        {$\mathbf{z}=[R,v,\theta_{\mathrm{az}},\theta_{\mathrm{el}}]\to p_z$};
      \end{scope}
    \end{tikzpicture}}
  \caption{UL-SRS sensing geometry with the multistatic extension. Each nrUE
    transmits SRS; the single gNB receives, for each transmitter, the direct path
    and the UAV echo, and measures delay \emph{relative to the direct path}, so the
    per-pair observable is the excess bistatic path
    $\Delta R_{\mathrm{bi},i} = R_T + R_{R,i} - L_i$ (the tracker restores the total
    path by adding the known baseline $L_i$). Azimuth $\theta_{\mathrm{az}}$ is the angle of
    arrival at the gNB's own 4-element ULA, from the $+\hat{\mathbf{x}}$ boresight
    towards $+\hat{\mathbf{y}}$ (Section~\ref{sec:coords}), subtended at the
    \emph{array phase center} at antenna height $h_{\mathrm{gNB}}$. The baseline
    single pair (nrUE$_1$) gives no elevation observability, so target height
    $h_{\mathrm{tgt}}$ is a prior. Two independent upgrades can make it observable.
    Adding a second nrUE (nrUE$_2$) forms a second bistatic pair sharing the gNB
    receiver, whose range diversity fixes altitude through geometry alone
    (Section~\ref{sec:multistatic}; measured, and the most accurate of the
    configurations reported). Alternatively, a planar array (UPA) at the gNB
    resolves the \emph{orthogonal} vertical plane shown in the inset: it measures an
    elevation angle $\theta_{\mathrm{el}}$, extending the per-pair measurement to
    $[R, v, \theta_{\mathrm{az}}, \theta_{\mathrm{el}}]$ so the 3-D EKF estimates
    altitude $p_z$ directly (Section~\ref{sec:upa}). Either way $h_{\mathrm{tgt}}$
    is estimated rather than assumed.}
  \label{fig:sysmodel}
\end{figure*}

\emph{Per-pair form.} Section~\ref{sec:multistatic} adds a second transmitter,
so~\eqref{eq:excess} is written in the indexed form used from here on and in
Fig.~\ref{fig:sysmodel}. With $i$ indexing the transmitters, all sharing the one gNB
receiver,
\begin{equation}
  \Delta R_{\mathrm{bi},i} = R_T + R_{R,i} - L_i,
  \qquad L_i = |\mathbf{p}_G - \mathbf{p}_{U_i}|,
  \label{eq:excess_i}
\end{equation}
where $R_T$ --- the gNB-to-target leg --- is \emph{common} to every pair, while the
target-to-receiver leg $R_{R,i}$ and the baseline $L_i$ are not. That asymmetry is
what makes a second transmitter informative about altitude: two excess-path
measurements sharing $R_T$ constrain the target to the intersection of two
ellipsoids with a common focus, which has vertical curvature that one ellipsoid alone
does not. The single-pair sensor of this section is~\eqref{eq:excess_i} at $i=1$, and
every quantity in \eqref{eq:excess} carries over unchanged.

The bistatic range rate follows by differentiating~\eqref{eq:rbi} directly. Let
\begin{equation}
  \hat{\mathbf{u}}_G = \frac{\mathbf{p}_T-\mathbf{p}_G}{|\mathbf{p}_T-\mathbf{p}_G|},
  \qquad
  \hat{\mathbf{u}}_U = \frac{\mathbf{p}_T-\mathbf{p}_U}{|\mathbf{p}_T-\mathbf{p}_U|}
  \label{eq:ugu}
\end{equation}
be the unit vectors pointing from the gNB and from the nrUE \emph{towards the
  target} --- the position gradients of the two range terms in~\eqref{eq:rbi}.
Differentiating~\eqref{eq:rbi} gives the rate of change of the bistatic path,
$\dot{R}_{\mathrm{bi}} = \mathbf{v}_T\cdot(\hat{\mathbf{u}}_G+\hat{\mathbf{u}}_U)$:
the \emph{sum} appears, because either range term grows as the target recedes
from its station.

We report Doppler in the usual radar sense, positive for a \emph{closing} target,
so the reported observable is the negated path rate:
\begin{align}
  v_{\mathrm{bi}} \triangleq -\dot{R}_{\mathrm{bi}}
   & = -\,\mathbf{v}_T \cdot
  \left(\hat{\mathbf{u}}_G + \hat{\mathbf{u}}_U\right) \nonumber \\
   & = -2\,|\mathbf{v}_T|\,\cos\psi\,\cos(\beta/2)
  \;=\; \lambda f_D ,
  \label{eq:vbi}
\end{align}

\noindent with $\beta$ the bistatic angle subtended at the target (between
$\hat{\mathbf{u}}_G$ and $\hat{\mathbf{u}}_U$) and $\psi$ the angle between
$\mathbf{v}_T$ and the bisector of those two directions. Equivalently, in terms
of the propagation directions --- the incident ray
$\hat{\mathbf{k}}_{\mathrm{in}}$ (UE$\to$UAV) and the scattered ray
$\hat{\mathbf{k}}_{\mathrm{out}}$ (UAV$\to$gNB), for which
$\hat{\mathbf{u}}_U = \hat{\mathbf{k}}_{\mathrm{in}}$ and
$\hat{\mathbf{u}}_G = -\hat{\mathbf{k}}_{\mathrm{out}}$ ---
\begin{equation}
  v_{\mathrm{bi}} = \mathbf{v}_T\cdot
  \left(\hat{\mathbf{k}}_{\mathrm{out}} - \hat{\mathbf{k}}_{\mathrm{in}}\right).
  \label{eq:vbi_k}
\end{equation}
Both forms are given because the sign turns entirely on which is meant, and a
Doppler reported without its convention is not interpretable; the
scattered-ray form~\eqref{eq:vbi_k} is included only for physical intuition,
while the projected form~\eqref{eq:vbi} is the one used throughout the
implementation. Under this
convention the transiting target of Section~\ref{sec:results} approaches
($v_{\mathrm{bi}}>0$), passes through zero at the broadside crossing, and recedes
($v_{\mathrm{bi}}<0$).

Throughout, $v_{\mathrm{bi}}$ denotes this \emph{projected} quantity, not the
target's speed: recovering $|\mathbf{v}_T|$ from a single pair would
require the geometry factor $2\cos\psi\cos(\beta/2)$, and $\psi$ is not
observable from one bistatic measurement. As the target approaches the baseline,
$\beta \to 180^\circ$ and $\cos(\beta/2) \to 0$: the geometry degenerates, and
neither range nor Doppler is observable there. This near-baseline blind zone is
a property of the geometry, not of the processing, and coherent processing
intervals (CPIs) falling inside it
are excluded from the measurement statistics of
Section~\ref{sec:results}.

After clutter-subspace deflation and isolation of
$\hat{h}^{\mathrm{UAV}}[k,l]$, a 2D OFDM Range-Doppler map is
computed by a zero-padded inverse discrete Fourier transform (DFT) over the $K$ subcarriers
(range) followed by a DFT over the $N_{\mathrm{occ}}$ SRS occasions (Doppler)
within one CPI:

\begin{equation}
  \mathcal{X}[m,n] = \sum_{l=0}^{N_{\mathrm{occ}}-1}\!\left(\sum_{k=0}^{K-1}
  \hat{h}^{\mathrm{UAV}}[k,l]\,e^{+j2\pi\frac{mk}{K}}\right)\!
  e^{-j2\pi\frac{nl}{N_{\mathrm{occ}}}}.
  \label{eq:rd_map}
\end{equation}

\noindent The inner transform is an inverse DFT because the delay/range
profile of a frequency-domain channel is $\mathrm{IDFT}\{\hat{h}\}$.

The resulting resolutions are:

\begin{equation}
  \Delta R_{\mathrm{res}} = \frac{c}{B_{\mathrm{SRS}}} \approx \SI{8.0}{\metre}
  \quad\text{(bistatic range resolution)},
  \label{eq:dr}
\end{equation}

\noindent where $B_{\mathrm{SRS}}$ is the \emph{occupied SRS frequency aperture}
--- the span from the first to the last transmitted SRS tone --- and not the
nominal channel bandwidth. The sounding signal carries $M_{\mathrm{SRS}} = 624$
comb-2 tones spaced $k_{\mathrm{TC}}\Delta f = \SI{60}{\kilo\hertz}$, so
$B_{\mathrm{SRS}} = (M_{\mathrm{SRS}}-1)\,k_{\mathrm{TC}}\Delta f \approx
  \SI{37.4}{\mega\hertz}$. Five quantities are kept distinct here: the
\SI{40}{\mega\hertz} nominal channel bandwidth, the \SI{38.16}{\mega\hertz}
carrier occupied bandwidth (106~PRB), the \SI{37.4}{\mega\hertz} SRS aperture
that actually sets resolution, the \SI{1.22}{\metre} zero-padded FFT bin spacing
(below), and the resulting \SI{8.0}{\metre} physical range resolution. Deriving
the cell from the SRS aperture rather than the nominal \SI{40}{\mega\hertz} is
what moves it from the \SI{7.5}{\metre} of earlier versions to \SI{8.0}{\metre}.

\begin{equation}
  \Delta v = \frac{\lambda}{T_{\mathrm{CPI}}}
  = \frac{0.0903}{0.64}
  \approx \SI{0.14}{\metre\per\second},
  \label{eq:dv}
\end{equation}

\noindent the bistatic Doppler resolution,
where $\lambda = c/f_c = \SI{0.0903}{\metre}$ and
$T_{\mathrm{CPI}} = N_{\mathrm{occ}} \cdot T_{\mathrm{SRS}} = 64 \times
  \SI{10}{\milli\second} = \SI{640}{\milli\second}$ is the coherent
processing interval.

\emph{Resolution conventions and the absence of a factor of two.} A delay resolution
$\Delta\tau = 1/B$ maps to a resolution in the measured quantity through
$\Delta R_{\mathrm{res}} = c\,\Delta\tau = c/B$, because
$R_{\mathrm{bi}}$~\eqref{eq:rbi} is a total path length. The familiar
$\Delta R = c/(2B)$ is a \emph{monostatic} convention: it applies when the
measured delay is a round trip that is then halved to a one-way range, which is
not the case here. The same reasoning applies to Doppler, where the bistatic
formula likewise carries no factor of 2 since the UE transmitter is not
co-located with the gNB receiver.

Resolution and bin spacing are distinct quantities, and are easily conflated. The range transform is zero-padded to $N_{\mathrm{FFT}} = 4096$
points over tones spaced $k_{\mathrm{TC}}\Delta f = \SI{60}{\kilo\hertz}$
(comb-2 at \SI{30}{\kilo\hertz} SCS), giving a bin spacing of
$\delta_r = c/(N_{\mathrm{FFT}} k_{\mathrm{TC}} \Delta f) = \SI{1.22}{\metre}$.
That is an \emph{interpolation} step, not a resolution: it sets how finely the
peak of the range response can be located, whereas $\Delta R_{\mathrm{res}}$
sets how far apart two scatterers must be to appear as separate peaks.
Consequently a range accuracy \emph{better} than
$\Delta R_{\mathrm{res}}$ is possible in principle for an isolated, high
signal-to-noise ratio (SNR)
scatterer. In practice, however, the measured range accuracy is set by
the transit geometry rather than by peak-location precision, and lands at about
one resolution cell (Section~\ref{sec:results}).

\subsection{CFAR Detection}
\label{sec:cfar}

Cell-averaging constant false-alarm rate (CA-CFAR)~\cite{skolnik_radar_handbook,mahafza_radar_systems} detection is applied to
$|\mathcal{X}[m,n]|^2$ to detect UAV targets with a controlled
false alarm rate $P_{\mathrm{FA}}$. The detection threshold for
guard windows of $(N_{g,R}, N_{g,D})$ cells and
training windows of $(N_{t,R}, N_{t,D})$ cells in the
(range, Doppler) dimensions is

\begin{equation}
  \eta = \alpha_{\mathrm{CA}} \cdot
  \frac{1}{N_{\mathrm{ref}}}\sum_{(m',n') \in \mathcal{W}}
  |\mathcal{X}[m',n']|^2,
  \label{eq:cfar}
\end{equation}

\noindent where $N_{\mathrm{ref}} = (2N_{t,R}+1)(2N_{t,D}+1)
  - (2N_{g,R}+1)(2N_{g,D}+1)$ and, for the cell-averaging estimator under an
exponentially-distributed noise power, the scaling factor that attains a
design false-alarm rate $P_{\mathrm{FA}}$ is
$\alpha_{\mathrm{CA}} = N_{\mathrm{ref}}
  \left(P_{\mathrm{FA}}^{-1/N_{\mathrm{ref}}} - 1\right)$.

The implementation, however, uses the order-statistic variant
(OS-CFAR)~\cite{rohling_oscfar}: it replaces the training-cell mean
in~\eqref{eq:cfar} with the $k$-th order statistic of the training ring
($k$ at the $75^{\mathrm{th}}$ percentile), so that a target's own main-lobe
leakage into the ring cannot inflate the threshold. The scaling factor $\alpha_{\mathrm{CA}}$ does not carry over: OS-CFAR has its own scaling
law, $P_{\mathrm{FA}} = \prod_{i=0}^{k-1}
  \frac{N_{\mathrm{ref}}-i}{N_{\mathrm{ref}}-i+\alpha_{\mathrm{OS}}}$, and applying
the cell-averaging factor to an order-statistic threshold does not realize the
nominal rate. The threshold used here is therefore set empirically rather than
from a closed-form $P_{\mathrm{FA}}$, and the $10^{-4}$ figure in
Table~\ref{tab:params} should be read as the design target of the underlying
CA formulation, not as a calibrated false-alarm rate. The realized
$P_{\mathrm{FA}}$ on target-absent data, a calibrated variant that attains the
nominal rate, and the detection probability $P_{\mathrm{D}}$ against echo SNR
and range are characterized in Appendix~\ref{sec:detchar}.

\subsection{Bistatic EKF Tracker}

Detections $(R_{\mathrm{bi}}, v_{\mathrm{bi}}, \theta)$ are
associated and tracked using an Extended Kalman
Filter (EKF)~\cite{barshalom_tracking_estimation} with
Cartesian constant-velocity state
$\mathbf{x} = [p_x,\; p_y,\; v_x,\; v_y]^\top$. The nonlinear
measurement function $\mathbf{h}: \mathbb{R}^4 \to \mathbb{R}^3$
maps state to predicted bistatic observables:

\begin{align}
  h_1(\mathbf{x}) & = |\mathbf{p} - \mathbf{p}_G|
  + |\mathbf{p} - \mathbf{p}_U|, \label{eq:h1}          \\
  h_2(\mathbf{x}) & = -\,\mathbf{v} \cdot
  \bigl(\hat{\mathbf{u}}_G(\mathbf{p})
  + \hat{\mathbf{u}}_U(\mathbf{p})\bigr), \label{eq:h2} \\
  h_3(\mathbf{x}) & = \mathrm{atan2}(p_y - p_{G,y},\;
  p_x - p_{G,x}), \label{eq:h3}
\end{align}

\noindent where $h_1$ takes the total path length --- the measured excess range
of~\eqref{eq:excess} having had the baseline restored ---
$\mathbf{p} = [p_x, p_y, h_{\mathrm{tgt}}]^\top$
augments the 2-D state with a fixed height prior $h_{\mathrm{tgt}}$
(the 2-D EKF does not observe elevation), and the azimuth $h_3$ follows the ENU
convention of~\eqref{eq:az_conv}, referenced to the gNB position
$\mathbf{p}_G$ and matching the gNB's own 4-element Rx ULA, at which the
bearing is estimated as an angle of arrival (AoA) on the receive side.
The EKF predict-update cycle follows the standard formulation

\begin{align}
  \mathbf{x}_{t|t-1} & = \mathbf{F}\,\mathbf{x}_{t-1|t-1}, \label{eq:kf_pred} \\
  \mathbf{P}_{t|t-1} & = \mathbf{F}\,\mathbf{P}_{t-1|t-1}\mathbf{F}^\top
  + \mathbf{Q}, \label{eq:kf_pcov}                                            \\
  \mathbf{K}_t       & = \mathbf{P}_{t|t-1}\mathbf{H}_t^\top
  \!\left(\mathbf{H}_t\mathbf{P}_{t|t-1}\mathbf{H}_t^\top
  + \mathbf{R}\right)^{-1}, \label{eq:kf_gain}                                \\
  \mathbf{x}_{t|t}   & = \mathbf{x}_{t|t-1}
  + \mathbf{K}_t\bigl(\mathbf{z}_t
  - \mathbf{h}(\mathbf{x}_{t|t-1})\bigr), \label{eq:kf_upd}
\end{align}

\noindent where $\mathbf{F}$ is the constant-velocity transition
matrix with $\Delta t = T_{\mathrm{CPI}} = \SI{640}{\milli\second}$,
$\mathbf{H}_t = \partial\mathbf{h}/\partial\mathbf{x}
  \big|_{\mathbf{x}_{t|t-1}}$ is the numerical Jacobian evaluated
at the predicted state, $\mathbf{Q}$ the process noise covariance
(acceleration model, $\sigma_a = \SI{0.5}{\metre\per\second^2}$),
$\mathbf{R} = \mathrm{diag}(\sigma_R^2, \sigma_v^2, \sigma_\theta^2)$
the measurement noise and $\mathbf{z}_t = [\hat{R}_{\mathrm{bi}},\;
  \hat{v}_{\mathrm{bi}},\; \hat{\theta}]^\top$ the detection.
The range noise is set to approximately one resolution cell,
$\sigma_R = \SI{8.0}{\metre} \approx \Delta R_{\mathrm{res}}$; the range-rate noise
$\sigma_v = \SI{0.89}{\metre\per\second}$ is set from the measured range-rate
residual over a full transit (Section~\ref{sec:filter_consistency}), some six
times the one-Doppler-cell $\Delta v = \SI{0.14}{\metre\per\second}$; and
the azimuth noise is $\sigma_\theta = \ang{9}$. The last was chosen \emph{a priori} as a
conservative value for a 4-element aperture, and the measured bearing error over
a full transit (\ang{\CAMPflatAzR}, Section~\ref{sec:results}) sits a little below
it --- so the prior is appropriately conservative without being pessimistic.
Note that $\sigma_R$ follows
the total-path convention of~\eqref{eq:dr}; the monostatic $c/(2B)$ value would
make the filter over-trust range by a factor of two, and we report the effect of
that mis-setting as an ablation in Section~\ref{sec:results}.
Data association uses the Hungarian algorithm with a Mahalanobis
gating threshold $\gamma = 16$ and SNR fingerprint
consistency as a secondary cost term.

%% FIG 1: System model / bistatic geometry

%% =========================================================
\section{Testbed Architecture}
\label{sec:testbed}

\subsection{Overview}

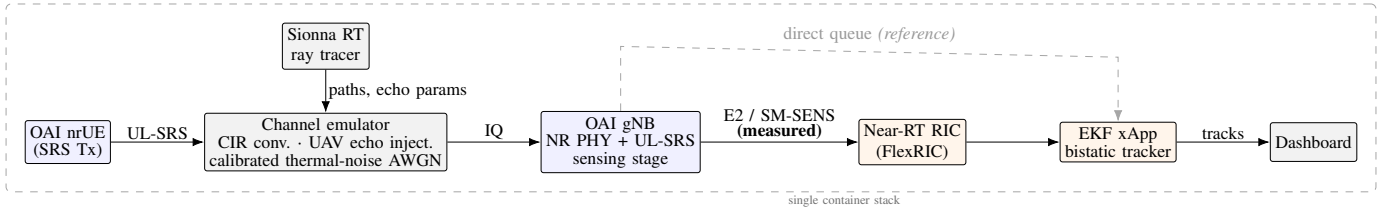
\begin{figure*}[!tbp]
  \centering
  \resizebox{\textwidth}{!}{%
    \begin{tikzpicture}
      %% --- one even row: nrUE -> emulator -> gNB -> RIC -> xApp,
      %%     with Sionna above the emulator and the dashboard below the xApp.
      \node[phy] (ue) {OAI nrUE\\[-1pt]\scriptsize (SRS Tx)};
      \node[ext, right=13mm of ue] (emu)
      {Channel emulator\\[-1pt]\scriptsize CIR conv.\ $\cdot$ UAV echo inject.\\[-1pt]\scriptsize calibrated thermal-noise AWGN};
      \node[ext, above=6mm of emu] (rt) {Sionna RT\\ray tracer};
      \node[phy, right=13mm of emu] (gnb)
      {OAI gNB\\[-1pt]\scriptsize NR PHY + UL-SRS\\[-1pt]\scriptsize sensing stage};
      %% wider gap here: the two-line E2 label is the widest edge label in the row
      \node[ric, right=22mm of gnb] (ric) {Near-RT RIC\\(FlexRIC)};
      \node[ric, right=13mm of ric] (xapp)
      {EKF xApp\\[-1pt]\scriptsize bistatic tracker};
      \node[ext, right=13mm of xapp] (dash) {Dashboard};

      %% flows (uniform arrows; all labels share the lbl style)
      \draw[flow] (rt)  -- node[lbl, right] {paths, echo params} (emu);
      \draw[flow] (ue)  -- node[lbl, above] {UL-SRS} (emu);
      \draw[flow] (emu) -- node[lbl, above] {IQ} (gnb);
      %% The measured path is SOLID; the direct in-memory queue is DASHED and
      %% grey. The manuscript's transport claim lives or dies on this
      %% distinction, so the figure states it rather than implying it.
      \draw[flow] (gnb) -- node[lbl, above, align=center]
      {E2 / SM-SENS\\[-2pt]\textbf{(measured)}} (ric);
      \draw[flow] (ric) -- (xapp);
      \draw[flow] (xapp) -- node[lbl, above] {tracks} (dash);
      %% Reference queue: the gNB bypasses the RIC. Routed above the row as two
      %% strictly vertical risers joined by one level horizontal run (explicit
      %% coordinates, so the run cannot end up skewed), with the label carried
      %% clear of the line rather than sitting on it.
      \coordinate (qL) at ($(gnb.north)+(0,9mm)$);
      \coordinate (qR) at ($(xapp.north)+(0,9mm)$);
      \draw[proto] (gnb.north) -- (qL) -- (qR) -- (xapp.north);
      \node[lbl, gray!80, above=0.8mm] at ($(qL)!0.5!(qR)$)
      {direct queue \emph{(reference)}};

      %% --- enclosing stack ----------------------------------------------
      \begin{scope}[on background layer]
        \node[draw, dashed, black!35, rounded corners=2pt,
          fit=(rt)(ue)(gnb)(ric)(xapp)(dash)(qL)(qR),
          inner sep=2.6mm, label={[slbl, black!55]below right:
              single container stack}] {};
      \end{scope}
    \end{tikzpicture}}
  \caption{Testbed architecture. The nrUE's UL-SRS is propagated through a
    ray-traced channel with UAV echo injection and calibrated thermal noise, and
    the gNB's PHY sensing stage produces per-CPI detections. Solid arrows
    carry the measured results: each CPI's detection report is sent as a RIC
    INDICATION for the SM-SENS service model, through the Near-RT RIC, to the
    subscribing EKF xApp. A direct in-memory queue (dashed, gray) is retained as a
    low-latency reference for the transport comparison of
    Appendix~\ref{sec:e2e_latency}.}
  \label{fig:arch}
\end{figure*}

Fig.~\ref{fig:arch} illustrates the end-to-end testbed comprising
five integrated components: (1)~an OAI gNB carrying the ISAC PHY sensing stage,
(2)~an OAI nrUE acting as the SRS transmitter, (3)~FlexRIC Near-RT RIC with ISAC
EKF xApp, (4)~Sionna RT channel emulator, and (5)~real-time
sensing dashboard. All components run as Docker containers
orchestrated as a single container stack, so that a run is reproducible from
the parameters and seeds reported in Section~\ref{sec:repro}.

%% FIG 2: Testbed architecture block diagram (single-row propagation-to-tracking
%% chain, matching the layout used in the multistatic manuscript)

\subsection{OAI gNB and nrUE}

The gNB is instantiated using OpenAirInterface~5G NR
\cite{kaltenberger_oai} in monolithic CU+DU mode. It operates on
Band~n78 (\SI{3.32}{\giga\hertz}) with a \SI{40}{\mega\hertz}
channel bandwidth, 30~kHz subcarrier spacing (numerology~1), and
106~PRBs. The time-division duplex (TDD) frame structure is
\texttt{DDDDDDSUU} (7 DL + 2 UL slots per 5~ms half-frame). The
gNB is equipped with $N_{\mathrm{rx}} = 4$ receive antenna ports
modeled as a half-wavelength ULA.

The nrUE is the \emph{illuminator}: a ground-level, network-registered UE that
transmits UL-SRS as part of normal uplink operation. The
UAV is not a network subscriber --- the UAV is a passive, non-cooperative
scatterer that reflects the UE's SRS towards the gNB. Nothing is mounted on the
target, which is what makes the geometry bistatic and the sensing passive.
It is configured with $N_{\mathrm{tx}} = 4$ transmit antennas
forming a half-wavelength ULA (4-element nrUE TX ULA), and
transmits periodic UL-SRS with a period of
$T_{\mathrm{SRS}} = \SI{10}{\milli\second}$ (one burst per 20
slots) across the full \SI{40}{\mega\hertz} bandwidth.

An ISAC sensing hook inserted into the OAI gNB PHY
uplink channel-estimation stage intercepts per-SRS
occasion channel estimates $\hat{H}[n_{\mathrm{rx}},k]$ for each
of the $N_{\mathrm{rx}}$ receive ports after minimum mean-square error (MMSE) channel
estimation. These are accumulated in a CPI ring buffer until
$N_{\mathrm{occ}} = 64$ bursts are collected, then forwarded to the sensing
pipeline (Section~\ref{sec:sensing}).

\subsection{Detection Report and Transport}
\label{sec:transport}

FlexRIC~\cite{schmidt_flexric, polese_oai_flexric} provides the
Near-RT RIC hosting the ISAC EKF xApp. Each CPI, the gNB emits a
\emph{detection report}: a schema version, a detection count, a CPI timestamp,
and one record per detection. Table~\ref{tab:report} specifies the fields, their
units and their valid ranges, so that the report can be reproduced without
reference to our implementation.

\begin{table}[!tbp]
  \centering
  \caption{Per-CPI detection report. One header per CPI, one record per detection.}
  \label{tab:report}
  \footnotesize
  \begin{tabular}{@{}lllp{0.30\linewidth}@{}}
    \toprule
    Field                 & Units                  & Range        & Notes                     \\
    \midrule
    \multicolumn{4}{@{}l}{\textit{Header (once per CPI)}}                                     \\
    schema version        & ---                    & ---          & report format revision    \\
    detection count       & ---                    & 0--16        & CFAR cap per CPI          \\
    CPI timestamp         & \si{\micro\second}     & monotonic    & anchors the latency chain \\
    \midrule
    \multicolumn{4}{@{}l}{\textit{Per detection}}                                             \\
    excess bistatic range & \si{\metre}            & $0$--$85$    & $\Delta R_{\mathrm{bi}}$,
    \eqref{eq:excess}; \SI{1.22}{\metre} bin; scene-gated                                     \\
    bistatic range rate   & \si{\metre\per\second} & $\pm 4.5$    &
    signed; $\pm$Nyquist at the \SI{100}{\hertz} pulse repetition frequency (PRF)             \\
    azimuth (AoA at gNB)  & \si{\degree}           & $-90$--$+90$ &
    ENU convention, \eqref{eq:az_conv}                                                        \\
    target-cell echo SNR  & \si{\decibel}          & ---          &
    per-detection range-Doppler cell SNR (echo power over local noise); median
    $\CAMPpdOpSnr$~dB at the operating point, characterized in
    Appendix~\ref{sec:detchar} (Fig.~\ref{fig:detpd})                                         \\
    \bottomrule
  \end{tabular}
\end{table}

\emph{Transport.} We define \textit{SM-SENS}, an E2 service model that carries
the report of Table~\ref{tab:report} as a RIC INDICATION from the gNB's E2 agent,
through the Near-RT RIC, to the subscribing xApp. The xApp issues a RIC
SUBSCRIPTION for the SM-SENS RAN function, and the agent then emits one
indication per CPI over the standard E2 Application Protocol (E2AP) subscription/indication machinery ---
the same path every other FlexRIC service model uses. This is the path that
carries the detections evaluated in this paper.

A second, non-E2 delivery path exists --- a direct in-memory queue between the
gNB and the xApp --- retained as a low-latency reference against which the E2
path can be compared (Appendix~\ref{sec:e2e_latency}). The two are selectable at
run time, and we report both. To verify that the E2 path genuinely carries the
data rather than merely being subscribed, we disable the queue entirely and
confirm that the number of indications the xApp receives equals the number of
CPIs the gNB detected in: with the queue off, delivery is one-for-one.

The scope of SM-SENS should be clear. It follows the
structural pattern of an existing service model and uses the same E2AP
procedures, but it has no ASN.1 specification (its payload uses the plain
encoding FlexRIC's other custom service models use), and it has not been tested
for interoperability against any other RIC or E2 node. It should be read as a
working demonstration of what a sensing service model must carry --- the field
set in Table~\ref{tab:report} --- rather than as a standards contribution. An
ASN.1-specified, interoperability-tested E2SM remains future work.

The xApp decodes the report, runs the bistatic EKF tracker
(Section~\ref{sec:sysmodel}), and publishes per-track state to
a key-value store at each CPI for the dashboard to poll.

\subsection{Sionna RT Channel Emulation}

A Sionna RT~\cite{hoydis_sionna} server computes, once per
simulation step, a ray-traced static channel impulse response (CIR)
between the gNB and nrUE (direct/multipath taps: per-path delay and
complex gain) together with per-UAV bistatic echo parameters
(bistatic delay, Doppler shift, complex amplitude), serving both via
a REST API to the \emph{channel emulator} container. Uplink and
downlink share the same underlying ray trace rather than requiring a
second PathSolver run: by TDD reciprocity the downlink CIR is the
transpose of the uplink CIR, $H_{\mathrm{dl}} = H_{\mathrm{ul}}^{T}$,
computed once and cached alongside the uplink result.

The channel emulator acts as an rfsimulator proxy between the nrUE
and gNB, running independently on the uplink and downlink IQ streams.
Each direction
convolves the clean transmitted signal through its static CIR tap(s)
and adds the UAV echoes (delay, Doppler, complex amplitude) before an
absolute link-budget calibration stage injects thermal-noise-referenced
AWGN, applies a headroom-safe automatic gain, and quantizes to the
wire's fixed-point (16-bit) in-phase/quadrature (IQ) format (Table~\ref{tab:simtarget} and
Section~\ref{sec:discussion}). UAV motion is modeled using a 3D
kinematic state (position, velocity, radar cross-section [RCS]) with Gauss-Markov hovering
and constant-velocity transit modes.

\subsection{Sensing Dashboard}

A real-time web dashboard (Python/FastAPI with an HTML5 canvas frontend) provides
operator situational awareness: a live top-down tracking map of UAV ground-truth
positions, EKF track estimates and velocity vectors; an EKF position-error
time series ($\Delta x$, $\Delta y$ per UAV); and a parameter summary table,
refreshed by HTTP polling against a state endpoint. All state is exchanged through
the key-value store, so the dashboard is stateless, and it supports live debugging
and post-run playback of the recorded state history.

%% =========================================================
\section{UL-SRS Sensing Pipeline}
\label{sec:sensing}

The full sensing pipeline executes in two stages: (i) the
\emph{gNB PHY sensing module} (C, FFTW, runs per CPI at
$T_{\mathrm{CPI}} = \SI{640}{\milli\second}$) and
(ii) the \emph{ISAC xApp EKF tracker} (Python, runs on the
Near-RT RIC per SM-SENS indication). This split keeps
computationally intensive signal processing close to the PHY
while allowing algorithmic sensing logic to evolve independently
as an xApp.

\subsection{SRS Resource Configuration}

UL-SRS is configured with periodicity
$T_{\mathrm{SRS}} = \SI{10}{\milli\second}$ (one SRS burst per
20 TDD slots at 30~kHz SCS). The coherent processing interval
spans $N_{\mathrm{occ}} = 64$ consecutive SRS bursts, yielding
$T_{\mathrm{CPI}} = N_{\mathrm{occ}} \cdot T_{\mathrm{SRS}} = \SI{640}
  {\milli\second}$ and a bistatic Doppler resolution of $\approx \SI{0.14}{\metre\per\second}$
at $f_c = \SI{3319.68}{\mega\hertz}$.

\subsection{Clutter-Subspace Deflation}
\label{sec:clutter}

Let $\mathbf{H} \in \mathbb{C}^{N_{\mathrm{occ}} \times K}$ collect one receive
port's channel estimates over a CPI, its rows the $N_{\mathrm{occ}}$ slow-time
occasions and its columns the $K$ packed subcarriers. The disturbance --- the
direct path and the static ground return, which after inter-occasion phase-ramp
carrier-frequency-offset (CFO) compensation are constant across slow time --- occupies the dominant
left singular subspace of $\mathbf{H}$. It is removed by deflating the leading
$K_c$ such directions:
\begin{align}
  \mathbf{H}^{(0)} & = \mathbf{H}, \nonumber       \\
  \mathbf{u}_i     & = \arg\max_{\|\mathbf{u}\|=1}
  \bigl\| \mathbf{u}^{\mathsf{H}} \mathbf{H}^{(i-1)} \bigr\|_2 ,
  \label{eq:defl_u}                                \\
  \mathbf{H}^{(i)} & = \mathbf{H}^{(i-1)}
  - \mathbf{u}_i \bigl( \mathbf{u}_i^{\mathsf{H}} \mathbf{H}^{(i-1)} \bigr),
  \quad i = 1,\dots,K_c ,
  \label{eq:defl_H}
\end{align}

\noindent each $\mathbf{u}_i$ being obtained by power iteration on
$\mathbf{H}^{(i-1)}\mathbf{H}^{(i-1)\mathsf{H}}$ without forming the covariance
explicitly. The result $\mathbf{H}^{(K_c)}$ is the clutter-canceled estimate
$\hat{H}^{\mathrm{UAV}}$ used in~\eqref{eq:rd_map}. Cancellation operates on the
\emph{current} CPI only, with no memory across CPIs.

\emph{Relation to ECA.} This is a reference-free subspace deflation, and not as
the Extensive Cancellation Algorithm (ECA) of the
passive-radar literature~\cite{colone_eca_passive_radar}. ECA projects the
surveillance channel onto the orthogonal complement of a subspace spanned by
delay- and Doppler-shifted replicas of a \emph{separately acquired reference
  channel}; here the disturbance subspace is instead estimated from the
surveillance data itself, as its dominant slow-time singular subspace. The two
share a goal --- suppressing the zero-Doppler direct and static returns --- but
not a mechanism. We adopt the subspace form because the gNB has no separate
reference receive chain to supply ECA's replica bank.

In the
flat-LOS single-target scene the rank must be held at $K_c=1$: the
UAV echo is the second-strongest return, so a rank $K_c \ge 2$ deflation
subtracts the target itself and leaves only a smeared residual
(Section~\ref{sec:discussion}). Scenes with additional strong static
multipath (e.g.\ building bounces) require a higher, scene-dependent
rank, which we do not exercise in the flat-LOS evaluation here.

\subsection{Range-Doppler Map and CFAR}

The 2D map of~\eqref{eq:rd_map} is computed per receive antenna using
zero-padded transforms (range: 4096 points, Doppler: 128 points) and
non-coherently combined:

\begin{equation}
  \mathcal{X}_{\mathrm{comb}}[m,n] =
  \sum_{i=1}^{N_{\mathrm{rx}}} |\mathcal{X}_i[m,n]|^2.
  \label{eq:nc_combine}
\end{equation}

The range transform is an \emph{inverse} DFT, since the delay/range
profile of a frequency-domain channel is $h(\tau)=\mathrm{IDFT}\{H(f)\}$.
Order-statistic CFAR (OS-CFAR~\cite{rohling_oscfar}, $75^{\mathrm{th}}$-percentile training
statistic, robust to the target's own main-lobe leakage) is applied
with guard window $(N_{g,R}, N_{g,D}) = (14, 5)$ cells and training
window $(N_{t,R}, N_{t,D}) = (28, 8)$ cells, $P_{\mathrm{FA}} = 10^{-4}$,
up to a maximum of 16 detections per CPI. A single-peak non-maximum
suppression (NMS) --- which keeps only the locally strongest range-Doppler cell
and discards weaker cells within a fixed radius around it --- with a range
radius of $18$ bins then groups each target's
comb-least-squares window sidelobes into one detection, so a single
physical target yields a single detection per CPI.

\subsection{Azimuth Estimation: Single-Snapshot Interferometric AoA}
\label{sec:aoa}

\emph{Estimator.} Let $\mathbf{a} = [a_0,\dots,a_{N_{\mathrm{rx}}-1}]^\top$
denote the complex range-Doppler cell values of a detected target across the
$N_{\mathrm{rx}} = 4$ receive ports. For a single dominant scatterer on a
half-wavelength ULA, $a_p = s\,e^{\,j\pi p \sin\theta} + n_p$: the
inter-element phase step is $\pi\sin\theta$, and averaging it over the three
available baselines gives the bearing in closed form,
\begin{equation}
  \hat{\theta} = \arcsin\left(\frac{1}{\pi}
  \arg\left( \sum_{p=0}^{N_{\mathrm{rx}}-2} a_{p+1}\, a_p^{*} \right)\right).
  \label{eq:aoa}
\end{equation}
This is a single-snapshot interferometric (adjacent-element phase-difference)
estimator. For a single source on a uniform linear array with unit inter-element
displacement it coincides with the least-squares ESPRIT (estimation of signal
parameters via rotational invariance techniques) solution over
maximally-overlapping subarrays~\cite{roy_esprit}; we describe it as
ESPRIT-type only in that reduced sense and do not invoke the subspace machinery.
It is unambiguous for $|\sin\theta| < 1$ and, at high SNR away from endfire,
approximately unbiased; its bias grows toward endfire, as the error breakdown
below and in Section~\ref{sec:results} shows. It produces every bearing reported
in this paper.

\emph{Array geometry.} The four spatial channels are the gNB's own 4-element
half-wavelength Rx ULA, read directly from its four receive chains. The array
lies along $\hat{\mathbf{y}}$ with boresight $+\hat{\mathbf{x}}$
(Section~\ref{sec:coords}), which gives a signed, monotonic phase ramp as the
target crosses broadside and resolves the sign ambiguity an endfire orientation
would suffer over this target corridor. Element geometry is encoded by the ray
tracer's own per-element path solving.

\emph{Limitations.} The 4-element aperture is modest: we assign
$\sigma_\theta \approx \ang{9}$ in the filter (against $\approx \ang{30}$ for a
2-element ULA), which comfortably bounds the measured bearing error of \ang{\CAMPflatAzR}
over a full transit (Section~\ref{sec:results}). The error is not
uniform across the transit: it is smallest near broadside and grows as the target
approaches endfire, where the inter-element phase ramp of~\eqref{eq:aoa} becomes
insensitive to bearing. Cross-range uncertainty
therefore grows linearly with range (Section~\ref{sec:discussion}). Because the
array is one-dimensional, \eqref{eq:aoa} resolves azimuth only and cannot
disambiguate elevation, so target height is not sensed and is instead supplied
as a prior (Table~\ref{tab:params}).

\subsection{EKF xApp Integration}

Detections passing the CFAR threshold are assembled into a per-CPI detection
report (Table~\ref{tab:report}) and delivered to the Near-RT RIC over the
transport described in Section~\ref{sec:transport}. On receipt the xApp first
restores the baseline, $\hat{R}_{\mathrm{bi}} = \Delta\hat{R}_{\mathrm{bi}} + L$
(\eqref{eq:excess}), converting the reported excess range into the total path
length that~\eqref{eq:h1} expects. It then performs Hungarian
data association with a Mahalanobis gate of $\gamma = 16$,
and runs one EKF predict-update cycle per track per CPI.
Track management follows 1-of-1 initiation (any ungated detection
spawns a track) and a coast limit of 5 missed CPIs before
deletion. Because any ungated detection spawns a new track, CPIs carrying more
than one detection can seed spurious tracks; this initiation policy is the direct
cause of the \CAMPflatTrackIds{} track identifiers per run reported in
Table~\ref{tab:single_uav_meas}, and it is analyzed in Section~\ref{sec:results}.
The active track state is written to the key-value store for dashboard
display.

%% FIG 3: Sensing pipeline flowchart
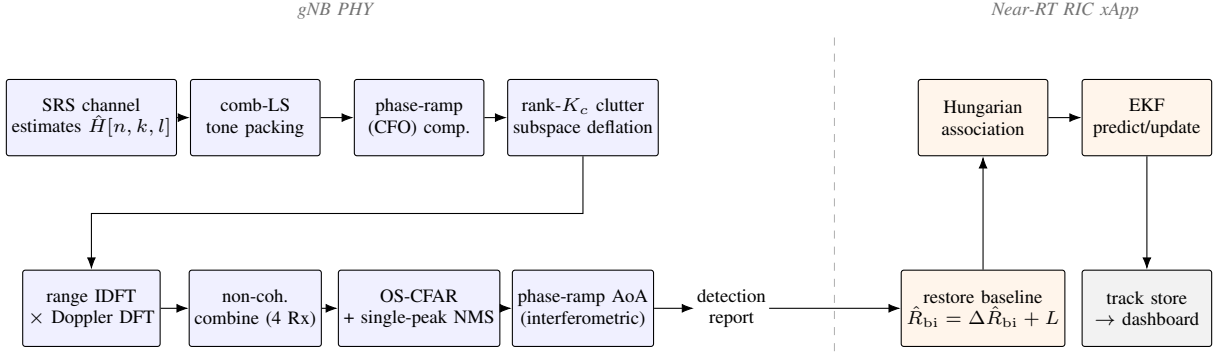
\begin{figure*}[!tbp]
  \centering
  \resizebox{0.88\textwidth}{!}{%
    \begin{tikzpicture}[blk/.append style={minimum height=10mm, minimum width=17mm}]
      %% Explicit column x-centres (\cA..\cD gNB PHY, \rL/\rR RIC) and the two
      %% row levels \yT/\yB, SHARED by both blocks so the rows line up exactly.
      \def\cA{0}\def\cB{2.15}\def\cC{4.30}\def\cD{6.45}
      \def\cDet{8.4}\def\rL{11.7}\def\rR{13.85}
      \def\yT{0}\def\yB{-2.5}

      %% ---- gNB PHY (left): top row, left -> right ----------------------
      \node[phy] (h)   at (\cA,\yT) {SRS channel\\estimates $\hat{H}[n,k,l]$};
      \node[phy] (ls)  at (\cB,\yT) {comb-LS\\tone packing};
      \node[phy] (cfo) at (\cC,\yT) {phase-ramp\\(CFO) comp.};
      \node[phy] (eca) at (\cD,\yT) {rank-$K_c$ clutter\\subspace deflation};
      %% ---- gNB PHY (left): bottom row, left -> right -------------------
      \node[phy] (rd)   at (\cA,\yB) {range IDFT\\$\times$ Doppler DFT};
      \node[phy] (nc)   at (\cB,\yB) {non-coh.\\combine (4 Rx)};
      \node[phy] (cfar) at (\cC,\yB) {OS-CFAR\\+ single-peak NMS};
      \node[phy] (aoa)  at (\cD,\yB) {phase-ramp AoA\\(interferometric)};

      \draw[flow] (h) -- (ls); \draw[flow] (ls) -- (cfo); \draw[flow] (cfo) -- (eca);
      %% folded-snake return, routed in the inter-row gap (no box crossings)
      \draw[flow] (eca.south) -- ++(0,-0.75) -| (rd.north);
      \draw[flow] (rd) -- (nc); \draw[flow] (nc) -- (cfar); \draw[flow] (cfar) -- (aoa);

      %% ---- bridge: STRAIGHT horizontal on the bottom row ---------------
      \node[lbl] (det) at (\cDet,\yB) {detection\\report};
      \draw[flow] (aoa) -- (det);

      %% ---- Near-RT RIC xApp (right): 2 rows, ALIGNED to the PHY rows ----
      \node[ric] (ing)   at (\rL,\yB) {restore baseline\\$\hat R_{\mathrm{bi}}=\Delta\hat R_{\mathrm{bi}}+L$};
      \node[ric] (assoc) at (\rL,\yT) {Hungarian\\association};
      \node[ric] (ekf)   at (\rR,\yT) {EKF\\predict/update};
      \node[ext] (store) at (\rR,\yB) {track store\\$\rightarrow$ dashboard};

      \draw[flow] (det) -- (ing);          % straight, crosses the lane divider
      \draw[flow] (ing) -- (assoc); \draw[flow] (assoc) -- (ekf);
      \draw[flow] (ekf) -- (store);

      %% ---- vertical lane separator, centred in the clear PHY|RIC gap ---
      %% (\cDet detection report ends ~9.05; RIC restore box begins ~10.5)
      \def\sx{9.75}
      \draw[black!30, dashed] (\sx,1.05) -- (\sx,-3.15);
      %% region labels sit centred over their own blocks, not on the divider
      \node[font=\scriptsize\itshape, black!55] at (3.2,1.4) {gNB PHY};
      \node[font=\scriptsize\itshape, black!55] at (12.78,1.4) {Near-RT RIC xApp};
    \end{tikzpicture}}
  \caption{UL-SRS sensing pipeline. The gNB PHY stage runs once per CPI: SRS
    channel estimates are packed gap-free across the comb, phase-ramp (CFO)
    compensated, and the dominant static (direct-path) subspace is deflated; a
    zero-padded range IDFT and slow-time Doppler DFT form the range-Doppler map,
    which is combined non-coherently across the four receive ports before OS-CFAR
    detection with single-peak grouping, and the bearing of the detected cell is
    estimated in closed form. The xApp restores the baseline to convert the
    measured \emph{excess} bistatic range to a total path length before
    association and EKF update.}
  \label{fig:pipeline}
\end{figure*}

%% =========================================================
\section{Evaluation and Results}
\label{sec:results}

\subsection{Simulation Setup}

\begin{table}[!tbp]
  \centering
  \caption{Radio and Sensing Configuration}
  \label{tab:params}
  \footnotesize
  \begin{tabular}{@{}p{0.40\linewidth}p{0.52\linewidth}@{}}
    \toprule
    \textbf{Parameter}                           & \textbf{Value}                                                                                                                                                                                                                                                                                                         \\
    \midrule
    NR Band (carrier)                            & n78, $f_c = \SI{3319.68}{\mega\hertz}$                                                                                                                                                                                                                                                                                 \\
    Subcarrier spacing                           & 30~kHz (numerology~1)                                                                                                                                                                                                                                                                                                  \\
    Channel bandwidth                            & \SI{40}{\mega\hertz} nominal; 106~PRBs $=$ \SI{38.16}{\mega\hertz} occupied                                                                                                                                                                                                                                            \\
    TDD pattern                                  & DDDDDDSUU (7+2 per 5~ms)                                                                                                                                                                                                                                                                                               \\
    gNB Rx antennas ($N_{\mathrm{rx}}$)          & 4 (half-$\lambda$ ULA)                                                                                                                                                                                                                                                                                                 \\
    nrUE Tx antennas ($N_{\mathrm{tx}}$)         & 4 (half-$\lambda$ ULA)                                                                                                                                                                                                                                                                                                 \\
    Antenna ULA orientation                      & gNB Rx: array along $y$-axis, boresight $+x$ (east, broadside to the target corridor, giving a signed and resolvable azimuth); nrUE Tx: array along $x$-axis, boresight $+y$ (north)                                                                                                                                   \\
    Antenna element pattern                      & Isotropic; both gNB Rx and nrUE Tx                                                                                                                                                                                                                                                                                     \\
    \midrule
    SRS occupied tones ($M_{\mathrm{SRS}}$)      & 624 (comb-2, every 2nd subcarrier over 106~PRBs; 1272 subcarriers total)                                                                                                                                                                                                                                               \\
    SRS periodicity ($T_{\mathrm{SRS}}$)         & \SI{10}{\milli\second}                                                                                                                                                                                                                                                                                                 \\
    CPI length ($N_{\mathrm{occ}}$)              & 64~occasions                                                                                                                                                                                                                                                                                                           \\
    CPI duration ($T_{\mathrm{CPI}}$)            & \SI{640}{\milli\second}                                                                                                                                                                                                                                                                                                \\
    Range resolution ($\Delta R_{\mathrm{res}}$) & \SI{8.0}{\metre} ($c/B_{\mathrm{SRS}}$, $B_{\mathrm{SRS}} \approx \SI{37.4}{\mega\hertz}$ SRS aperture, total bistatic path); zero-padded bin \SI{1.22}{\metre}                                                                                                                                                        \\
    Doppler resolution ($\Delta v$)              & \SI{0.14}{\metre\per\second} ($\lambda/T_{\mathrm{CPI}}$); \SI{0.07}{\metre\per\second} zero-padded bin                                                                                                                                                                                                                \\
    Wavelength ($\lambda$)                       & \SI{0.0903}{\metre}                                                                                                                                                                                                                                                                                                    \\
    \midrule
    Clutter-deflation rank ($K_c$)               & 1 (direct path is the sole strong static return in the flat-LOS single-target scene)                                                                                                                                                                                                                                   \\
    CFAR type                                    & Order-statistic (OS-CFAR), $75^{\mathrm{th}}$-percentile training                                                                                                                                                                                                                                                      \\
    CFAR guard (range, Doppler)                  & 14, 5 cells (comb-LS range response)                                                                                                                                                                                                                                                                                   \\
    CFAR training (range, Doppler)               & 28, 8 cells                                                                                                                                                                                                                                                                                                            \\
    CFAR $P_{\mathrm{FA}}$                       & $10^{-4}$                                                                                                                                                                                                                                                                                                              \\
    Single-peak NMS range radius                 & 18 range bins (groups a target's window sidelobes into one detection)                                                                                                                                                                                                                                                  \\
    Max detections per CPI                       & 16                                                                                                                                                                                                                                                                                                                     \\
    \midrule
    EKF $\sigma_R$                               & \SI{8.0}{\metre} ($\approx$ one resolution cell)                                                                                                                                                                                                                                                                       \\
    EKF $\sigma_v$                               & \SI{0.89}{\metre\per\second} (measured range-rate residual)                                                                                                                                                                                                                                                            \\
    EKF $\sigma_\theta$                          & $\ang{9}$ (4-el gNB Rx ULA, phase-ramp AoA; conservative vs.\ the measured \ang{\CAMPflatAzR} bearing error)                                                                                                                                                                                                           \\
    EKF $\sigma_a$ (process noise)               & \SI{0.5}{\metre\per\second^2}                                                                                                                                                                                                                                                                                          \\
    Mahalanobis gate ($\gamma$)                  & 16                                                                                                                                                                                                                                                                                                                     \\
    Coast limit                                  & 5~CPIs                                                                                                                                                                                                                                                                                                                 \\
    Height prior $h_{\mathrm{tgt}}$ (2-D EKF)    & \SI{55}{\metre} --- a \emph{deployment} parameter (the expected UAV altitude band), not a quantity inferred from the measurements: a single bistatic pair has no elevation observability. Here it matches the scenario's transit altitude; the cost of a mismatched prior is the ablation in Section~\ref{sec:results} \\
    \bottomrule
  \end{tabular}
\end{table}

\begin{table}[!tbp]
  \centering
  \caption{Simulation Scene and Target}
  \label{tab:simtarget}
  \footnotesize
  \begin{tabular}{@{}p{0.40\linewidth}p{0.52\linewidth}@{}}
    \toprule
    \textbf{Parameter}                     & \textbf{Value}                                                                                                                                                                                                                                                          \\
    \midrule
    gNB / nrUE TX power (calibrated label) & 33 / 23~dBm --- \texttt{rfsimulator}'s baseband IQ loopback has no physical RF stage and transmits at a fixed digital amplitude regardless of configured power, so these are calibration-constant labels assigned to that invariant waveform, not power-control targets \\
    gNB / nrUE receiver noise figure       & 5 / 9~dB; thermal noise ($k_BT_0B$) is injected as calibrated AWGN at this NF (plus a configurable interference-over-thermal margin, 0~dB here) after the CIR convolution stage (Section~\ref{sec:testbed})                                                             \\
    \midrule
    UAV-A (transit)                        & $x = 27$~m, $z = \SI{55}{\metre}$, $v_y = \SI{2}{\metre\per\second}$, $y: -40 \rightarrow +42$~m (south-to-north transit through the baseline broadside), sphere target, diffuse scattering $0.9$                                                                       \\
    gNB position                           & $(-75, 0, 10)$~m                                                                                                                                                                                                                                                        \\
    UE position (bistatic baseline)        & $(55, 0, 1.5)$~m, $L = \SI{130}{\metre}$                                                                                                                                                                                                                                \\
    Scene boundary ($x$, $y$)              & $[-100, 100]$~m, $[-60, 60]$~m                                                                                                                                                                                                                                          \\
    Channel model                          & Sionna RT (ray-traced echo injection; low-permittivity dry-ground floor, buildings removed for the flat-LOS benchmark)                                                                                                                                                  \\
    \bottomrule
  \end{tabular}
\end{table}

\begin{figure*}[!tbp]
  \centering
  \includegraphics[width=\textwidth,height=0.34\textheight,keepaspectratio]{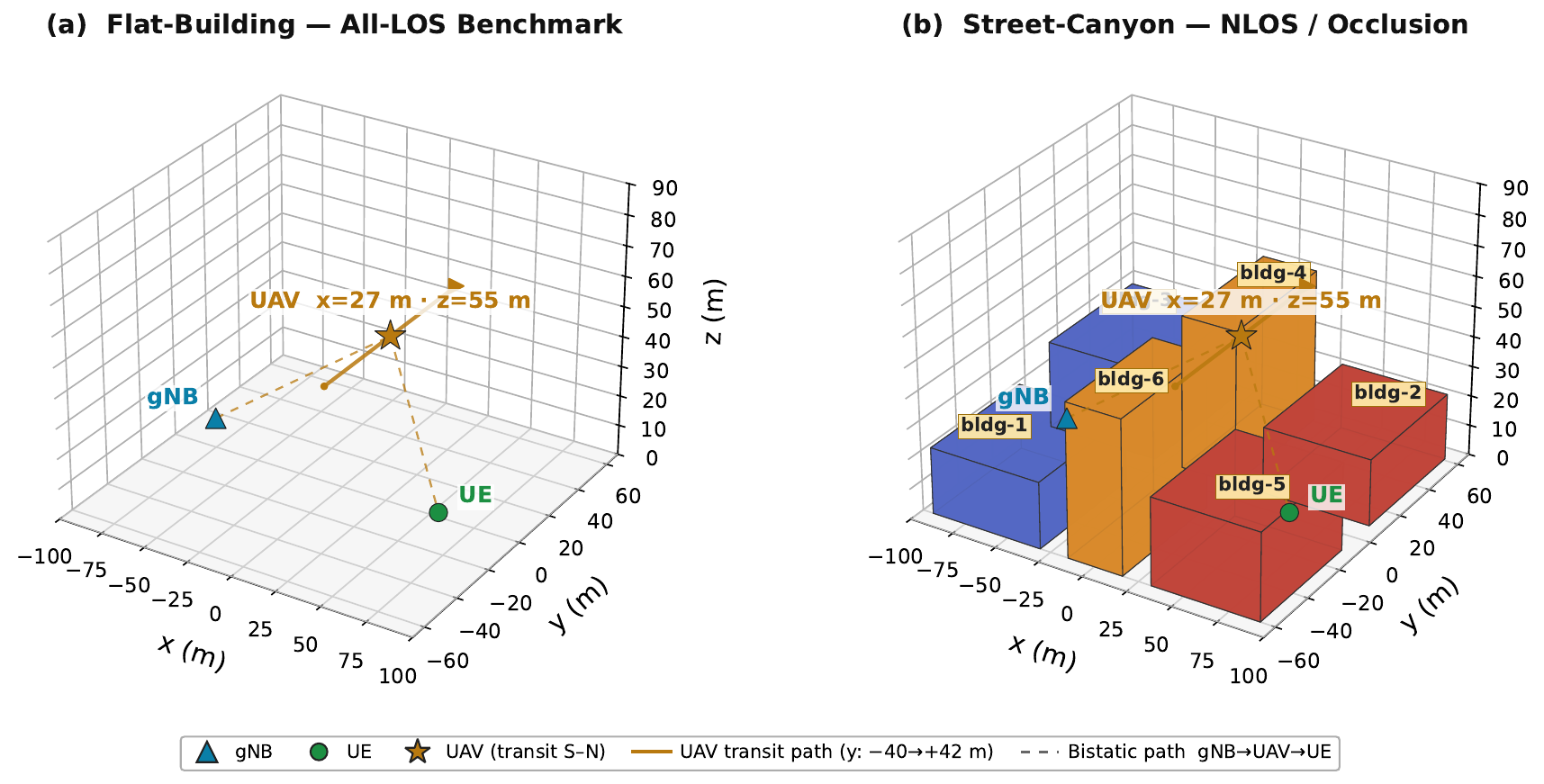}
  \caption{The two evaluation scenes, drawn to a common viewpoint and marker set.
    In both, the gNB sits at $(-75,0,10)$~m, the nrUE at $(55,0,1.5)$~m, and the
    UAV transits south-to-north (along $+y$) at $x=27$~m, $z=55$~m through the
    bistatic broadside. \textbf{(a)}~Flat-LOS benchmark --- the ray-traced canyon
    with all buildings removed, giving unobstructed line-of-sight throughout the
    transit. \textbf{(b)}~Street-canyon NLOS/occlusion scene --- the same geometry
    with the six buildings restored. The gNB at \SI{10}{\metre} looks up past
    buildings reaching \SI{51}{\metre}, so the gNB$\rightarrow$UAV ray is
    intercepted for much of the transit even though the UAV flies above the
    rooftops at \SI{55}{\metre}: the occlusion is set by the gNB height, not the
    target's.}
  \label{fig:scenes}
\end{figure*}

Experiments are conducted using the parameters listed in
Tables~\ref{tab:params} and~\ref{tab:simtarget}. The gNB is placed at $(-75, 0, 10)$~m and the
nrUE at $(55, 0, 1.5)$~m, giving a bistatic baseline
$L = \SI{130}{\metre}$ along the main street corridor. A UAV
performs a horizontal transit at fixed $x = 27$~m and altitude
$z = \SI{55}{\metre}$ (above rooftop level, line-of-sight), flying
south-to-north along $+y$ at $\SI{2}{\metre\per\second}$ from
$y = -40$~m to $y = +42$~m. This trajectory carries the target through
the array broadside near $y = 0$, so a single run exercises the full
signature of interest: the bistatic range sweeps to a minimum and back,
the radial velocity reverses sign through the crossing, and the azimuth
sweeps through boresight. The UAV is modeled as a diffuse sphere
(radius $3$~m, scattering coefficient $0.9$), whose mean bistatic radar
cross-section measures \SI{+13.9}{\dBsm} --- about \SI{0.6}{\decibel} below its
$\pi r^2$ optical limit of \SI{+14.5}{\dBsm}. This is a deliberately strong
reference target; the RCS ladder of Fig.~\ref{fig:detpd} scales it down to the
realistic small-UAV band ($-20$ to $-10$~dBsm) at which $P_{\mathrm{D}}$ is
reported. Ground
truth positions are provided by the channel emulator's UAV
kinematic model and published to the key-value store for real-time comparison
with EKF estimates, and the full run is additionally recorded
as an annotated trajectory video (ground truth vs. EKF estimate)
for qualitative inspection. The evaluation scene is the flat-LOS
benchmark (Fig.~\ref{fig:scenes}a); results use the Sionna
RT ray-traced channel model, providing ray-traced echo
coefficients (bistatic delay, Doppler shift, complex amplitude)
for the UAV at every simulation step.

Two counts appear in this paper and are kept strictly separate: $N$ is
always a number of runs, and $n$ is always a number of pooled per-CPI samples.
$n$ is used only for latency, where the per-CPI distribution --- not the
run-to-run distribution --- is what matters. A single transit yields
$\approx 79$ observable CPIs.

\textbf{Metric definitions.} Three reported quantities need stating precisely.

\emph{Detection coverage} is the fraction of \emph{observable} CPIs in which the
pipeline reports at least one detection. A CPI is observable when the ray tracer
returns a UAV echo at all and the target lies outside the near-baseline blind
zone. \emph{Line-of-sight visibility} is the fraction of blind-zone-excluded CPIs
that are observable in that sense; in free space it is unity, and in the street
canyon it is the occlusion penalty. The two are reported separately, since a CPI
in which the target is physically occluded is not a detection failure.

\emph{Track continuity} is the fraction of a target's observable CPIs in which it
is tracked at all --- matched by at least one of its (possibly several) track
fragments. With $\mathcal{C}_{\mathrm{obs}}$ the set of observable CPIs and
$\mathcal{C}_{\mathrm{match}} \subseteq \mathcal{C}_{\mathrm{obs}}$ those in which
some fragment of the target is associated,
\begin{equation}
  \mathrm{continuity} =
  \frac{\lvert \mathcal{C}_{\mathrm{match}} \rvert}{\lvert \mathcal{C}_{\mathrm{obs}} \rvert} .
  \label{eq:continuity}
\end{equation}
Scored over the \emph{union} of the target's fragments (as the aggregator
computes it, consistent with the position rows of
Table~\ref{tab:single_uav_meas}), continuity is coverage-complete: a target
handed between several identifiers still counts as tracked at each CPI, so the
shortfall below \SI{100}{\percent} measures CPIs in which no fragment of the
target is associated --- an association/detection loss --- not identifier churn.
Identifier fragmentation is a \emph{distinct} effect, reported separately as the
per-run track count (Section~\ref{sec:frag}); together the two say how often the
target is tracked and across how many identifiers. Unless stated otherwise
continuity counts only detection-associated CPIs; the coast-bridged variant
(Section~\ref{sec:frag}) additionally counts a CPI spanned by a confirmed track's
predicted (coasted) state, and is thus a \emph{track-level} figure that includes
bounded predicted coverage between measurements.

Finally, CPIs falling inside the near-baseline blind zone are excluded from the
measurement statistics. As the target crosses the baseline the bistatic angle
approaches $180^\circ$, the excess range approaches zero and the geometry
degenerates (Section~\ref{sec:sysmodel}); reporting a range or Doppler error
there would measure the geometry's singularity, not the pipeline's accuracy.

\subsection{Calibration Validation}
\label{sec:calib_validation}

The link budget underlying Table~\ref{tab:simtarget}'s noise-floor and
TX-power entries is cross-checked against closed-form references
before being trusted for the results below. For the direct
UE$\to$gNB path ($d = \SI{130}{\metre}$ at
$f_c = \SI{3.32}{\giga\hertz}$), free-space path loss gives
$\mathrm{FSPL} \approx \SI{85.1}{\decibel}$, which against the
\SI{23}{\dBm} UE TX power predicts a received level of
$\approx \SI{-62}{\dBm}$ -- consistent with the emulator's calibrated
digital-to-dBm mapping (Section~\ref{sec:testbed}). The
injected thermal noise floor is set from
$-174 + 10\log_{10}(\SI{38.16}{\mega\hertz}) + \mathrm{NF}$, with
$\mathrm{NF} = \SI{5}{\decibel}$ at the gNB, giving
$\approx \SI{-93.2}{\dBm}$; a representative UAV echo
($R_T \approx \SI{110}{\metre}$ from gNB to the above-rooftop target,
$R_R \approx \SI{60}{\metre}$ from target to UE) lands well below the
per-sample noise floor and is recovered only via the $\sim\SI{50}{dB}$
coherent processing gain of the \SI{640}{\milli\second} CPI --
the expected regime for a physically-calibrated ISAC link. These three anchors (FSPL, thermal floor, echo level) are
computed analytically from the same constants the emulator uses at
runtime; a full numerical
cross-validation against the Sionna RT-reported path gains and an
in-loop dBm audit at each proxy stage (analogous for RSRP
consistency) are left as follow-up verification work.

\subsection{Filter Consistency: NIS and NEES}
\label{sec:filter_consistency}

Calibrating the link budget establishes that the \emph{measurements} are
physically sound; it says nothing about whether the tracker's assumed
uncertainty matches its actual error. The two standard consistency statistics
test that directly, checking the filter's self-reported confidence against its
observed error. The normalized innovation squared,
$\mathrm{NIS} = \tilde{\mathbf{y}}^{\top} \mathbf{S}^{-1} \tilde{\mathbf{y}}$, uses only the innovation
$\tilde{\mathbf{y}}$ and its covariance $\mathbf{S} = \mathbf{H}\mathbf{P}\mathbf{H}^{\top} + \mathbf{R}$ and therefore needs no
ground truth; under a correctly specified filter it is $\chi^2$-distributed with
$\dim(\mathbf{z}) = 3$ degrees of freedom, so $\mathbb{E}[\mathrm{NIS}] = 3$. The
normalized estimation error squared,
$\mathrm{NEES} = \mathbf{e}^{\top} \mathbf{P}^{-1} \mathbf{e}$ with $\mathbf{e} = \hat{\mathbf{x}} - \mathbf{x}_{\mathrm{true}}$,
requires ground truth and tests the state covariance itself, with
$\mathbb{E}[\mathrm{NEES}] = \dim(\mathbf{x}) = 4$.

Measured over the flat-LOS campaign, the two disagree in \emph{opposite}
directions. NIS averages \CAMPflatNis\ against an expected 3 --- \CAMPflatNisVerdict,
indicating that the assumed measurement covariance is if anything too large for
the innovations actually observed. NEES, by contrast, averages \CAMPflatNees\
against an expected 4 --- \CAMPflatNeesVerdict, the state covariance roughly six
times smaller than the estimation error it is supposed to describe. A filter
whose $R$ is merely mis-scaled cannot produce this combination, because rescaling
$R$ moves the two statistics in the same sense: inflating $R$ to pull NEES down
drives NIS further below its expectation, and shrinking $R$ to raise NIS inflates
NEES. The filter is already run with $\sigma_v$ set from the
measurement-minus-truth residuals --- the range-rate error over a full transit is
\SI{0.89}{\metre\per\second} ($N=5$), about six times the one-Doppler-cell
\SI{0.14}{\metre\per\second} a naive model would assume --- yet NEES stays at
\CAMPflatNees, far above 4. No single choice of $R$ makes the filter
self-consistent: the residual over-confidence is structural (unmodelled cross-axis
correlation and the heavy-tailed azimuth error of Section~\ref{sec:results}), not
a scale error.

The residual over-confidence is therefore not a mis-set covariance but
\emph{structural}, from two effects the two-dimensional filter cannot represent.
First, the bistatic range estimate carries a consistent bias of about
\SI{-8}{\metre} --- roughly one range cell, against a residual spread of only
\SI{2.3}{\metre} --- so the position error has a systematic component that
repeated, mutually consistent measurements make the filter increasingly
confident about rather than reveal. Second, a constant-velocity dynamic model
with a fixed height prior in a two-dimensional state cannot represent the true
motion and geometry closely enough for the covariance to be believed, however
$R$ is chosen. A filter that
is optimistic by this margin should not have its uncertainty used for gating or
fusion decisions downstream. It also identifies where the fix must come from:
elevation is unobservable from a single bistatic pair with azimuth-only bearing,
so removing the height prior requires multistatic geometry rather than a larger
state vector (Section~\ref{sec:future}).

\subsection{Range-Doppler Map}
\label{sec:rdmap}

\begin{figure*}[!tbp]
  \centering
  \subfloat[Before clutter cancellation]{\includegraphics[width=0.49\textwidth]{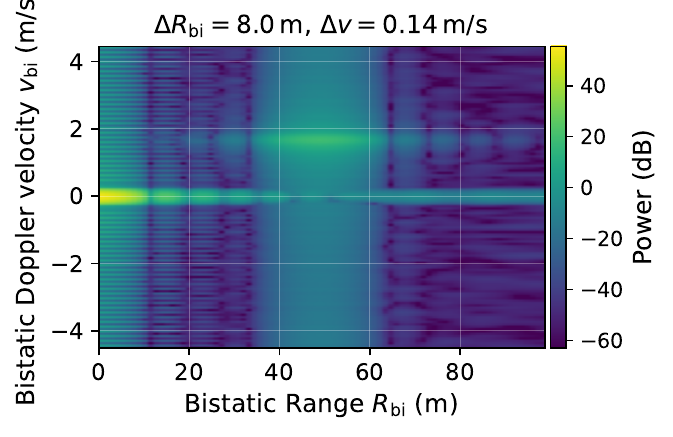}}\hfil
  \subfloat[After deflation + single-peak detection]{\includegraphics[width=0.49\textwidth]{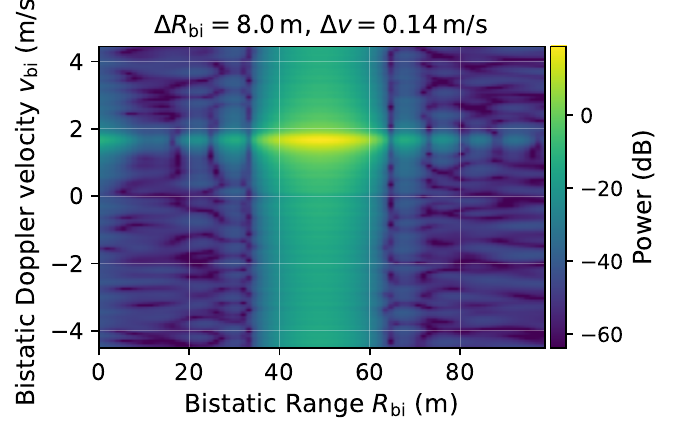}}
  \caption{Range-Doppler map of the same CPI --- the first of the transit
    (CPI~0, UAV at $y \approx \SI{-39}{\metre}$, approaching) --- before and after
    clutter cancellation, at $\Delta R_{\mathrm{res}} = \SI{8.0}{\metre}$ range and
    $\Delta v = \SI{0.14}{\metre\per\second}$ Doppler resolution over a
    $64\times\SI{10}{\milli\second}$ CPI. (a)~The static direct-path/ground ridge
    dominates at zero Doppler; (b)~after rank-1 clutter-subspace deflation and the
    single-peak corrections of Section~\ref{sec:rdmap}, the moving UAV survives as
    a single compact peak at $(R_{\mathrm{bi}}, v_{\mathrm{bi}}) \approx
      (\SI{54}{\metre}, +\SI{1.6}{\metre\per\second})$, matching the ray-traced
    ground truth ($\SI{59.5}{\metre}$, $+\SI{1.74}{\metre\per\second}$) to within
    the range bias of Section~\ref{sec:results}.}
  \label{fig:rdmap}
\end{figure*}

Fig.~\ref{fig:rdmap} shows the Range-Doppler (RD) map of a representative
mid-transit CPI before and after clutter suppression. Before clutter
cancellation (Fig.~\ref{fig:rdmap}(a)) a strong zero-Doppler ridge --
the static direct UE$\to$gNB path plus the ground bounce -- dominates
across all ranges. After a rank-1 clutter-subspace deflation pass
(Fig.~\ref{fig:rdmap}(b)) this ridge is removed and the moving UAV
emerges as a single compact peak at its true bistatic range and Doppler.

Obtaining this clean single peak required two signal-processing
corrections to the gNB pipeline, each of which we found could otherwise
fragment a single physical target into several competing detections
(Section~\ref{sec:discussion}):
(i)~the deflation rank is held at $K_c=1$ so that only the direct-path/static
subspace is removed --- in this flat-LOS single-target scene the target
is the second-strongest return, and a higher rank subtracts the target
itself; and
(ii)~a single-peak non-maximum-suppression radius wide enough to absorb
the comb-least-squares window sidelobes ($\pm 18$ range bins) collapses
the target's main-lobe and sidelobes into one detection. With these in place, the RD
map contains one dominant target peak per CPI over the whole transit.

\subsection{Detection and Tracking Performance}

%% UAV measurement signature over the transit (Det.0 vs ground truth)
\begin{figure}[!tbp]
  \centering
  \includegraphics[width=\columnwidth]{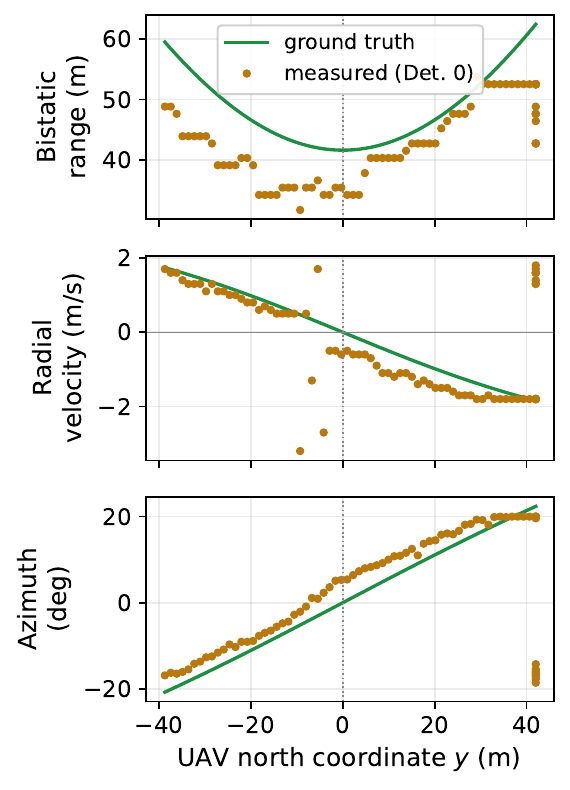}
  \caption{UAV flat-LOS transit: the strongest detection per CPI
    (Det.\,0) vs.\ ray-traced ground truth, over the UAV's north coordinate
    $y$. The bistatic range (top) sweeps to a minimum at the $y=0$ broadside
    crossing and back; the radial velocity (middle) passes through zero and
    \emph{reverses sign} (approach $\rightarrow$ recede); and the azimuth
    (bottom) sweeps monotonically through boresight. The green curve is the
    signed ground truth; orange markers are the measurements.}
  \label{fig:track}
\end{figure}

\begin{table}[!tbp]
  \centering
  \caption{UAV flat-LOS transit. Mean $\pm$ sample std over $N = \CAMPflatN$
    independent runs differing only in the noise seed. Measurement rows compare
    the strongest detection per CPI against ray-traced ground truth; position rows
    are the downstream 2-D EKF. CPIs inside the near-baseline blind zone are
    excluded (Section~\ref{sec:sysmodel}). OSPA and GOSPA are the (generalized) optimal sub-pattern assignment distances.}
  \label{tab:single_uav_meas}
  \footnotesize
  \begin{tabular}{@{}lc@{}}
    \toprule
    Quantity                                           & Value ($N = \CAMPflatN$ runs)             \\
    \midrule
    \multicolumn{2}{@{}l}{\textit{Measurement accuracy (per CPI, vs.\ ground truth)}}              \\
    Excess bistatic range RMSE (m)                     & \CAMPflatRangeRmse                        \\
    Bistatic range-rate RMSE (m/s)                     & \CAMPflatVelRmse                          \\
    Azimuth RMSE (deg), 4-element gNB Rx ULA           & \CAMPflatAzRmse                           \\
    Range-rate sign agreement (\%)                     & \CAMPflatSignAgree                        \\
    \midrule
    \multicolumn{2}{@{}l}{\textit{Detection}}                                                      \\
    Detection coverage (\% of observable CPIs)         & \CAMPflatCoverage                         \\
    Single-detection CPIs (\%)                         & \CAMPflatSingleDet                        \\
    CFAR margin (dB above threshold)                   & \CAMPflatMargin                           \\
    \midrule
    \multicolumn{2}{@{}l}{\textit{End-to-end 2-D EKF position (target, over all track fragments)}} \\
    Range ($\Delta x$) RMSE (m)                        & \CAMPflatDx                               \\
    Cross-range ($\Delta y$) RMSE (m)                  & \CAMPflatDy                               \\
    Track continuity (\%)                              & \CAMPflatCont                             \\
    \quad with coast-bridging (Section~\ref{sec:frag}) & \CAMPflatContBridge                       \\
    Track identifiers per run                          & \CAMPflatTrackIds                         \\
    \quad with coast-bridging                          & \CAMPflatTrackIdsBridge                   \\
    OSPA ($c=\SI{20}{\metre}$, $p=2$) (m)              & \CAMPflatOspa                             \\
    GOSPA ($\alpha=2$, $c=\SI{20}{\metre}$) (m)        & \CAMPflatGospa                            \\
    \bottomrule
  \end{tabular}
\end{table}

Table~\ref{tab:single_uav_meas} reports the pipeline's performance on
the flat-LOS transit, separating the raw \emph{measurement} accuracy
(the strongest detection per CPI compared against ray-traced
ground truth) from the downstream 2-D EKF \emph{position} accuracy. The two
answer different questions --- what the gNB measured, and where the tracker put
the target.

At the measurement level the pipeline is clean. It produces a detection in every
observable CPI of the transit, and the strongest detection tracks the target's
excess bistatic range to \CAMPflatRangeR~m root-mean-square error (RMSE), its bistatic range rate to
\CAMPflatVelR~m/s, and its azimuth to \ang{\CAMPflatAzR}. All three
characteristic signatures of a broadside transit are recovered as a single
smooth track: the range sweeps to a minimum near the $y=0$ crossing and back
out; the range rate passes through zero and \emph{reverses sign}
(approach~$\rightarrow$~recede); and the azimuth sweeps monotonically through
boresight (Fig.~\ref{fig:track}).

\begin{figure}[!tbp]
  \centering
  \includegraphics[width=\columnwidth]{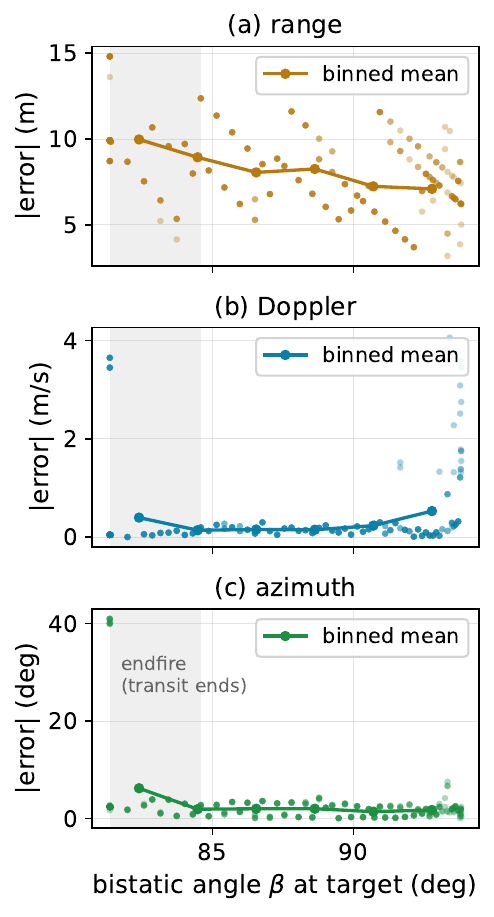}
  \caption{Absolute measurement error versus bistatic angle $\beta$ (angle at the
    target between the target--gNB and target--nrUE rays), pooled over five seeds
    ($368$ target-matched CPIs; scatter plus binned mean). Range~(a) and
    azimuth~(c) error grow toward endfire --- the low-$\beta$ transit ends (shaded)
    --- with endfire/broadside ratios of $1.34\times$ and $1.94\times$. Doppler~(b)
    is the exception, degrading toward broadside where the sign-mirror ambiguity is
    strongest.}
  \label{fig:err_angle}
\end{figure}

The range figure is the one governed by the corrected
definition of Section~\ref{sec:sysmodel}. A range RMSE of
\CAMPflatRangeR~m sits at about one \SI{8.0}{\metre} resolution cell
of~\eqref{eq:dr} --- slightly above it. This is not sub-resolution-cell accuracy: the error is dominated not by how precisely an
isolated peak can be located (which the \SI{1.22}{\metre} zero-padded bin spacing
would allow to be far finer) but by the geometry at the ends of the transit,
where the target approaches endfire and both the range and the bearing estimate
degrade. Averaged over the whole transit the pipeline therefore delivers
about one resolution cell of range accuracy, and the run-to-run spread
(Table~\ref{tab:single_uav_meas}) reflects how much of the poor-geometry tail a
given noise draw happens to survive. Fig.~\ref{fig:err_angle} makes this
geometry dependence explicit: binned against the bistatic angle $\beta$ (at the
target, between the target--gNB and target--nrUE rays), the range error grows by
$1.34\times$ from broadside toward the low-$\beta$ transit ends and the azimuth
error by $1.94\times$ (correlations $-0.51$ and $-0.24$ over $368$ target-matched
CPIs). Because the target is high ($z = \SI{55}{\metre}$) and the baseline
single, $\beta$ itself stays within $81$--$94^\circ$ --- a narrow angular
diversity that is a limitation of the single-pair geometry and a further
motivation for the multistatic extension of Section~\ref{sec:future}.

%% Measurement error vs bistatic angle

Downstream, the 2-D EKF tracks the target to \CAMPflatDxM~m in range
($\Delta x$) and \CAMPflatDyM~m in cross-range ($\Delta y$). To avoid rewarding
identifier fragmentation (below), each figure is computed over the target's
\emph{entire} set of track fragments --- at every CPI the estimate closest to
ground truth is scored; this makes the numbers coverage-complete and reproducible run to run.

Those two figures therefore describe the accuracy of the \emph{union} of the
target's tracks, and that is a different quantity from the accuracy of any one
of them. Because the union is scored per CPI at whichever fragment is closest,
its accuracy improves with the number of fragments. Scored that way, over the one track that carries the target for the
largest share of the transit, the same runs give a 2-D position RMSE of
\CAMPFlatPrimaryRmse~m. Both are reported here deliberately: the pooled figure
answers ``how well was the target covered'', the single-identifier figure
answers ``how good is the track a consumer acts on'', and the gap between them
is a direct measure of the fragmentation quantified below.
The cross-range figure is about one \SI{8.0}{\metre} range cell, at the
measurement floor. The range figure is several cells larger, and it is
\emph{not} additional measurement error: the measurement-level range RMSE is
only \CAMPflatRangeR~m. It is the amplification incurred when the 2-D filter
converts the bistatic slant range to a ground-plane $x$ under a single elevation
prior --- largest near the transit ends, where the target approaches array
endfire and the ellipse-to-ground mapping is most sensitive to residual range
and bearing error. A filter with elevation observability (a 3-D state, or a
second receiver) would relax this term; within the 2-D filter it dominates
$\Delta x$.

This qualifier is why $\sigma_R$ is treated as a
result rather than a detail. Because a single bistatic pair provides no elevation
observability, the 2-D filter must convert the measured slant range to a ground
range using the height prior $h_{\mathrm{tgt}}$ (Table~\ref{tab:params}), and it
must weigh that range against the bearing using $\sigma_R$. Get either wrong and
the position degrades even though the \emph{measurement} is untouched ---
which is precisely what Table~\ref{tab:htgt_ablation} shows.

\begin{figure}[!tbp]
        \centering
        \includegraphics[width=\columnwidth]{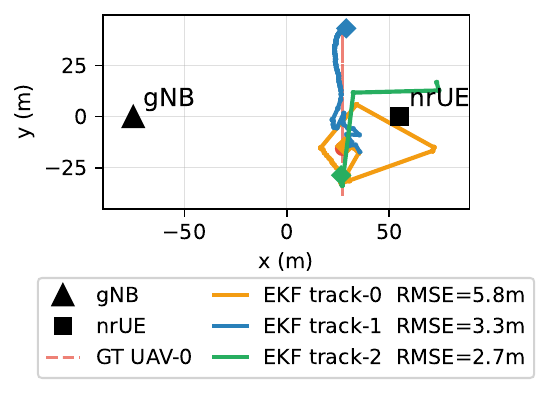}
        \caption{Flat-LOS transit in the ground plane: EKF position estimates
          (solid) against the ray-traced ground-truth UAV path (dashed), with the
          gNB and nrUE at their scene positions. The estimates follow the
          south--north transit over its full extent. The transit is covered by several
          track identifiers because the association gate re-spawns a track through the
          multi-detection CPIs; the fragments all lie on the same true trajectory and
          are scored together, so the reported position accuracy and continuity reflect
          the target's full observed track rather than a single fragment. Only states
          updated by a real detection are drawn (baseline tracker; the coast-bridging pass
          of Section~\ref{sec:frag}, which republishes predicted states through
          detection gaps, is disabled here). Representative run (seed 1001).}
        \label{fig:track_overlay}
      \end{figure}

\textbf{Track fragmentation.}
\label{sec:frag}
Fig.~\ref{fig:track_overlay} overlays the EKF state estimates on
the ground-truth transit in the ground plane. The estimated track follows
the south--north transit over its full extent. The transit is
covered by \CAMPflatTrackIds{} track identifiers rather than one: the tracker's
association gate opens a new identifier whenever a run of CPIs yields more
than one detection and the incumbent track is not the one updated. This is an
\emph{identifier} limitation, not a coverage one --- the fragments all lie on
the same true trajectory, and scored over all of them (as
Table~\ref{tab:single_uav_meas}'s position rows are) the target is tracked for
\CAMPflatContM\% of its observable transit. As single-number multi-target
scores the transit yields an optimal sub-pattern assignment (OSPA) of
\CAMPflatOspa~m and a generalized OSPA (GOSPA) of
\CAMPflatGospa~m ($c=\SI{20}{\metre}$, $p=2$); the GOSPA decomposition ---
localization \CAMPflatGospaLoc~m, missed \CAMPflatGospaMiss~m, false
\CAMPflatGospaFalse~m --- shows the error is split between localization and the
missed CPIs of an intermittent track, with a negligible false-track term, i.e.\
the fragmentation costs identifier continuity rather than spawning spurious
targets. The tracker already applies the
standard consolidation stack --- $M$-of-$N$ track confirmation, intra-CPI
detection clustering, predicted-measurement (Mahalanobis) gating, adaptive
coast, and track merging with duplicate removal --- which suppresses spurious
identifiers but does not, on its own, hold a single identifier across the
periodic multi-detection CPIs. The time-driven maintenance pass described next
recovers most of this continuity by keeping a confirmed track alive across
detection gaps, but it does \emph{not} reduce the identifier count --- a
republished coasted track is not merged with the fragment that later re-confirms,
so continuity and identity move independently. Collapsing the transit onto a
single stable identifier remains for future work: graph-based or joint
probabilistic data association (JPDA), or admitting more than one detection per
track per CPI, would consolidate the fragments directly.

\begin{figure}[!tbp]
        \centering
        \includegraphics[width=\columnwidth]{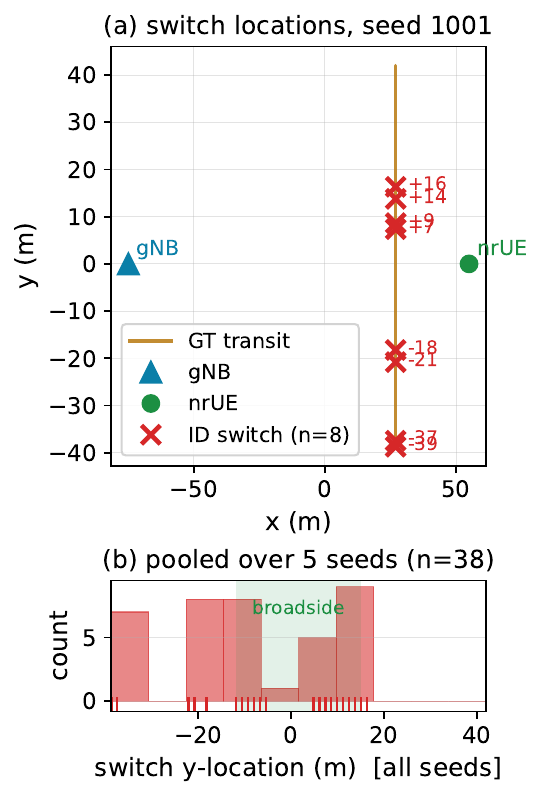}
        \caption{Track-ID-switch locations for the single-UAV transit, baseline tracker
          (coast-bridging disabled).
          (a)~Ground-plane view (seed 1001): the ground-truth transit with gNB/nrUE, each
          \textcolor{red}{$\times$} marking a CPI where the target's matched track
          identifier changes. (b)~Switch locations pooled over all five seeds ($N=\CAMPflatIdSwitch$).
          Switches occur throughout the transit and, if anything, concentrate near
          broadside rather than at the endfire ends --- fragmentation is driven by
          per-CPI detection dropouts, not a specific geometry. Coast-bridging
          (Section~\ref{sec:frag}) restores continuity but not single-identity tracking:
          the per-run identifier count is \CAMPflatTrackIds{} without bridging and
          \CAMPflatTrackIdsBridge{} with it.}
        \label{fig:id_switches}
      \end{figure}

Two properties of the identifier count reported here should be read carefully.
It is measured over a fixed pool of three track slots, so it is bounded above by
three; and because a released slot is immediately reusable, two tracks that are
disjoint in time can share one identifier and be counted once. The figure is
therefore a \emph{lower bound} on the number of distinct tracks the transit is
split into, and the identity metrics of the multi-object-tracking literature
(identity switches and IDF1, which are invariant to identifier relabelling) are
the sounder basis for comparison. Fig.~\ref{fig:id_switches}
locates the switches for the baseline tracker (coast-bridging disabled): pooled over five
seeds they number $\CAMPflatIdSwitch$ ($\sim$$\CAMPflatIdSwitchPer$ per transit) and, rather than clustering at
      the endfire ends where the geometry is worst, they fall throughout the transit
      and if anything concentrate near broadside --- consistent with the cause being
      per-CPI detection dropouts (a CPI carrying two returns re-spawns the track)
      rather than a geometry-specific failure.

      \input{coast_bridging}

      %% Track-ID-switch locations along the transit

      %% 2-D top-down EKF track overlay (flat-LOS)

      \textbf{Two ablations.} Table~\ref{tab:htgt_ablation} isolates the two tracker
      parameters that could plausibly be suspected of doing the work, re-running the
      same five seeds with each changed in turn. In both, the measurement rows are
      unchanged within run-to-run variation: both parameters live
      in the tracker, not in the gNB, so neither can affect what is detected.

      \emph{Height prior.} With the prior mismatched to a low-altitude class
      ($h_{\mathrm{tgt}} = \SI{15}{\metre}$ against the true \SI{55}{\metre}), the
      range ($\Delta x$) RMSE degrades from \CAMPflatDxM~m to \CAMPhlowDxM~m. The
      mechanism is geometric: with too low an assumed height, $h(\mathbf{x})$ can only
      reconcile the measured slant range by placing the target at a farther
      \emph{ground} range, biasing $\hat{x}$ outward. Cross-range degrades by a comparable factor
      (\CAMPhlowDyM~m vs \CAMPflatDyM~m). Scored over the target's full track
      (Section~\ref{sec:frag}) the mis-specified prior therefore degrades the estimate
      in \emph{both} dimensions rather than biasing range alone: the outward-biased
      state also weakens the association of subsequent detections, so the penalty does
      not stay confined to the radial direction. The radial bias remains the primary
      mechanism --- it is what $h(\mathbf{x})$ can misresolve --- but its cost is paid
      across the whole track, which is why the height prior is reported as a
      first-order deployment parameter rather than a tuning detail.

      This sensitivity is the clearest motivation for removing the prior altogether: a
      3-D EKF that carries an altitude state, fed by either a vertical receive aperture
      or a second bistatic pair, makes the position independent of any assumed height.
      Sections~\ref{sec:upa} and~\ref{sec:multistatic} realize exactly those two routes
      and measure both live.

      \emph{Range noise.} Setting $\sigma_R$ to the monostatic $c/(2B) =
    \SI{4.0}{\metre}$ --- the value the corrected definition of~\eqref{eq:dr} rules
      out --- makes the filter over-trust range by a factor of two. Over the full
      transit this turns out \emph{not} to move the position estimate significantly:
    $\Delta x$ RMSE is \CAMPsigoldDxM~m against \CAMPflatDxM~m at the correct
    $\sigma_R$, a difference well inside the run-to-run spread. The position error here is
      set by the height prior and by the bearing, not by how much the filter trusts
      range.

      \begin{table*}[!tbp]
        \centering
        \caption{Ablations, paired by seed: each column repeats the same
          $N = \CAMPflatN$ runs with a single parameter changed. Left, the height prior
          is mismatched to a low-altitude class (true altitude \SI{55}{\metre}); right,
          the EKF range noise is set to the erroneous monostatic $c/(2B)$ value. The
          height prior and $\sigma_R$ both live in the tracker, downstream of the gNB
          measurement, so the measurement rows are identical across the three columns by
          construction; we report them once and vary only the tracker rows below. (The
          three independent seeded runs agree on the measurement to within run-to-run
          noise, confirming the ablated parameter does not propagate upstream.)}
        \label{tab:htgt_ablation}
        \footnotesize
        \begin{tabular}{@{}lccc@{}}
          \toprule
          Quantity                        & $h_{\mathrm{tgt}}=\SI{15}{\metre}$  & $\sigma_R=\SI{4.0}{\metre}$
                                          & Baseline                                                                                    \\
                                          & (mismatched)                        & (monostatic $c/2B$)         & (correct)               \\
          \midrule
          \multicolumn{4}{@{}l}{\textit{Measurement (gNB) --- tracker-independent, shared across columns}}                              \\
          Excess bistatic range RMSE (m)  & \multicolumn{3}{c}{\CAMPflatRangeR}                                                         \\
          Bistatic range-rate RMSE (m/s)  & \multicolumn{3}{c}{\CAMPflatVelR}                                                           \\
          Azimuth RMSE (deg)              & \multicolumn{3}{c}{\CAMPflatAzR}                                                            \\
          Detection coverage (\%)         & \multicolumn{3}{c}{\CAMPflatCov}                                                            \\
          \midrule
          \multicolumn{4}{@{}l}{\textit{2-D EKF position --- tracker-dependent}}                                                        \\
          Range $\Delta x$ RMSE (m)       & \CAMPhlowDxM                        & \CAMPsigoldDxM              & \textbf{\CAMPflatDxM}   \\
          Cross-range $\Delta y$ RMSE (m) & \CAMPhlowDyM                        & \CAMPsigoldDyM              & \textbf{\CAMPflatDyM}   \\
          Track continuity (\%)           & \CAMPhlowContM                      & \CAMPsigoldContM            & \textbf{\CAMPflatContM} \\
          \bottomrule
        \end{tabular}
      \end{table*}

      \subsection{Reproducibility}
      \label{sec:repro}

      Every quantity needed to rebuild this experiment is reported in the paper: the scene geometry and trajectory, the full
      parameter set (Table~\ref{tab:params}), the detection report format
      (Table~\ref{tab:report}), the link budget and noise figures used to calibrate
      the injected AWGN (Section~\ref{sec:calib_validation}), and the statistical
      protocol below.

      Runs differ \emph{only} in the seed of the injected thermal noise. The five
      flat-LOS runs use seeds $1001$--$1005$, and the NLOS and ablation campaigns
      reuse the same five seeds, which makes those comparisons paired: a difference
      between two configurations is then attributable to the parameter changed rather
      than to the noise draw. The UAV kinematics and the ray-traced propagation are
      deterministic given the scene, so the seed is the only stochastic input.
      Statistics are reported as mean $\pm$ sample standard deviation over the
    $N=\CAMPflatN$ runs, except for latency, which is pooled over per-CPI samples
      and reported with percentiles; we keep the two counts distinct and label them
    $N$ and $n$ respectively.

      \input{upa_elevation}
      \input{multistatic}

      %% =========================================================
      \section{Discussion}
      \label{sec:discussion}

      The integration of UL-SRS-based sensing within the OAI gNB PHY
      and O-RAN Near-RT RIC xApp demonstrates the feasibility of
      repurposing existing NR uplink signals for UAV surveillance
      without spectrum or waveform changes. Key design insights are:

      \textbf{PHY vs.\ xApp split:} Placing clutter cancellation, the
      Range-Doppler transforms, CFAR and AoA estimation inside the gNB PHY gives them
      access to calibrated per-port channel estimates, while the xApp handles only
      association and EKF tracking, keeping the Near-RT RIC interface insensitive to
      the heavy signal processing. The cost, however, is real: the
      PHY stage runs in \CAMPlatpipe~ms per CPI
      (Table~\ref{tab:e2e_latency_agg}), \emph{not} sub-millisecond. It fits inside the
      \SI{640}{\milli\second} CPI accumulation window, but it is dominated by the
    $4096\times128$ zero-padded transforms across four ports and is not optimized
      for speed. Architecturally, this split --- heavy processing inside the gNB PHY, detection
      results lifted to the xApp via E2 --- mirrors the dApp
      model~\cite{polese_dapp,oran_isac_dapp}, where inference microservices run at
      the O-DU with direct IQ access and report compact outputs upward. The difference
      is lifecycle: the current implementation is a static OAI PHY patch; a
      production-grade sensing dApp would be a containerized service with a
      standardized IQ ingestion interface (Open Fronthaul or E3) deployable across
      O-DU vendors.

      \emph{Three senses of ``real time''.} These three senses are easily conflated
      and are separated here. (i)~\emph{Radio time}: the pipeline consumes SRS at
      the rate the air interface delivers it, and every rate in this paper is computed
      against radio time. (ii)~\emph{In-CPI deadline}: each stage completes within one
      CPI accumulation window, which is the property that makes the pipeline
      non-accumulating. (iii)~\emph{Wall-clock real time}: this we do \emph{not}
      claim. The evaluation host runs the ray-traced channel emulator, the full NR
      stack and the RIC concurrently, and the emulator is the bottleneck, so a
      \SI{640}{\milli\second} CPI of simulated time takes substantially longer than
      \SI{640}{\milli\second} of wall-clock time to produce. Senses (i) and (ii) are
      properties of the sensing pipeline and carry over to a hardware deployment;
      sense (iii) is a property of this simulation host and does not. A quantitative
      real-time budget --- host utilization, missed-deadline count and frame-drop rate
      against a wall-clock reference --- is not reported here and is left as
      verification work alongside the hardware evaluation
      (Section~\ref{sec:future}). The dashboard's polling interval is a display-side
      refresh and is excluded from every latency figure in
      Table~\ref{tab:e2e_latency_agg}, which measures only the sensing and E2 path.

      \textbf{Per-detection SNR.} Each detection reports
      the range-Doppler cell SNR --- the echo power at the detected cell over the
      local CFAR noise estimate --- which is a per-target radar SNR (median
    $\CAMPpdOpSnr$~dB at the operating point) and is the quantity against which
      detection probability is characterized in Appendix~\ref{sec:detchar}. This is
      distinct from the wideband SRS \emph{channel-estimate} SNR
      ($10\log_{10}(P_{\mathrm{signal}}/P_{\mathrm{noise}})$ over the received SRS
      samples, $\approx 46$--$51$~dB), which measures the
      nrUE$\leftrightarrow$gNB direct-link reception quality common to all detections
      in a CPI; we retain that quantity only as a link-quality diagnostic (the
      loaded-link comparison of Appendix~\ref{sec:comms}). The raw echo sits below the
      per-sample noise floor and is lifted into the cell by the
    $\sim\SI{50}{\decibel}$ coherent processing gain of the \SI{640}{\milli\second}
      CPI (Section~\ref{sec:calib_validation}).

      \textbf{Detection-to-track association.} The association needs further work,
      though the time-driven coast-bridging already improves track continuity.
      Extending the per-CPI optimal (Hungarian) assignment to multiple simultaneous
      UAVs is a natural next step once the single-target geometry limits above are
      addressed.

      \textbf{Coverage and range.} The evaluated geometry is a compact bistatic
      triangle --- gNB--target and target--UE legs of about \SI{111}{} and \SI{60}{\metre}
      across a \SI{130}{\metre} baseline. What sets the detection range there is the
      \emph{echo}-cell SNR, not the strong direct link: a realistic small-UAV return
      ($\sim$$-15$~dBsm) is received at about \SI{27}{\decibel} of range-Doppler cell
    SNR, only a few decibels above the \CAMPpdOpSnr~dB needed for
  $P_{\mathrm{D}}=0.9$. Because the bistatic cell SNR falls as the fourth power of range, that
    few-decibel echo margin lets a single pair reach only modestly beyond the tested
    legs for a small target; materially larger coverage follows from a larger target
    RCS, a shorter baseline, longer coherent integration, or --- most effectively ---
    a second bistatic pair that both widens the range-product envelope and fills the
    Doppler blind angles (the multistatic direction of Section~\ref{sec:future}).
    The gNB's four-element array is used for angle estimation rather than receive
    beamforming, so this budget already reflects the true, non-beamformed
    receive aperture rather than an idealized array gain. Three effects other than
    raw SNR further bound the envelope: a near-baseline clutter dead zone of a few
    tens of meters (the direct-path skirt the range gate rejects); a bistatic Doppler
    ambiguity of $\pm\lambda\,\mathrm{PRF}/2 \approx 4.5$~m/s at the
    \SI{10}{\milli\second} SRS period (the speed sweep confirms robust detection to
    the tested \CAMPspdHi~m/s, but faster targets alias unless the period is
    shortened, at the cost of coherent-integration gain); and the bistatic-angle
    projection $v_{\mathrm{bi}} \propto \cos(\beta/2)$, which hides targets crossing the
    baseline bisector until a second pair views them from another angle.

    \textbf{UL-SRS power budget.} The wideband sounding that sets the range
    resolution is not free in uplink power. Sounding all 106~PRBs at
    \SI{30}{\kilo\hertz} SCS drives the UE power amplifier (PA) essentially
    continuously through the SRS symbol, and at the \SI{23}{\dBm} UE maximum,
    repeating that every \SI{10}{\milli\second} leaves little headroom --- most
    acutely for a cell-edge target already power-limited by path loss. Several
    standard mechanisms trade this budget against sensing performance without new
    hardware. \emph{Aperiodic (type-1) SRS}, triggered by radio resource control
    (RRC) only when the gNB needs an update, removes the standing duty cycle at the
    cost of a downlink control information (DCI) order latency of about one uplink
    slot ($\sim$\SI{5}{\milli\second}); \emph{adaptive periodicity} that relaxes the
    SRS period from \SI{10}{\milli\second} to \SIrange{40}{80}{\milli\second} while no
    target is tracked cuts the duty cycle $4$--$8\times$ at a proportional drop in
    update rate. \emph{Partial-band SRS} that hops $N$ sub-bands over successive CPIs
    and stitches them in the Doppler domain recovers the full-aperture resolution
    over $N\,T_{\mathrm{SRS}}$ while cutting per-CPI power by $1/N$. On a
    multi-antenna UE, \emph{precoded SRS} beamformed toward the gNB raises the
    effective isotropic radiated power (EIRP) by 3--6~dB for the same PA power, and
    standard NR open-loop power control,
  $P_{\mathrm{SRS}} = \min\!\big(P_{\max},\, P_0 + \alpha\,\mathrm{PL} +
  10\log_{10} M_{\mathrm{SRS}} + f\big)$, can lower $P_0$ for a close-range target
    to spare budget for a distant one. Our testbed runs the fixed
    \SI{10}{\milli\second} full-band operating point; selecting among these per
    deployment --- by range, target dynamics, and UE capability --- is a practical
    extension for the hardware phase.

    \section{Future Work}
    \label{sec:future}

    \textbf{Detector characterization.} Appendix~\ref{sec:detchar} calibrates the
    realized $P_{\mathrm{FA}}$ to its nominal value and reports $P_{\mathrm{D}}$
    against the measured echo-cell SNR and bistatic range; what remains is threefold.
    First, the single flat transit spans only a narrow band of bistatic angles, so
    the angular dependence of $P_{\mathrm{D}}$ needs multi-geometry campaigns to
    separate it from range. Second, at echo levels well above the modeled UAV RCS the per-CPI detection
    breaks down and $P_{\mathrm{D}}$ settles onto a floor (Appendix~\ref{sec:detchar});
    this whole-CPI dropout is deflation-rank-independent (higher deflation rank makes
    it worse, not better), so the fix is an upstream diagnosis of the
    CFO/deflation/CFAR-reference interaction under echo over-drive rather than a
    deflation-rank change. Third, the tracking results here
    still run on the empirical below-noise operating point; folding the calibrated
    threshold into the end-to-end pipeline, and confirming the tracking metrics hold
    at a controlled $P_{\mathrm{FA}}$, is the natural consolidation.

    \textbf{Tracking.} A single visible target should sustain one identifier. The
    time-driven coast-bridging pass (Section~\ref{sec:frag}) already closes most of
    the gap between near-complete detection coverage and track continuity, raising
    the latter from \CAMPflatCont\% to \CAMPflatContBridge\%; the residual
  $100 - \CAMPflatContBridgeM \approx \SI{12}{\percent}$ is track initiation and
    re-confirmation latency --- the CPIs before a confirmed track first exists, which
    a pass operating only on already-confirmed tracks cannot fill --- and \emph{not}
    detection dropouts, whose longest run (\SI{\CAMPflatDetGapMax}{\second}) sits
    well inside the \SI{\CAMPflatCoastMaxS}{\second} bridging horizon.
    It does not, however, improve identity: the per-run identifier
    count is \CAMPflatTrackIds{} without bridging and \CAMPflatTrackIdsBridge{} with
    it, because a republished coasted track is never merged with the fragment that
    later re-confirms. What remains is collapsing the transit onto a \emph{single}
    identifier --- better peak grouping, a wider association gate informed by the
    predicted track, and track management that absorbs a multi-detection CPI without
    re-spawning (e.g.\ JPDA) --- and shortening the dropouts themselves through the
    detection layer.

    \textbf{Geometry.} Removing the height prior was the single largest structural
    improvement available, and this paper realizes it two ways and measures both live at
  $N=\MScampaignN$: a planar receive array adding a vertical aperture
    (Section~\ref{sec:upa}) and a two-user multistatic geometry
    (Section~\ref{sec:multistatic}). The multistatic route is the stronger of the two on
    altitude --- pinned and unbiased --- while the planar array wins horizontal position;
    neither subsumes the other, and a hybrid combining both is the configuration the
    offline analysis favors but that we cannot yet measure (Section~\ref{sec:multistatic}).
    What remains on this axis is resolving the \SI{\MhbiasC}{\metre} elevation bias in
    the vertical-aperture route and improving single-gNB cross-range precision --- which stays
    bound to the eight-antenna receive aperture, with a wider radio unit or a multi-gNB
    deployment being the routes.

    \textbf{Scenario coverage.} The evaluation uses one deterministic trajectory under
    five noise realizations. Multiple trajectories, speeds, UAV altitudes and gNB
    mounting heights (with a visibility heatmap over the two), and RCS values --- and
    more than one simultaneous target, exercising the Hungarian assignment that is
    presently trivial --- are needed before the tracking claim generalizes.

    \textbf{Channel realism and hardware.} A richer multipath rendering (more static
    taps, and the UAV as an extended scatterer rather than a single collapsed echo)
    would tighten the NLOS bound, and rotor micro-Doppler is absent altogether.
    Hardware-in-the-loop validation with USRP radios and a physical UAV remains the
    decisive test.

    \textbf{Standardization.} An ASN.1-specified, interoperability-tested E2SM-SENS is
    a natural follow-on, extending the payload prototype of Table~\ref{tab:report}
    from an illustration of what a sensing E2SM must carry into a standards-track
    proposal. Packaging the PHY sensing stage as a formal dApp~\cite{polese_dapp,ngrg_dapp}
    --- with standardized IQ access via Open Fronthaul metadata and a lifecycle
    managed by the O-RAN Service Management and Orchestration framework --- is the
    natural productization path, and the sensing dApp architecture
    of~\cite{oran_isac_dapp} provides a directly compatible reference design.

    \textbf{Command-and-control integration.} The pipeline presently terminates at a
    visualization dashboard; an operational counter-UAS or urban-air-mobility (UAM)
    deployment would instead consume the tracks over a standard sensor-to-C2
    interface. Emitting the EKF output in an open format --- for example the Dstl
    SAPIENT autonomous-sensor-to-C2 standard, or an ASTERIX/UTM (unmanned traffic management) airspace feed ---
    would let the O-RAN sensing plane drive third-party command-and-control and
    traffic de-confliction without bespoke integration, and is a natural step from
    this digital twin toward an operational counter-UAS or UTM trial.

    %% =========================================================
    \section{Conclusion}
    \label{sec:conclusion}

    We have presented an end-to-end O-RAN testbed for 5G ISAC-based
    UAV detection and tracking using UL-SRS. The testbed integrates
    OAI gNB (with ISAC PHY sensing patch), FlexRIC Near-RT RIC
    (EKF xApp), Sionna RT channel emulation, and a real-time
    live dashboard into a single containerized platform built on open-source
    components. The UL-SRS sensing pipeline achieves
    \SI{8.0}{\metre} bistatic range resolution and
    \SI{0.14}{\metre\per\second} Doppler resolution over a \SI{37.4}{\mega\hertz}
    SRS aperture with a \SI{640}{\milli\second} CPI,
    validated end-to-end in simulation. Over $N = \CAMPflatN$ independent seeded
    runs of a UAV performing a line-of-sight transit through the bistatic footprint,
    the pipeline detects the target in every observable CPI, measuring its excess
    bistatic range to \CAMPflatRangeR~m RMSE --- slightly above one bistatic
    resolution cell, the error set by the endfire geometry --- its bistatic range rate to
    \CAMPflatVelR~m/s (recovering the sign reversal through the crossing), and its
    azimuth to \ang{\CAMPflatAzR}.
    The bistatic EKF tracks the target to \CAMPflatDxM~m in range and
    \CAMPflatDyM~m in cross-range. Because a single bistatic
    pair leaves elevation unobservable, we additionally establish altitude
    observability two ways --- a planar receive array supplying a vertical aperture,
    and a two-transmitter multistatic geometry (two UEs, one gNB) --- the latter
    achieving a fully pinned 3-D track with no height prior.
    Detections reach the tracker as RIC INDICATIONs over an E2 sensing service
    model, at a transport cost of \CAMPlattx~ms --- negligible against the
    \SI{640}{\milli\second} CPI, and the actual sensing pipeline processing. And the
    sensing is not obtained on an idle carrier: with the same link concurrently
    serving a \SI{10}{\mega\bit\per\second} uplink flow (\CAMPloadMcs{} mean MCS over
    the full 106~PRBs), detection coverage is unaffected --- the SRS occupies its own
    comb, so user traffic does not displace the sounding signal
    (Appendix~\ref{sec:comms}).
    Future work will therefore extend to multi-static (multi-cell)
    cooperative sensing together with NLOS-robust detection
    (adaptive clutter-subspace rank and reflection-aware ranking) and richer
    multipath modeling, alongside
    multi-UAV tracking, hardware-in-the-loop USRP validation, and AI-based
    UAV classification xApps for real-time low-altitude air surveillance such as for C-UAS and UAM applications.

    %% =========================================================
    \section*{Acknowledgements}
    The authors would like to thank the OpenAirInterface Software Alliance (OSA),
    the FlexRIC, and NVIDIA Sionna RT teams for the open-source projects.

    Note that AI tools (Anthropic Claude and Code, OpenAI ChatGPT)
    were used to assist in the preparation of this work, including
    platform building, testing and data postprocessing, as well as
    writing assistance, language editing, and manuscript refinement.

    %% =========================================================

    %% =========================================================
    %% Supporting studies. Moved out of the main Evaluation at v0.9: each one
    %% characterises the platform rather than supporting either headline finding,
    %% and the section had grown to ten subsections. Nothing is cut -- the labels
    %% move with the text, so every cross-reference still resolves.
    \appendices

    \section{Detector Characterization}
    \label{sec:detchar}

    Section~\ref{sec:cfar} noted that the order-statistic threshold is set
    empirically and that the $10^{-4}$ design figure is not, on its own, a
    calibrated false-alarm rate. This subsection closes that gap in two steps: it
    measures the realized $P_{\mathrm{FA}}$ on target-absent data, and it reports
    the detection probability $P_{\mathrm{D}}$ as a function of the echo-cell SNR
    actually presented to the detector and of the target's bistatic range. Both are
    measured on the calibrated detector (OS-CFAR with the closed-form
  $\alpha_{\mathrm{OS}}$ of Section~\ref{sec:cfar} plus a single scene-independent
    offset), so a single operating point underlies both the false-alarm and the
    detection results.

    \subsubsection{Realized false-alarm rate}
    We drive the pipeline with a target-absent replay of the flat scene --- static
    clutter and the emulator's thermal noise, no UAV echo --- and accumulate the
    per-cell CFAR margin (test-cell power minus local threshold) over
  $\CAMPpfaCells$ cell tests across $\CAMPpfaN$ noise seeds. The realized
  $P_{\mathrm{FA}}$ at any design value is then the fraction of cells whose margin
    is positive at that design threshold, read directly off the accumulated
    histogram. The empirical threshold the tracking results use sits far below the
    noise reference, so essentially every tested cell crosses it: its realized
  $P_{\mathrm{FA}}$ is $\CAMPpfaBefore$, four orders of magnitude above the
    nominal $10^{-4}$. False-alarm control in that configuration is delegated
    entirely to the downstream selectivity (Doppler and range gates, the
    keep-strongest cap, single-peak non-maximum suppression, and the detection
    cap), not to the CFAR probability. Substituting the correct order-statistic
    multiplier $\alpha_{\mathrm{OS}}$ moves the realized rate to $\CAMPpfaAlpha$ at
    a nominal $10^{-4}$; the residual factor is the gap between the single-look
    exponential assumption behind the closed form and the $4$-port non-coherently
    combined range-Doppler cells, which are approximately $\chi^2$ with a higher
    degree count. A single scene-independent offset of $\CAMPpfaCalDb$~dB, obtained
    once from the target-absent margin histogram, removes that residual and attains
  $\CAMPpfaAfter$ --- the nominal rate, now calibrated rather than nominal.

    \begin{figure}[!tbp]
      \centering
      \includegraphics[width=\columnwidth]{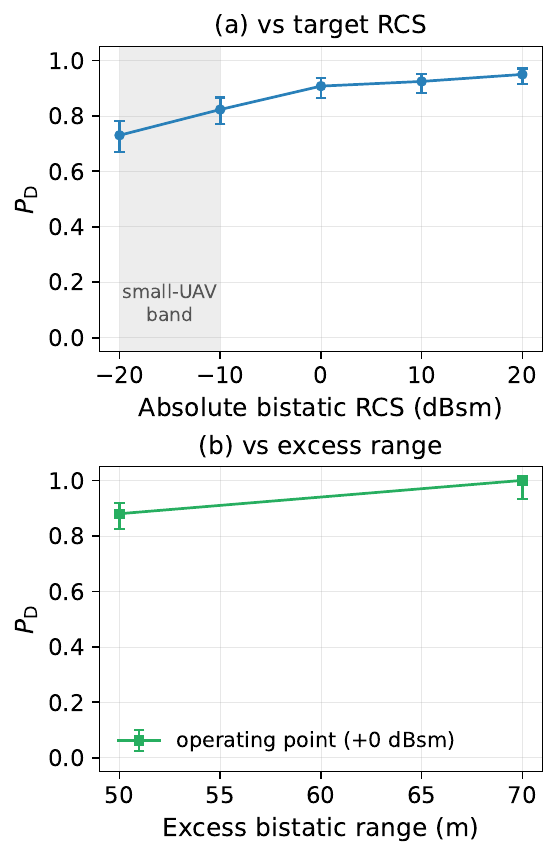}
      \caption{Detector characterization on the flat-LOS transit, calibrated
        OS-CFAR ($\CAMPpdLevels$ RCS levels spanning $\CAMPpdRcsSpan$~dB, over
        $\CAMPpdSeeds$ noise seeds, $\CAMPpdCpis$ observable CPIs per level; error bars
        are Wilson score intervals). (a)~$P_{\mathrm{D}}$ versus absolute bistatic RCS,
        rising monotonically from $\CAMPpdRcsLoPd$ at $\CAMPpdRcsLo$~dBsm to
        $\CAMPpdRcsHiPd$ at $\CAMPpdRcsHi$~dBsm; the shaded band marks the realistic
        small-UAV regime ($-20$ to $-10$~dBsm). RCS is used as the abscissa rather
        than the measured echo-cell SNR because rank-one deflation compresses
        $\CAMPpdRcsSpan$~dB of RCS into $\CAMPpdSnrSpan$~dB of measured SNR and
        reorders it (text). (b)~$P_{\mathrm{D}}$ versus excess bistatic range at the
        operating point ($+0$~dBsm); the far-range falloff is an SNR effect, not a
        geometric blind zone. The transit spans only $\sim$40--80~m of excess range,
        so this covers a correspondingly narrow span.}
      \label{fig:detpd}
    \end{figure}

    \subsubsection{Detection probability versus echo SNR}
    To trace $P_{\mathrm{D}}$ across the detector's operating range we replay the
    same transit with the UAV echo scaled by a constant factor $g$, which scales the
    received echo power by $g^2$ and is equivalent to scaling the target RCS by
  $g^2$ while leaving the static clutter and injected noise untouched. Each level is
    solved so that its \emph{absolute} mean bistatic RCS lands on a requested value in
    dBsm (by inverting the bistatic radar equation on the rendered echo), so the sweep
    is parameterized by target size rather than by a bare amplitude factor, and the
    curve is binned by that physical quantity. Fig.~\ref{fig:detpd} reports
  $P_{\mathrm{D}}$ over $\CAMPpdLevels$ RCS levels spanning $\CAMPpdRcsSpan$~dB,
  $\CAMPpdSeeds$ seed each ($\CAMPpdCpis$ observable CPIs per level), with Wilson
    score intervals.

  $P_{\mathrm{D}}$ rises monotonically with target RCS across the whole ladder,
    from $\CAMPpdRcsLoPd$ at $\CAMPpdRcsLo$~dBsm to $\CAMPpdRcsHiPd$ at
  $\CAMPpdRcsHi$~dBsm. The canonical operating point --- the unscaled transit ---
    detects in $\CAMPpdOpPd$ of observable CPIs.

    \paragraph*{Why RCS and not measured echo-cell SNR}
    Earlier versions binned this curve by the \emph{measured} echo-cell SNR. That
    axis is not well posed for this detector. The rank-one clutter-subspace deflation
    of Section~\ref{sec:clutter} is data-adaptive, so it captures a growing share of
    the target as the echo strengthens: across the $\CAMPpdRcsSpan$~dB of RCS spanned
    here the median measured echo-cell SNR moves by only $\CAMPpdSnrSpan$~dB, and it
    is not even monotonic in RCS. A rank sweep of the unscaled transit makes the
    mechanism explicit --- echo-cell SNR falls from $32$ to $17$ to $15$~dB at
    deflation rank $0$, $1$ and $2$, with rank~$\ge 2$ over-canceling the target,
    which is why the single-target operating choice is $K_c=1$. Binning by an axis
    that compresses $\CAMPpdRcsSpan$~dB of physical variation into
  $\CAMPpdSnrSpan$~dB, and reorders it, produced a curve that appeared to saturate
    and then floor. We therefore report $P_{\mathrm{D}}$ against absolute bistatic
    RCS, which is monotone and physically interpretable, and retain the measured
    echo-cell SNR only as a per-level diagnostic ($\CAMPpdOpSnr$~dB at the operating
    point) with this compression noted.

    \subsubsection{Detection probability versus range}
    At the operating point (Fig.~\ref{fig:detpd}b) $P_{\mathrm{D}}$ falls from
  $\CAMPpdRangeNear$ at a $\CAMPpdRangeNearBin$ excess bistatic range to
  $\CAMPpdRangeFar$ beyond $\CAMPpdRangeFarEdge$~m, as the target recedes from the
    broadside crossing toward the ends of the transit. This falloff is an SNR
    effect, not a geometric blind zone: pooling the reliably-detected (higher
    echo-SNR) levels, where SNR is not the limiting factor, $P_{\mathrm{D}}$ is flat
    at $\CAMPpdRangeFlat$ across the same range span, so the operating-point drop is
    the $1/(R_T R_R)$ bistatic power law weakening the echo at
    the transit extremes rather than a loss of detectability at longer range. The
    single flat transit spans only a narrow band of bistatic angles, so a
    bistatic-angle dependence is not separable here and is left to the
    multi-geometry campaigns of Section~\ref{sec:future}.

    \subsubsection{Detection probability versus speed}
\begin{figure}[!tbp]
      \centering
      \includegraphics[width=0.82\columnwidth]{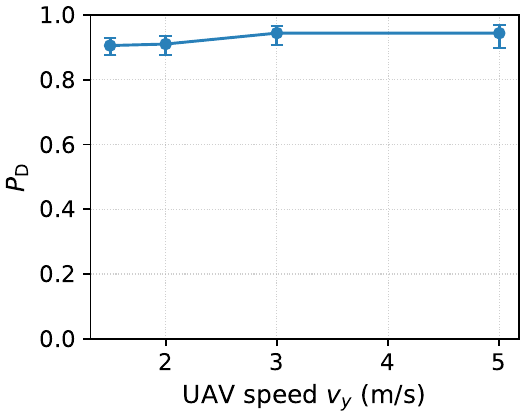}
      \caption{Detection probability versus UAV crossing speed $v_y$ over
        $\CAMPspdN$ speeds ($N=5$ seeds each, calibrated OS-CFAR at the operating echo
        level; error bars are Wilson score intervals). $P_{\mathrm{D}}$ is essentially
        flat, rising slightly from the slowest $v_y$ toward $\CAMPspdHi$~m/s as the
        larger bistatic Doppler separates the target more cleanly from the clutter
        ridge.}
      \label{fig:detspeed}
    \end{figure}

    To probe sensitivity to target dynamics we repeat the operating transit at four
    crossing speeds, $v_y \in \{1.5, 2, 3, 5\}$~m/s, regenerating the ray-traced
    channel at each speed (3~m sphere, operating echo level, $N=5$ seeds) and sizing
    each run to a full transit. Fig.~\ref{fig:detspeed} reports $P_{\mathrm{D}}$
    against $v_y$. Detection is essentially flat across the span and, if anything, improves
    slightly with speed: $P_{\mathrm{D}} = \CAMPspdMinPd$ at the slowest
  $\CAMPspdMinPdV$~m/s and $\CAMPspdOpPd$ at the $v_y = 2$~m/s operating point,
    rising to $\CAMPspdMaxPd$ at $3$ and $5$~m/s. The trend is mild but in the
    expected direction: a faster crossing carries a larger bistatic Doppler, which
    separates the moving target more cleanly from the near-zero-Doppler clutter
    ridge, so the slowest transit --- not the fastest --- is the hardest. Even at
  $\CAMPspdHi$~m/s the per-CPI Doppler migration stays within the
    coherent-integration window of the \SI{640}{\milli\second} CPI, so there is no
    high-speed roll-off. Across the $\CAMPspdLo$--$\CAMPspdHi$~m/s span the detector
    holds $P_{\mathrm{D}} \geq \CAMPspdMinPd$, so within the low-altitude UAV regime
    of interest the sensing is not speed-limited.

    %% P_D vs UAV crossing speed (regenerated channel per speed, calibrated OS-CFAR)

    \subsubsection{Comparison of CFAR estimators}
\begin{figure}[!tbp]
      \centering
      \includegraphics[width=0.86\columnwidth]{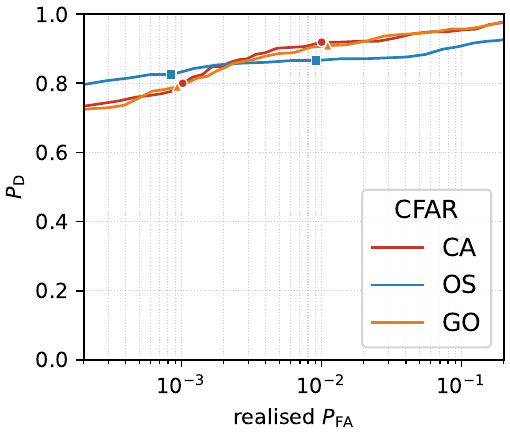}
      \caption{CFAR estimator comparison on the operating transit: detection
      probability versus \emph{realized} false-alarm probability for cell-averaging
      (CA), order-statistic (OS), and greatest-of (GO) references, sweeping a common
      threshold offset over $\CAMPcfarObsCpi$ observable CPIs. Markers denote nominal
      $P_{\mathrm{FA}}\in\{10^{-2},10^{-3},10^{-4}\}$. At the operating echo level the
      three references are indistinguishable in $P_{\mathrm{D}}$
      ($\approx\CAMPcfarOSPdEmFour$, curves overlapping), since detection there is
      dropout- rather than CFAR-limited; they separate only on calibration
      (Table~\ref{tab:cfar}), where OS holds the nominal rate to
      $\CAMPcfarOSPfa\times$ versus $\CAMPcfarCAPfa\times$ for CA.}
      \label{fig:cfarroc}
    \end{figure}

    The characterization so far uses the order-statistic (OS) CFAR that the pipeline
    operates. To justify that choice we replay both the target-absent calibration
    and the target-present operating transit under three reference estimators ---
    cell-averaging (CA), order-statistic (OS), and greatest-of (GO) --- holding every
    other stage fixed and sweeping a common threshold offset to trace each one's
    receiver operating characteristic (ROC)
    (Fig.~\ref{fig:cfarroc}) and its realized false-alarm calibration
    (Table~\ref{tab:cfar}), over the same $\CAMPcfarObsCpi$ observable
    operating-transit CPIs.

    The two estimators separate on \emph{calibration}, not on detection. On the
    realized $P_{\mathrm{FA}}$ delivered at a designed nominal the ordering is
    OS~$\prec$~GO~$\prec$~CA: at a nominal $10^{-4}$ the CA mean reference,
    contaminated by target and clutter energy leaking into its training cells,
    over-delivers by $\CAMPcfarCAPfa\times$, GO by $\CAMPcfarGOPfa\times$, and OS by
    only $\CAMPcfarOSPfa\times$. On detection at a matched realized
  $P_{\mathrm{FA}}=10^{-4}$ (the ROC), by contrast, the three nearly coincide: over
    the $\CAMPcfarObsCpi$ observable CPIs they reach
  $P_{\mathrm{D}}=\CAMPcfarCAPdEmFour$ (CA), $\CAMPcfarOSPdEmFour$ (OS), and
  $\CAMPcfarGOPdEmFour$ (GO) --- a spread of a few percentage points. At the deployed
    echo level the target sits far enough above threshold that detection is limited
    by whole-CPI dropout, not by the CFAR reference, so the choice of estimator
    barely moves $P_{\mathrm{D}}$. (This ROC $P_{\mathrm{D}}$ is the dropout-free
    efficiency at a matched false-alarm rate, and so runs higher than the
    operating-point $P_{\mathrm{D}}=\CAMPpdOpPd$ of Appendix~\ref{sec:detchar}, which
    is measured from the reported detections at the \emph{calibrated} threshold and
    therefore includes the whole-CPI dropouts; the two describe the same target
    through different apertures.)

    The operating choice is therefore settled by calibration alone. A sensing
    detector must hold a \emph{designed} false-alarm rate, and OS is the only one of
    the three that does --- to within $\CAMPcfarOSPfa\times$ of nominal, against
  $\CAMPcfarGOPfa\times$ (GO) and $\CAMPcfarCAPfa\times$ (CA) --- while costing
    nothing in detection at the operating point. OS-CFAR is retained.

    %% CFAR comparison ROC (calibrated CA/OS/GO on the operating transit)

    %% CFAR comparison table (\input generated by make_cfar_comparison.py)
    \begin{table}[!tbp]
      \centering
      \caption{CFAR estimators at matched realized $P_{\mathrm{FA}}$: the
        realized-to-nominal $P_{\mathrm{FA}}$ ratio (calibration, lower is better) and
        $P_{\mathrm{D}}$ at three operating points, over $\CAMPcfarObsCpi$ CPIs. OS is
        the tightest-calibrated; GO is dominated by OS.}
      \label{tab:cfar}
      \input{cfar_table}
    \end{table}

    %% (no FloatBarrier here: with the tracking floats still queued it forced a
    %%  \clearpage and stranded a near-empty page. Floats land beside their refs.)

    \section{Street-Canyon Occlusion Stress Test}
    \label{sec:nlos}

    To bound how far the flat-LOS result carries into an obstructed environment,
    we repeat the identical run --- same trajectory, same altitude, same pipeline
    configuration and the same $h_{\mathrm{tgt}} = \SI{55}{\metre}$ prior ---
    with the six buildings of the street-canyon scene restored
    (Fig.~\ref{fig:scenes}b), so the only change is the propagation
    environment. Table~\ref{tab:nlos} compares the two.

    %% Street-canyon scene: same geometry as the flat scene, buildings restored

    \begin{table}[!tbp]
      \centering
      \caption{Flat-LOS vs.\ street-canyon (NLOS). Identical trajectory, pipeline
        configuration and height prior, and the same $N = \CAMPflatN$ noise seeds ---
        only the buildings differ, so the comparison is paired. Dominant track per run
        for the position rows.}
      \label{tab:nlos}
      \footnotesize
      \begin{tabular}{@{}lcc@{}}
        \toprule
        Quantity                              & Flat-LOS        & Street-canyon         \\
        \midrule
        UAV line-of-sight visibility (\%)     & \CAMPflatLos    & \textbf{\CAMPnlosLos} \\
        Detection coverage, visible CPIs (\%) & \CAMPflatCov    & \CAMPnlosCov          \\
        Excess bistatic range RMSE (m)        & \CAMPflatRangeR & \CAMPnlosRangeR       \\
        Single-detection CPIs (\%)            & \CAMPflatSgl    & \textbf{\CAMPnlosSgl} \\
        Range $\Delta x$ RMSE (m)             & \CAMPflatDxM    & \CAMPnlosDxM          \\
        Cross-range $\Delta y$ RMSE (m)       & \CAMPflatDyM    & \CAMPnlosDyM          \\
        Track continuity (\%)                 & \CAMPflatContM  & \CAMPnlosContM        \\
        \bottomrule
      \end{tabular}
    \end{table}

    Two distinct mechanisms degrade the result, and they act on different
    quantities. First, \emph{occlusion}: the UAV passes behind the buildings for
    much of the transit, and the ray tracer returns no UAV echo at all in more than
    half of the CPIs --- a missed detection there is not a signal-processing failure
    but a geometric impossibility. We therefore separate the two questions the single
    word ``coverage'' would otherwise blur: \emph{visibility} is the fraction of CPIs
    in which an echo exists to be found (\CAMPnlosLos\% here), while \emph{detection
      coverage} is the fraction of \emph{those} CPIs in which the pipeline found it
    (\CAMPnlosCov\%). The pipeline does not miss visible targets; it is simply
    denied a signal for most of the transit. Notably the
    occlusion is driven by the \emph{gNB height} rather than the UAV's: with the
    gNB at \SI{10}{\metre} and buildings up to \SI{51}{\metre}, the shallow
    gNB$\rightarrow$target ray is intercepted even for an above-rooftop target,
    so raising the UAV does not recover visibility. This is a siting observation from
    a single tested geometry: in this scene it is the gNB mounting height, not the
    flight ceiling, that governs coverage. A gNB-height$\times$UAV-altitude sweep
    (Section~\ref{sec:future}) is needed before generalizing it into a placement
    rule. Second, \emph{building
      reflections}: the surviving specular paths produce clutter that a rank-1
    deflation --- tuned for a scene whose only strong static return is the direct
    path --- does not remove, so CFAR reports several detections per CPI instead of
    one and the single-detection rate collapses (Table~\ref{tab:nlos}). The
    strongest detection is then frequently a reflection rather than the
    target, which is what propagates into the position and continuity rows.
    Where the target \emph{is} visible the measurement itself remains usable, so the
    sensing front end degrades gracefully; it is the \emph{selection} of the
    target among reflections that fails.

    These figures are a lower bound on the true NLOS penalty. The
    ray-traced channel retained here keeps only the two strongest static
    (non-UAV) taps per CPI and collapses the UAV's own multi-bounce return into a
    single echo tap. A richer multipath rendering would add further reflected
    clutter and further degrade detection selection. A full NLOS
    characterization --- together with NLOS-robust countermeasures (an adaptive
    or higher deflation rank matched to the scene's clutter subspace,
    reflection-aware detection ranking, and rooftop gNB siting to reduce occlusion)
    --- is left as future work (Section~\ref{sec:discussion}).

    \section{Sensing on a Loaded Link}
    \label{sec:comms}

    The results above establish the sensing (``S'') of ISAC but not the
    communication (``C''): a sensing result obtained on an otherwise idle carrier
    would be a weak basis for claiming integration, since SRS is a reference signal,
    and with no user-plane traffic the scheduler leaves the uplink at its minimum
    grant, so the sensing stage would be observing a link that is doing nothing else.
    We therefore repeat
    the flat-LOS evaluation with the same UE simultaneously carrying a
    \SI{10}{\mega\bit\per\second} uplink data flow through its protocol data unit (PDU) session --- real
    user-plane traffic, traversing the same PHY whose SRS the sensing stage taps.

\begin{figure}[!tbp]
      \centering
      \includegraphics[width=\columnwidth]{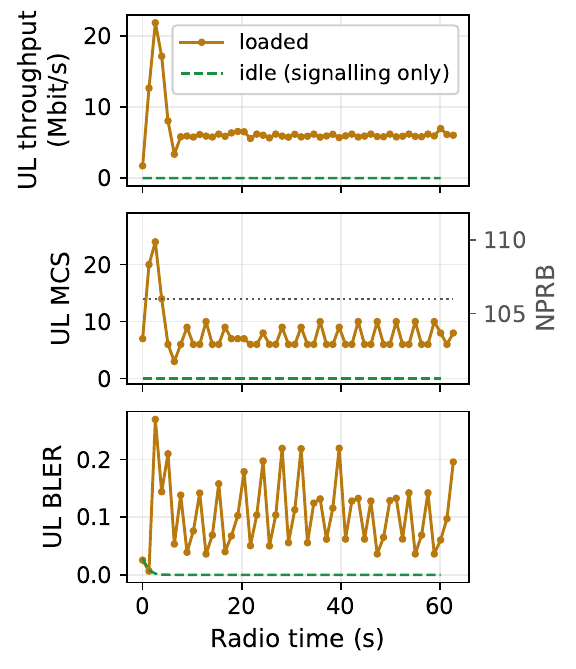}
      \caption{Communication-link KPIs while the sensing pipeline runs, against
        radio time. Loaded (solid), the scheduler allocates the full 106~PRBs and
        raises the MCS to carry a \SI{10}{\mega\bit\per\second} uplink flow; idle
        (dashed), it never leaves MCS~0 and the link carries only signaling. Rates
        are computed against radio time rather than wall-clock time, since the
        simulator runs slower than real time and a wall-clock denominator would
        understate the rate the air interface actually carried.}
      \label{fig:comms}
    \end{figure}

    Fig.~\ref{fig:comms} shows the effect on the communication link. Idle, the
    scheduler sits at MCS~0 on a minimum grant, with throughput at the
    \si{\kilo\bit\per\second} level: nothing but signaling. Loaded, it allocates the
    full 106~PRBs, drives the MCS to a mean of $\sim$\CAMPloadMcs{}, and carries the offered load. The
    sensing stage is running throughout.

    Table~\ref{tab:comms} reports what the load does to the sensing. Detection
    coverage is not reduced by the load, and structurally one should not expect it
    to be: the SRS occupies its own comb and its own symbols, so physical uplink
    shared channel (PUSCH) data never overlaps the sounding resource. The load
    competes with the sensing for the scheduler's resources, not for the sounding
    signal itself.

    An earlier version of this experiment appeared to show accuracy that was
    better under load. That effect was an artifact of the emulator's output
    stage rather than a benefit of ISAC --- a trap inherent to any fixed-point
    channel emulator, and recorded here so it can be avoided. The emulator
    applies a headroom-safe automatic gain before quantizing to the wire's 16-bit
    format, and that gain is capped by a running estimate of the signal peak. With
    the uplink carrying only signaling the waveform is weak and a large gain is
    applied; loaded, the waveform is stronger and the gain is correspondingly
    smaller. The loaded arm therefore sat at a more favorable point in the
    quantizer's range and its channel estimate saw a higher effective SNR --- an
    advantage conferred by the arithmetic, not by the air interface.

    Both arms of
    Table~\ref{tab:comms} are re-run with the emulator's uplink output gain
    pinned to the same fixed value, so the quantizer operating point is
    identical whether or not the link carries data; the two arms use the same scene,
    the same seeds and the same pipeline, and differ only in the presence of the
    user-plane flow. We confirm that neither arm clips, so the pinned gain is not
    itself distorting either case. Under this control the apparent advantage
    disappears: excess range RMSE is \CAMPidleRangeR~m idle versus
    \CAMPloadRangeR~m loaded, range-rate RMSE \CAMPidleVelR versus \CAMPloadVelR~m/s,
    and azimuth RMSE \CAMPidleAzR versus \CAMPloadAzR~degrees --- differences smaller
    than the seed-to-seed spread in every case. Loading the link neither improves
    nor degrades sensing accuracy.

    Uplink load therefore does not reduce detection coverage,
    because PUSCH traffic does not displace the sounding resource. Coverage is
    \CAMPidleCov\% idle and \CAMPloadCov\% loaded while the link carries
    \CAMPloadTput~Mbit/s at a mean MCS of \CAMPloadMcs. Because coverage is measured
    on the detections themselves, upstream of the tracker and independent of the
    output gain, it is unaffected by the artifact described above.

    \begin{table}[!tbp]
      \centering
      \caption{Sensing accuracy with and without a concurrent
        \SI{10}{\mega\bit\per\second} uplink data flow, under \emph{identical
          fixed-point scaling}. Same scene, same seeds ($N = \CAMPidleN$), same
        pipeline, and the emulator's uplink output gain pinned to the same value in
        both arms, so the quantizer operating point cannot differ between them; only
        the user-plane load differs. Neither arm clips. Every sensing difference here
        is smaller than the seed-to-seed spread: load neither improves nor degrades
        accuracy. Detection coverage --- measured on the detections, upstream of the
        tracker and independent of the output gain --- is not reduced by the load,
        because PUSCH does not displace the SRS resource.}
      \label{tab:comms}
      \footnotesize
      \begin{tabular}{@{}lcc@{}}
        \toprule
        Quantity                       & Idle link       & Loaded link     \\
        \midrule
        \multicolumn{3}{@{}l}{\textit{Communication}}                      \\
        UL throughput (Mbit/s)         & $\approx 0$     & \CAMPloadTput   \\
        UL MCS (mean)                  & 0               & \CAMPloadMcs    \\
        \midrule
        \multicolumn{3}{@{}l}{\textit{Sensing}}                            \\
        Detection coverage (\%)        & \CAMPidleCov    & \CAMPloadCov    \\
        Excess bistatic range RMSE (m) & \CAMPidleRangeR & \CAMPloadRangeR \\
        Bistatic range-rate RMSE (m/s) & \CAMPidleVelR   & \CAMPloadVelR   \\
        Azimuth RMSE (deg)             & \CAMPidleAzR    & \CAMPloadAzR    \\
        Single-detection CPIs (\%)     & \CAMPidleSgl    & \CAMPloadSgl    \\
        \bottomrule
      \end{tabular}
    \end{table}

    This is the sense in which the testbed is integrated rather than merely
    co-located: one carrier, one PHY, carrying user data and producing UAV
    detections at the same time.

    \section{End-to-End Latency}
    \label{sec:e2e_latency}

    All latency figures reported here were measured on the host
    machine actually running the testbed: an Apple MacBook Air
    (M5 chip, 16~GB RAM), with the full container stack running locally.
    The chain from UL-SRS reception to dashboard display is:
    (1)~gNB PHY accumulation, $T_{\mathrm{CPI}} = \SI{640}{\milli\second}$
    (deterministic, by construction);
    (2)~the sensing pipeline;
    (3)~transport of the detection report to the xApp;
    (4)~EKF update and track write;
    (5)~dashboard poll, $\leq \SI{1}{\second}$ (a fixed polling period, not a
    processing delay).
    Stages (2)--(4) are measured per CPI and reported in
    Table~\ref{tab:e2e_latency_agg} and illustrated in Fig.~\ref{fig:timing}.

    %% End-to-end timing diagram (SRS -> PHY -> E2 -> EKF -> dashboard)
    \begin{figure*}[!tbp]
      \centering
      \begin{tikzpicture}[
        font=\small,
        stage/.style={draw, rounded corners, minimum height=10mm, minimum width=26mm,
            align=center, fill=blue!5},
        lat/.style={font=\footnotesize, text=black!75},
        >={Stealth[round]}]
        \node[stage] (srs) {(1)~UL-SRS\\accumulation};
        \node[stage, right=9mm of srs] (phy) {(2)~gNB PHY\\{\footnotesize deflation$\cdot$RD$\cdot$CFAR$\cdot$AoA}};
        \node[stage, right=9mm of phy] (e2)  {(3)~E2 / SM-SENS\\transport};
        \node[stage, right=9mm of e2]  (ekf) {(4)~EKF update\\+ track write};
        \node[stage, right=9mm of ekf] (dash){(5)~dashboard\\poll};
        \foreach \a/\b in {srs/phy, phy/e2, e2/ekf, ekf/dash}
        \draw[->, thick] (\a) -- (\b);
        \node[lat, below=1.5mm of srs]  {$T_{\mathrm{CPI}}=\SI{640}{\milli\second}$};
        \node[lat, below=1.5mm of phy]  {\CAMPlatpipe~ms};
        \node[lat, below=1.5mm of e2]   {\CAMPlattx~ms};
        \node[lat, below=1.5mm of ekf]  {\CAMPlatekf~ms};
        \node[lat, below=1.5mm of dash] {$\leq\SI{1}{\second}$};
        \node[font=\footnotesize\itshape, above=1.5mm of srs] {radio time (deterministic)};
        \begin{scope}[on background layer]
          \node[draw=black!40, dashed, rounded corners, fit=(phy)(e2)(ekf),
            inner sep=3mm, label={[font=\footnotesize\itshape,black!60]above:in-CPI wall-clock budget}] {};
        \end{scope}
      \end{tikzpicture}
      \caption{End-to-end timing of the sensing chain, from UL-SRS reception to
        dashboard display. Stage~(1) is deterministic \emph{radio time} --- the
        \SI{640}{\milli\second} CPI that must elapse before a detection exists.
        Stages~(2)--(4) are the per-CPI \emph{wall-clock} processing budget measured in
        Table~\ref{tab:e2e_latency_agg}: the gNB PHY pipeline (\CAMPlatpipe~ms), E2/SM-SENS
        transport (\CAMPlattx~ms), and EKF update (\CAMPlatekf~ms). Stage~(5), the
        dashboard poll ($\leq\SI{1}{\second}$), is a fixed display period, not a
        processing delay, and is excluded from the budget.}
      \label{fig:timing}
    \end{figure*}

    \emph{Cost of the E2 path.} Because both transports of
    Section~\ref{sec:transport} carry the same gNB-side timestamp, stage~(3) is
    directly comparable between them. The direct in-memory queue is not free: it too
    incurs \CAMPlatqueue~ms (p95 \CAMPlatqueueP, max \CAMPlatqueueM~ms;
  $n=\CAMPlatqueueN$), most of it timestamping and report serialization common to
    both paths. The E2 service model's \emph{marginal} cost is therefore small --- in
    a both-transport reference run the same detections cost \CAMPlatqRefTx~ms via E2
    against \CAMPlatqueue~ms via the queue, a difference of under \SI{2}{\milli\second}
    for E2AP encoding, the Stream Control Transmission Protocol (SCTP) hop to the RIC,
    and decoding at the xApp, and consistent with the loaded-link E2 latency of
    \CAMPlattx~ms in Table~\ref{tab:e2e_latency_agg}. Set against the
    \SI{640}{\milli\second} CPI that must elapse before a detection exists at all,
    that overhead is negligible: routing sensing data through a standard E2AP-based
    service model rather than a private side-channel costs well under one percent of
    the sensing latency budget.

    \emph{Worst-case bounds.} We do not infer a worst case from the means. The
    means and standard deviations in Table~\ref{tab:e2e_latency_agg} describe the
    typical case; the maximum and the 95th/99th percentiles, reported alongside
    them, are the observed tail. Taking the measured maxima of stages (2)--(4)
    together with the deterministic $T_{\mathrm{CPI}}$ and the \SI{1}{\second}
polling period gives an \emph{observed} end-to-end worst case; because the
polling period alone contributes up to \SI{1}{\second} and the CPI another
\SI{640}{\milli\second}, an alert latency below \SI{2}{\second} is not
guaranteed by these measurements, and any deployment needing a hard bound should
shorten the polling period (or replace polling with a push) rather than rely on
the processing stages, which are not the dominant term.

\begin{table}[!tbp]
  \centering
  \caption{End-to-end latency by stage, flat-LOS scene with the link loaded.
    Statistics are over pooled per-CPI samples from the $N=\CAMPflatN$ seeded runs;
    $n$ is the number of pooled per-CPI samples entering each statistic --- a
    \emph{sample} count, not a run count. The max column is an observed worst case,
    not an inferred bound. The E2 row is the loaded-link measurement; the
    direct-queue row is a companion reference measured in a separate both-transport
    flat-LOS run (same pipeline and code), where both transports carried the same
    detections. The pipeline row is populated only on CPIs whose report carried the
    gNB's processing duration, which the sensing service model does not convey;
    its $n$ is therefore a subset of the E2AP transport samples, which is why the
    two rows differ by roughly an order of magnitude.}
  \label{tab:e2e_latency_agg}
  \footnotesize
  \begin{tabular}{@{}l@{\hskip 3pt}c@{\hskip 3pt}c@{\hskip 3pt}c@{\hskip 3pt}c@{\hskip 3pt}c@{}}
    \toprule
    Stage                      & Mean $\pm$ std (ms) & p95            & p99            & Max            & $n$            \\
    \midrule
    Sensing pipeline (gNB PHY) & \CAMPlatpipe        & \CAMPlatpipeP  & \CAMPlatpipeQ  & \CAMPlatpipeM  & \CAMPlatpipeN  \\
    \midrule
    \multicolumn{6}{@{}l}{\textit{Detection transport, gNB $\rightarrow$ xApp}}                                          \\
    \quad via E2 / SM-SENS     & \CAMPlattx          & \CAMPlattxP    & \CAMPlattxQ    & \CAMPlattxM    & \CAMPlattxN    \\
    \quad via direct queue     & \CAMPlatqueue       & \CAMPlatqueueP & \CAMPlatqueueQ & \CAMPlatqueueM & \CAMPlatqueueN \\
    \midrule
    EKF update + track write   & \CAMPlatekf         & \CAMPlatekfP   & \CAMPlatekfQ   & \CAMPlatekfM   & \CAMPlatekfN   \\
    \bottomrule
  \end{tabular}
\end{table}

The sensing pipeline dominates the processing chain and sits inside the
deterministic \SI{640}{\milli\second} CPI accumulation window, so the pipeline
keeps pace with the SRS stream. Both transports and the EKF update are
sub-\SI{15}{\milli\second} even at their maxima --- together they are a rounding
error against the CPI, which is what makes the choice of transport a design
question rather than a performance one.

\bibliographystyle{IEEEtran}
\bibliography{references}

\end{document}

%% file: campaign_numbers.tex
\newcommand{\CAMPflatLos}{100.0 $\pm$ 0.0}

\newcommand{\CAMPflatRangeRmse}{8.66 $\pm$ 0.04}
\newcommand{\CAMPflatVelRmse}{0.44 $\pm$ 0.12}
\newcommand{\CAMPflatAzRmse}{7.02 $\pm$ 0.01}
\newcommand{\CAMPflatCoverage}{92.7 $\pm$ 0.5}
\newcommand{\CAMPflatSingleDet}{87.7 $\pm$ 1.7}
\newcommand{\CAMPflatSignAgree}{96.4 $\pm$ 0.7}
\newcommand{\CAMPflatMargin}{15.2 $\pm$ 0.1}
\newcommand{\CAMPflatDx}{7.26 $\pm$ 1.88}
\newcommand{\CAMPflatDy}{5.32 $\pm$ 0.63}
\newcommand{\CAMPflatCont}{54.2 $\pm$ 3.7}

\newcommand{\CAMPflatTrackIds}{2.6 $\pm$ 0.5}
\newcommand{\CAMPflatN}{5}
\newcommand{\CAMPflatOspa}{13.60 $\pm$ 0.24}

\newcommand{\CAMPflatGospa}{10.95 $\pm$ 0.28}
\newcommand{\CAMPflatGospaLoc}{4.09 $\pm$ 0.60}
\newcommand{\CAMPflatGospaMiss}{6.57 $\pm$ 0.42}
\newcommand{\CAMPflatGospaFalse}{0.47 $\pm$ 0.18}
\newcommand{\CAMPflatNis}{0.55}

\newcommand{\CAMPflatNisVerdict}{conservative}
\newcommand{\CAMPflatNees}{23.29}

\newcommand{\CAMPflatNeesVerdict}{over-confident}
\newcommand{\CAMPflatRangeR}{8.7}
\newcommand{\CAMPflatVelR}{0.44}
\newcommand{\CAMPflatAzR}{7.0}
\newcommand{\CAMPflatCov}{93}
\newcommand{\CAMPflatSgl}{88}
\newcommand{\CAMPflatDxM}{7.3}
\newcommand{\CAMPflatDyM}{5.3}
\newcommand{\CAMPflatContM}{54.2}
\newcommand{\CAMPnlosLos}{29.1 $\pm$ 0.0}

\newcommand{\CAMPnlosRangeR}{7.5}

\newcommand{\CAMPnlosCov}{64}
\newcommand{\CAMPnlosSgl}{55}
\newcommand{\CAMPnlosDxM}{19.3}
\newcommand{\CAMPnlosDyM}{13.9}
\newcommand{\CAMPnlosContM}{17.0}

\newcommand{\CAMPhlowDxM}{33.7}
\newcommand{\CAMPhlowDyM}{8.5}
\newcommand{\CAMPhlowContM}{46.1}

\newcommand{\CAMPsigoldDxM}{10.1}
\newcommand{\CAMPsigoldDyM}{6.9}
\newcommand{\CAMPsigoldContM}{59.5}

\newcommand{\CAMPloadRangeR}{8.3}
\newcommand{\CAMPloadVelR}{0.34}
\newcommand{\CAMPloadAzR}{7.0}
\newcommand{\CAMPloadCov}{97}
\newcommand{\CAMPloadSgl}{82}

\newcommand{\CAMPidleN}{5}

\newcommand{\CAMPidleRangeR}{8.4}
\newcommand{\CAMPidleVelR}{0.46}
\newcommand{\CAMPidleAzR}{7.1}
\newcommand{\CAMPidleCov}{97}
\newcommand{\CAMPidleSgl}{81}

\newcommand{\CAMPlatpipe}{113.0 $\pm$ 5.7}
\newcommand{\CAMPlatpipeP}{121.7}
\newcommand{\CAMPlatpipeQ}{129.8}
\newcommand{\CAMPlatpipeM}{172.9}
\newcommand{\CAMPlatpipeN}{470}
\newcommand{\CAMPlattx}{5.8 $\pm$ 2.9}
\newcommand{\CAMPlattxP}{10.4}
\newcommand{\CAMPlattxQ}{12.0}
\newcommand{\CAMPlattxM}{14.9}
\newcommand{\CAMPlattxN}{3655}
\newcommand{\CAMPlatqueue}{5.3 $\pm$ 4.0}
\newcommand{\CAMPlatqueueP}{12.1}
\newcommand{\CAMPlatqueueQ}{12.1}
\newcommand{\CAMPlatqueueM}{12.1}
\newcommand{\CAMPlatqueueN}{44}
\newcommand{\CAMPlatekf}{2.1 $\pm$ 2.1}
\newcommand{\CAMPlatekfP}{6.7}
\newcommand{\CAMPlatekfQ}{10.2}
\newcommand{\CAMPlatekfM}{11.8}
\newcommand{\CAMPlatekfN}{3655}

\newcommand{\CAMPlatqRefTx}{6.9 $\pm$ 3.8}

\newcommand{\CAMPflatIdSwitch}{38}
\newcommand{\CAMPflatIdSwitchPer}{7.6}
\newcommand{\CAMPloadTput}{4.8 $\pm$ 0.0}
\newcommand{\CAMPloadMcs}{5.8 $\pm$ 0.1}

%% file: campaign_bridge_numbers.tex
\newcommand{\CAMPflatContBridge}{88.31 $\pm$ 0.90}
\newcommand{\CAMPflatContBridgeM}{88.3}

\newcommand{\CAMPflatTrackIdsBridge}{3.00 $\pm$ 0.00}

\newcommand{\CAMPflatRangeRmseBridge}{6.97 $\pm$ 0.69}
\newcommand{\CAMPflatXrangeRmseBridge}{4.44 $\pm$ 0.27}
\newcommand{\CAMPflatDetGapMed}{0.6}   % median inter-detection gap (s)
\newcommand{\CAMPflatDetGapMax}{1.9}    % max inter-detection gap (s)
\newcommand{\CAMPflatCoastMaxS}{8}       % coast-bridge horizon (s)

%% file: primary_accuracy.tex
\newcommand{\CAMPFlatPrimaryRmse}{14.28 $\pm$ 0.34}

%% file: detector_numbers.tex
\newcommand{\CAMPpdCpis}{237}

\newcommand{\CAMPpdLevels}{5}
\newcommand{\CAMPpdOpPd}{0.90}
\newcommand{\CAMPpdOpSnr}{23.1}
\newcommand{\CAMPpdRangeFar}{1.00}
\newcommand{\CAMPpdRangeFarEdge}{60}
\newcommand{\CAMPpdRangeFlat}{1.00}
\newcommand{\CAMPpdRangeNear}{0.87}
\newcommand{\CAMPpdRangeNearBin}{40--60~m}
\newcommand{\CAMPpdRcsHi}{+20}
\newcommand{\CAMPpdRcsHiPd}{0.95}
\newcommand{\CAMPpdRcsLo}{-20}
\newcommand{\CAMPpdRcsLoPd}{0.73}
\newcommand{\CAMPpdRcsSpan}{40}

\newcommand{\CAMPpdSeeds}{3}
\newcommand{\CAMPpdSnrSpan}{0.8}
\newcommand{\CAMPpfaAfter}{8.9\times10^{-5}}
\newcommand{\CAMPpfaAlpha}{4.9\times10^{-4}}
\newcommand{\CAMPpfaBefore}{0.98}
\newcommand{\CAMPpfaCalDb}{1.2}
\newcommand{\CAMPpfaCells}{6.2\times10^{5}}
\newcommand{\CAMPpfaN}{1}

%% file: cfar_numbers.tex
\newcommand{\CAMPcfarCAPfa}{76}

\newcommand{\CAMPcfarCAPdEmFour}{0.72}
\newcommand{\CAMPcfarOSPfa}{6}

\newcommand{\CAMPcfarOSPdEmFour}{0.77}
\newcommand{\CAMPcfarGOPfa}{50}

\newcommand{\CAMPcfarGOPdEmFour}{0.71}
\newcommand{\CAMPcfarObsCpi}{395}

%% file: speed_numbers.tex
\newcommand{\CAMPspdN}{4}
\newcommand{\CAMPspdLo}{1.5}
\newcommand{\CAMPspdHi}{5}
\newcommand{\CAMPspdOpPd}{0.91}
\newcommand{\CAMPspdMinPd}{0.91}
\newcommand{\CAMPspdMinPdV}{1.5}
\newcommand{\CAMPspdMaxPd}{0.94}

%% file: upa_numbers.tex
\newcommand{\UPArows}{2}
\newcommand{\UPAcols}{4}
\newcommand{\UPAelems}{8}
\newcommand{\UPAazAperture}{1.5}   % lambda, the long (4-element) baseline
\newcommand{\UPAelAperture}{0.5}   % lambda, the short (2-element) baseline

\newcommand{\UPAoffAzRmse}{0.9}
\newcommand{\UPAoffElRmse}{0.9}
\newcommand{\UPAoffHeightRmse}{1.8}
\newcommand{\UPAoffPosRmse}{2.4}
\newcommand{\UPAoffNees}{3.5}
\newcommand{\UPAoffAzRmseT}{0.92}
\newcommand{\UPAoffElRmseT}{0.86}

\newcommand{\UPAcampN}{\MScampaignN}
\newcommand{\UPAcampHeightRmse}{\MheightC}    % height RMSE (m), N=5 mean
\newcommand{\UPAcampHeightSD}{\MheightSDC}    % cross-seed std of that RMSE (m)
\newcommand{\UPAcampPosRmse}{\MposC}          % horizontal position RMSE (m)
\newcommand{\UPAcampHfinalSD}{\MhfinalSDC}    % cross-seed std of final altitude (m)

\newcommand{\UPAliveN}{5}
\newcommand{\UPAliveAz}{3.9}          % per-CPI azimuth RMSE, moving transit (deg)
\newcommand{\UPAliveAzSD}{1.2}        % cross-seed std (deg)
\newcommand{\UPAliveEl}{11.5}         % per-CPI elevation RMSE, moving transit (deg)
\newcommand{\UPAliveElSD}{2.0}        % cross-seed std (deg)
\newcommand{\UPAliveEndfireEl}{1.2}   % elevation RMSE, endfire approach (deg)
\newcommand{\UPAliveEndfireAz}{1.9}   % azimuth RMSE, endfire approach (deg)
\newcommand{\UPAliveBroadsideEl}{15.1}% elevation RMSE at broadside (deg)
\newcommand{\UPAliveBroadsideAz}{3.9} % azimuth RMSE at broadside (deg)
\newcommand{\UPAcohFloor}{0.946}      % live aoa_coh floor (no contamination on 2x4)
\newcommand{\UPAulaAz}{8}             % 4-element ULA azimuth RMSE (deg), reference

\newcommand{\UPAclimbTrackRMSE}{1.4}   % offline el-tracking RMSE over a vertical climb
\newcommand{\UPAclimbTrackCorr}{0.98}  % corr(est el, true el) over the climb

%% file: multistatic_numbers.tex
\newcommand{\MSheightTrue}{55}
\newcommand{\MScanonMaxR}{85}     % canonical single-pair range gate, m
\newcommand{\MSmaxR}{200}         % gate needed to admit strong-GDOP UE2, m

\newcommand{\MSliveSeeds}{5}
\newcommand{\MSseeds}{30}
\newcommand{\MSoffHeightRmse}{3.6}      % multistatic-only, good GDOP, m
\newcommand{\MSoffHeightWeakGdop}{19}   % near-colinear UE2 (bad GDOP), m
\newcommand{\MShybridPosRmse}{5.6}      % 2-pair + UPA elevation hybrid, m
\newcommand{\MShybridHeightRmse}{2.2}   % hybrid height RMSE, m
\newcommand{\MShybridNees}{4.3}         % hybrid NEES (ideal 6, consistent)

%% file: multistatic_campaign_numbers.tex
\newcommand{\MheightA}{9.0}\newcommand{\MposA}{18.5}
\newcommand{\MhfinalA}{55.9}\newcommand{\MhfinalSDA}{12.7}\newcommand{\McontinuityA}{97}
\newcommand{\MheightB}{2.2}\newcommand{\MposB}{5.6}
\newcommand{\MhfinalB}{55.3}\newcommand{\MhfinalSDB}{0.2}\newcommand{\McontinuityB}{100}
\newcommand{\MheightC}{5.3}\newcommand{\MheightSDC}{0.2}\newcommand{\MposC}{3.8}
\newcommand{\MhfinalC}{49.0}\newcommand{\MhfinalSDC}{0.2}\newcommand{\McontinuityC}{100}

\newcommand{\MScampaignN}{5}

\newcommand{\MhbiasC}{6.0}
\newcommand{\MpinRatio}{60}

%% file: coast_bridging.tex
% ---------------------------------------------------------------------------
% Coast-bridging (Stage 2c) methodology + result paragraph. Ready to \input or
% paste into the tracking results section of main.tex, AFTER the existing
% fragmentation/continuity discussion (near the \CAMPflatCont rows, ~line 1618).
% Requires: \input{campaign_bridge_numbers} in the preamble/numbers block.
% ---------------------------------------------------------------------------

\textbf{Time-driven coast-bridging.}
The tracker is \emph{detection-driven}: it advances a track only on a detection
that is both delivered over the E2SM-SENS service model \emph{and} associated to
that track. Per-CPI detection is near-complete --- the target is detected on the
large majority of CPIs, with a median inter-detection gap of only
\SI{\CAMPflatDetGapMed}{\second} and a longest gap of \SI{\CAMPflatDetGapMax}{\second}
--- so the baseline's $100 - \CAMPflatContM \approx \SI{46}{\percent}$ continuity
shortfall is \emph{not} a detection problem but a track-management one. Azimuth
scatter from the four-element aperture spreads the per-CPI cross-range estimate
enough that detections intermittently fall outside the association gate; the
track then receives no update, its published record expires, and it must
re-confirm before it reappears --- the $\CAMPflatIdSwitch$ identifier switches of
Fig.~\ref{fig:id_switches} are the visible symptom. Each expire--reconfirm cycle
blanks the target picture even though the filter still holds a valid posterior.

We close these update gaps with a light \emph{time-driven} maintenance pass that,
once per second, forward-extrapolates each confirmed track's posterior along its
constant-velocity model and republishes it, flagged as a coasted (predicted, not
measured) estimate. Bridging is capped at \SI{\CAMPflatCoastMaxS}{\second}: beyond
that a pure prediction is no longer trustworthy, so the record is allowed to lapse
and the gap is reported honestly as a loss of track rather than papered over. The
pass reads filter state only --- it never advances the filter --- so the next
associated detection updates from the clean posterior and measurement accuracy is
unchanged.

With coast-bridging enabled, flat-LOS track continuity rises from
\CAMPflatCont\% to \CAMPflatContBridge\% ($N=5$). Position accuracy, reported over
measured CPIs only (coasted predictions excluded), is unchanged within run-to-run
spread: range RMSE \CAMPflatRangeRmseBridge~m and cross-range
\CAMPflatXrangeRmseBridge~m. Bridging restores temporal continuity but not
single-identity tracking: the transit is still covered by \CAMPflatTrackIdsBridge{}
identifiers per run (\CAMPflatTrackIds{} without bridging), since republishing a
coasted track does not merge it with the fragment that later re-confirms ---
consolidating the fragments would need association-level work (e.g.\ JPDA), left to
future work. The residual $100 - \CAMPflatContBridgeM \approx \SI{12}{\percent}$ is
track initiation and re-confirmation latency --- the CPIs before a confirmed track
first exists, which a pass operating only on already-confirmed tracks cannot fill
--- not detection dropouts, whose longest run (\SI{\CAMPflatDetGapMax}{\second})
sits well inside the \SI{\CAMPflatCoastMaxS}{\second} bridging horizon.

%% file: upa_elevation.tex
% ============================================================================
% Elevation sensing with a planar array (UPA upgrade #1).
% Offline observability is solid (ideal Sionna channel). LIVE: the 3-D EKF resolves
% altitude to ~4.5 m over N=5, but the PER-CPI ANGLES ARE NOISY (az ~3.9, el ~11.5
% deg) and strongly geometry-dependent (el ~1.2 deg endfire vs ~15 deg broadside).
% Do NOT restate the angles as measurement accuracy matching the ideal channel --
% that framing was retracted 2026-07-27 as unreproducible. Numbers: upa_numbers.tex.
% Connects to
% Section~\ref{sec:multistatic}.
% ============================================================================
\section{Elevation Sensing with a Planar Array}
\label{sec:upa}

The height-prior dependence above is a direct consequence of the one-dimensional
aperture. We therefore replaced the horizontal ULA with a
$\UPArows\times\UPAcols$ ($\UPAelems$-element) uniform planar array (UPA) at the
gNB---$\UPAcols$ columns along the azimuth (horizontal) baseline and $\UPArows$
rows along the elevation (vertical) baseline, at $\lambda/2$ spacing---and
reduced the nrUE to a single SRS transmit port, so the angle estimate is formed
entirely at the gNB receive aperture. A rectangular array is separable: the
inter-element phase ramp of~\eqref{eq:aoa} applies independently along each
baseline, yielding azimuth from the horizontal ramp and elevation from the
vertical one at no additional estimator complexity. The tracker is extended to a
three-dimensional constant-velocity state $[\,p_x,p_y,p_z,v_x,v_y,v_z\,]$ with
measurement $[\,R_{\mathrm{bi}}, v_{\mathrm{bi}}, \theta_{\mathrm{az}},
  \theta_{\mathrm{el}}\,]$; the height prior $h_{\mathrm{tgt}}$ is removed and
altitude is estimated online.

\emph{Observability.} On the ray-traced $\UPAelems$-element channel---driven from
the clean per-CPI target echo, with no thermal noise, CFAR miss, or residual
CFO/timing---the separable estimator recovers both angles to
$\approx\!\ang{\UPAoffAzRmse}$ (azimuth) and $\approx\!\ang{\UPAoffElRmse}$
(elevation) RMSE, and the 3-D EKF---initialized with a deliberately wrong
altitude---converges to the true height with an RMSE of
\SI{\UPAoffHeightRmse}{\metre} (position \SI{\UPAoffPosRmse}{\metre},
NEES~$\approx\!\UPAoffNees$). Altitude is thus \emph{observable} from a single
bistatic pair once a vertical aperture is present: the prior is no longer
required. These two angle figures are a noiseless floor rather than an
aperture-limited accuracy, and should not be read as what the short vertical
baseline delivers: transposing the array to $\UPAcols\times\UPArows$ exchanges the
long and short baselines---which would move an aperture-limited error on each axis
by their length ratio, $\UPAazAperture\lambda$ against
$\UPAelAperture\lambda$, a factor of three---yet leaves both essentially unchanged
(azimuth \ang{\UPAoffAzRmseT}, elevation \ang{\UPAoffElRmseT}). What the ideal
channel establishes is that elevation is \emph{estimable at all} once a vertical
baseline exists; what it costs in practice is the live per-CPI figure reported
below, which is an order of magnitude larger.

\emph{Altitude in the live pipeline.} Driven end-to-end through the real gNB DSP---CFAR
detection, comb-LS channel estimation, ECA and the residual CFO/timing of the SDR
chain---the separable estimator is exercised end-to-end. Over $N=\UPAcampN$ seeded
transits (the same seed set and stack as the multistatic configurations of
Section~\ref{sec:multistatic}, so the comparison in
Table~\ref{tab:multistatic_cells} is paired), the 3-D EKF resolves altitude to a
height RMSE of $\UPAcampHeightRmse \pm \UPAcampHeightSD$~m and horizontal position
to \SI{\UPAcampPosRmse}{\metre}, against \SI{\MheightA}{\metre} height RMSE for the
same array without the vertical rows. The vertical aperture therefore improves
altitude \emph{accuracy} by roughly a third.

\emph{Repeatability without unbiasedness.} The
cross-seed spread of the final altitude is \SI{\UPAcampHfinalSD}{\metre} with the
vertical rows against \SI{\MhfinalSDA}{\metre} without them, a collapse of about
$\MpinRatio\times$. This is the observability signal --- the spread, not the RMSE,
is what distinguishes an observable state, because an unobservable height wanders
with the noise draw --- so the strict form of the claim is supported: the vertical
aperture makes altitude observable in the live chain.

The repeatability is not accuracy, and the two must not be conflated. All five seeds
settle between \SI{48.9}{\metre} and \SI{49.3}{\metre} against a
\SI{\MSheightTrue}{\metre} target: a \SI{\MhbiasC}{\metre} bias held to
\SI{0.2}{\metre}, i.e.\ an estimator that is confidently slightly wrong. Because the
bias is common to every seed it is not a noise or conditioning effect; it points to
elevation-scale calibration in the estimator itself. The two-transmitter
configuration of Section~\ref{sec:multistatic}, which reaches the same spread from
range diversity instead of a vertical baseline, does not exhibit it --- which
localizes the bias to the elevation path rather than to the filter, the geometry or
the alignment, all of which the two configurations share.

\emph{Role of the filter in delivering altitude.} The accuracy comes from the
filter integrating many measurements, not from the per-CPI angles, and the two are
easily conflated. Measured against ray-traced truth over the same $N=\UPAliveN$
transits, individual angle estimates are far from the ideal-channel bound:
azimuth $\ang{\UPAliveAz} \pm \ang{\UPAliveAzSD}$ and elevation
$\ang{\UPAliveEl} \pm \ang{\UPAliveElSD}$ RMSE per CPI. A single such elevation
sample, converted geometrically, would place the target tens of meters off in
height. The meter-level altitude above is therefore not per-measurement accuracy
but the product of the filter integrating excess range, range rate and many noisy
bearings across the transit, and why
the observability claim rests on the cross-seed spread rather than on any single
CPI.

\begin{figure}[!tbp]
  \centering
  \includegraphics[width=\columnwidth]{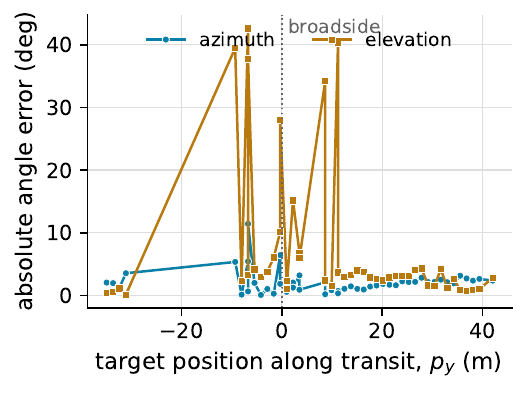}
  \caption{Per-CPI angle error along the transit (one representative seed of
    $N=\UPAliveN$). Elevation is accurate only where the geometry gives the two-row
    aperture leverage on the vertical: $\approx\ang{\UPAliveEndfireEl}$ near endfire
    against $\approx\ang{\UPAliveBroadsideEl}$ at broadside. Azimuth is comparatively
    flat. The whole-transit averages are dominated by broadside samples and represent
    neither regime; the filtered altitude of
    $\UPAcampHeightRmse \pm \UPAcampHeightSD$~m is obtained by integrating these
    measurements, not by any of them individually.}
  \label{fig:upa_angle_vs_geometry}
\end{figure}

\emph{Geometry dependence of elevation accuracy.} The per-CPI elevation
error is not uniform along the transit: it is
$\ang{\UPAliveEndfireEl}$ RMSE near endfire, where the vertical cut through the
target is most favorable, and degrades to $\ang{\UPAliveBroadsideEl}$ at
broadside, where the two-row aperture has least leverage on the vertical angle ---
an order of magnitude between the two regimes
(Fig.~\ref{fig:upa_angle_vs_geometry}). The whole-transit figure quoted above is
dominated by the broadside samples and understates what the array achieves where
the geometry suits it. Azimuth, by contrast, is comparatively flat across the
transit ($\ang{\UPAliveEndfireAz}$ to $\ang{\UPAliveBroadsideAz}$) and remains
comparable to the four-element azimuth-only ULA (\ang{\UPAulaAz}): adding the
vertical rows does not cost azimuth. Altitude is therefore sensed, not assumed, on
a commodity eight-element gNB---but the geometry under which it is sensed matters.

\emph{Robustness and scope.} The array response stays coherent across the transit
(a ground-truth-free inter-row phase-ramp coherence $\ge\!\UPAcohFloor$ on every
detection), so no per-detection angle down-weighting is required. Accuracy degrades
only when the target is \emph{stationary} at the scene boundary---where the near-zero
Doppler and the finite scene edge corrupt the snapshot---a regime outside the moving
transit these figures characterize. That the estimator itself is sound is confirmed
independently on a vertically \emph{climbing} target (true elevation swept
\ang{15.7}--\ang{33}), which the offline estimator tracks to \ang{\UPAclimbTrackRMSE}
RMSE (correlation \num{\UPAclimbTrackCorr}). The campaign covers one array shape
($\UPArows\times\UPAcols$) and one transit geometry; an aperture-shape and
placement sweep is the natural next consolidation.

Elevation is therefore \emph{observable in principle} (the ideal-channel result)
\emph{and usable in practice} (the live moving-transit result) from a single
bistatic pair with a commodity eight-element gNB UPA. The remaining single-gNB
limitation is cross-range: azimuth precision is bound to the receive aperture and is
not improved by the UPA's vertical rows. The multistatic geometry of
Section~\ref{sec:multistatic} adds a second bistatic pair, giving redundant altitude
observability through range diversity and filling the near-baseline/Doppler blind
zones---complementary to, not a replacement for, the UPA elevation cut.

%% file: multistatic.tex
% ============================================================================
% Multistatic (2x UE, 1x gNB) altitude observability — v0p1 new result.
% Honest tiered framing. Live tier is N=5 (seeds 1001,1003,2001-2003, 0 dBsm UAV);
% see multistatic_numbers.tex. \input{multistatic_numbers} in the preamble.
% Place after the UPA section (Section~\ref{sec:upa}).
% ============================================================================
\section{Multistatic Geometry: Altitude from a Second Bistatic Pair}
\label{sec:multistatic}

\begin{figure}[!tbp]
  \centering
  \includegraphics[width=\columnwidth]{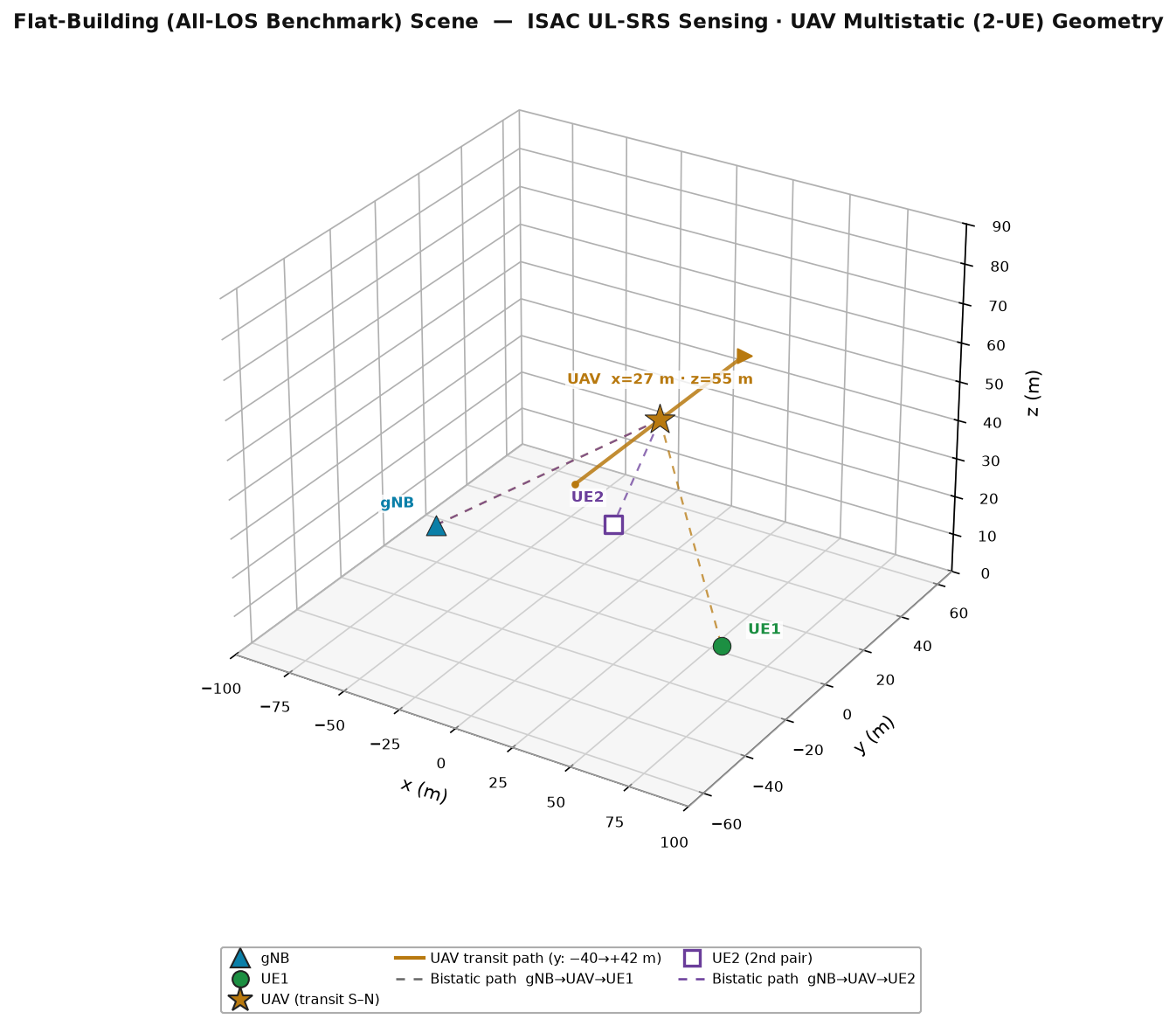}
  \caption{Two-transmitter geometry. A single gNB receiver is served by two
    spatially separated nrUEs, giving two bistatic paths
    (gNB$\rightarrow$UAV$\rightarrow$UE$i$) and therefore two iso-range surfaces
    through the target. UE2 sits on the \emph{opposite side} of the gNB from UE1, so
    the two surfaces intersect the vertical steeply --- the strong-GDOP placement
    the measurements use. A near-colinear UE2 dilutes that conditioning and degrades
    height accuracy by roughly $5\times$ (Section~\ref{sec:multistatic_gdop}).}
  \label{fig:multistatic_geometry}
\end{figure}

The planar array of Section~\ref{sec:upa} makes altitude observable by adding a
\emph{vertical aperture} at the receiver. A second, geometrically independent way
to the same end is to add a second \emph{bistatic pair}. A single gNB receiver
served by two spatially separated nrUE transmitters (Fig.~\ref{fig:multistatic_geometry})
observes the target over two distinct bistatic ellipsoids; where one pair constrains the target to an
iso-range surface, the intersection of two such surfaces --- together with the
gNB-side azimuth --- pins the target in three dimensions. Altitude then becomes
observable \emph{even when the gNB array is the plain one-dimensional
  azimuth-only array}, with no vertical aperture at all. This is attractive because
it needs no change to the radio unit: a second commodity UE, not a wider array.

\subsection{Platform Extension}
\label{sec:multistatic_platform}

Carrying a second pair end-to-end touches every layer of the testbed, which was
otherwise wired for a single transmit/receive pair. The changes are additive and
reduce to the single-pair path byte-for-byte when the second pair is disabled:

\begin{itemize}
  \item \emph{Channel and emulator.} The ray-traced channel exports one propagation
        link per transmitter; the emulator serves each link its own channel, and the
        rfsim server superposes them at the gNB, so the received signal is the true sum
        of both pairs' echoes.
  \item \emph{Waveform orthogonality.} The two UEs are assigned different SRS comb
        offsets, so their soundings are orthogonal in frequency and the gNB accumulates
        each pair's coherent processing interval independently rather than blending them.
  \item \emph{Detection transport.} Each detection carries a bistatic-pair
        identifier from the gNB PHY through the E2/redis transport to the xApp, so the
        tracker knows which transmitter produced each measurement. A build-time size
        assertion guards the shared detection structure against silent layout drift when
        the field is added.
  \item \emph{Tracker.} The xApp fuses detections from both pairs in a
        three-dimensional constant-velocity EKF. Each pair contributes its own nonlinear
        measurement model --- excess bistatic range, range-rate and (for the gNB-side
        pair) azimuth --- evaluated against that pair's known transmitter geometry, and
        the height prior is dropped.
\end{itemize}

\subsection{Fusion and Observability}
\label{sec:multistatic_fusion}

We evaluate the geometry at two levels of fidelity.

\emph{Fusion proof (ray-traced channel model).} An offline harness that mirrors
the production EKF fuses two analytic bistatic pairs over $\MSseeds$ noise
realizations. A single 3-D pair without a vertical aperture leaves height
unobservable (the altitude state diverges); fusing a second, well-separated pair
recovers it to \SI{\MSoffHeightRmse}{\metre} RMSE. Combining the two upgrades ---
a second pair \emph{and} the planar array of Section~\ref{sec:upa} --- is
decisively the best configuration, giving position RMSE
\SI{\MShybridPosRmse}{\metre} and height RMSE \SI{\MShybridHeightRmse}{\metre} at
a consistent filter (NEES~$\approx\!\MShybridNees$), and it is robust to where the
second UE is placed.

\begin{table}[!tbp]
  \centering
  \caption{Altitude observability across sensing configurations, $N=\MScampaignN$
    independent live transits per configuration (same seed set, identical stack;
    only the receive array and the number of transmitters change). ULA = the
    azimuth-only $\UPAcols$-element line array; UPA = the
    $\UPArows\times\UPAcols$ planar array of Section~\ref{sec:upa}; the two-transmitter
    rows are the geometry of this section. The observability signal is the
    \emph{cross-seed} standard deviation of the final altitude estimate, not the
    RMSE: an
    unobservable height wanders with the noise draw, an observable one does not.
    The azimuth-only baseline scatters by \SI{\MhfinalSDA}{\metre} across seeds,
    while both upgraded configurations hold \SI{\MhfinalSDC}{\metre} --- a factor of
    about $\MpinRatio$. Note that only the two-transmitter row is \emph{also}
    unbiased: the planar array is equally repeatable but settles
    \SI{\MhbiasC}{\metre} low, so its spread must not be read as accuracy.
    True altitude \SI{\MSheightTrue}{\metre}. Rows differ only in the receive array
    ($\UPAcols$-element azimuth-only line array vs.\ the
    $\UPArows\times\UPAcols$ planar array) and the number of transmitters; the
    waveform, seeds, stack and trace family are identical, and all rows come from a
    single build on a single host.}
  \label{tab:multistatic_cells}
  \footnotesize
  \setlength{\tabcolsep}{4pt}
  \begin{tabular}{@{}lcccc@{}}
    \toprule
                                & Height                                                  & Pos.     & Final alt.                  & Cont.         \\
    Configuration               & RMSE (m)                                                & RMSE (m) & (m)                         & (\%)          \\
    \midrule
    ULA, 1 UE \emph{(baseline)} & \MheightA                                               & \MposA   & $\MhfinalA \pm \MhfinalSDA$ & \McontinuityA \\
    ULA, 2 UE                   & \MheightB                                               & \MposB   & $\MhfinalB \pm \MhfinalSDB$ & \McontinuityB \\
    UPA, 1 UE                   & \MheightC                                               & \MposC   & $\MhfinalC \pm \MhfinalSDC$ & \McontinuityC \\
    UPA, 2 UE \emph{(hybrid)}   & \multicolumn{4}{c}{not measured --- see the Limitation}                                                          \\
    \bottomrule
  \end{tabular}
\end{table}

\begin{figure}[!tbp]
  \centering
  \includegraphics[width=\columnwidth]{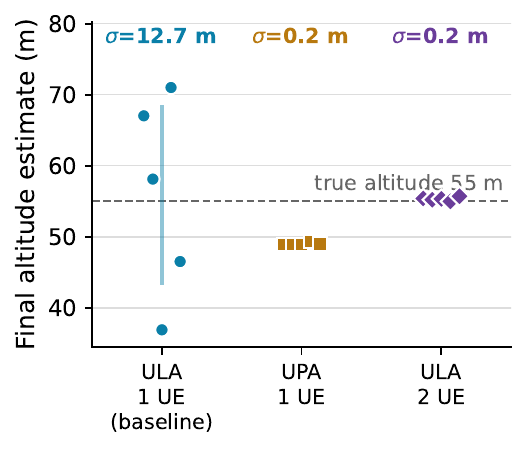}
  \caption{Final altitude estimate for every seed, by sensing configuration
    (azimuth-only four-element ULA vs.\ the $\UPArows\times\UPAcols$ planar UPA;
    one or two transmitters). Each mark is one independent transit; the bar spans
    $\pm1\sigma$ across seeds. Observability is the \emph{clustering}, not the
    distance to truth. The azimuth-only baseline does not cluster --- its five seeds
    span \SIrange{36.9}{71.0}{\metre} about a \SI{\MSheightTrue}{\metre} target, a
    spread of \SI{\MhfinalSDA}{\metre} --- while both upgraded configurations collapse
    to \SI{\MhfinalSDC}{\metre}. Reading clustering and distance separately is what
    distinguishes them: the two-transmitter marks cluster \emph{on} truth, the planar
    array clusters \SI{\MhbiasC}{\metre} below it. The baseline's near-correct mean is
    an artifact of averaging symmetric errors, not evidence of an observable state.}
  \label{fig:altitude_observability}
\end{figure}

Table~\ref{tab:multistatic_cells} and Fig.~\ref{fig:altitude_observability} report
both routes to altitude --- a second transmitter and a vertical aperture --- against
the baseline on the same seed set, build and host.

\paragraph*{Altitude observability and the two routes.} The
claim this section supports is not the RMSE but the \emph{cross-seed
  spread} of the final altitude: an unobservable height wanders with the noise draw,
an observable one does not. The baseline wanders by \SI{\MhfinalSDA}{\metre} --- its
five seeds span \SIrange{36.9}{71.0}{\metre} about a \SI{\MSheightTrue}{\metre}
target --- and both upgrades collapse that to \SI{\MhfinalSDC}{\metre}, a factor of
about $\MpinRatio$. Altitude is therefore observable live from either route, which is
what the ray-traced fusion proof predicts.

The two routes differ in a way the spread alone does not show. Range diversity from
a second transmitter is both pinned \emph{and} unbiased, settling at
\SI{\MhfinalB}{\metre} against a \SI{\MSheightTrue}{\metre} truth, and it gives the
lowest height RMSE of the three configurations (\SI{\MheightB}{\metre}). The vertical
aperture is equally repeatable but sits \SI{\MhbiasC}{\metre} low
(\SI{\MhfinalC}{\metre}); what it buys instead is horizontal accuracy
(\SI{\MposC}{\metre} against \SI{\MposB}{\metre} and \SI{\MposA}{\metre}). So the
planar array should be read as an upgrade to \emph{position} accuracy and to the
conditioning of the altitude estimate, not as the more accurate altitude sensor. A
residual elevation bias of this size is consistent with the calibration limits
discussed in Section~\ref{sec:upa} and is not explained by geometry, seeds or
alignment, all of which are shared with the two-transmitter row.

\paragraph*{Corroboration against the offline reference.} The live figures above are
consistent with the fusion proof of Section~\ref{sec:multistatic_gdop}, which drives the
same estimator and filter from a ray-traced channel model with no software-defined
radio, no sample transport and no real-time deadline: a single pair leaves height
unobservable while a second, well-separated pair recovers it to
\SI{\MSoffHeightRmse}{\metre} RMSE. Altitude observability from range diversity is
therefore established both in the signal-processing chain and end to end through the
live stack, and the live campaign's \SI{\MhfinalSDB}{\metre} cross-seed spread at a
\SI{\MhfinalB}{\metre} mean is what that prediction looks like when it is realized on a
real PHY.

\paragraph*{Two-transmitter configuration.}
All five two-UE runs in this campaign clear both criteria: median azimuth alignment
residual \SIrange{0.9}{1.2}{\degree} (against \SIrange{1.6}{2.1}{\degree} for the
single-transmitter rows), both pairs contributing throughout, full transit replayed,
\SI{100}{\percent} track continuity and association. This is now the best-conditioned
configuration we measure --- lowest height RMSE, unbiased final altitude, and the
tightest cross-seed spread jointly with the planar array.

\subsection{Geometry Dilution and the Range Gate}
\label{sec:multistatic_gdop}

Altitude observability from range diversity is strongly geometry-dependent. The
vertical conditioning is best when the second UE is well separated from the first
--- on the opposite side of the gNB, so the two bistatic ellipsoids intersect the
vertical steeply; a near-colinear second UE dilutes the geometry and the height
RMSE degrades by roughly $5\times$ (from \SI{\MSoffHeightRmse}{\metre} to
$\approx$\SI{\MSoffHeightWeakGdop}{\metre} in the fusion proof). This creates a
tension with detection gating: a well-separated second UE sits on a longer
bistatic path, so admitting its detections requires opening the range gate from
the canonical \SI{\MScanonMaxR}{\metre} to $\approx$\SI{\MSmaxR}{\metre}. A
gate-compatible near-side placement, by contrast, has vertical conditioning too
weak to survive the residual range bias of the ray-traced echo. Elevation
observability from multistatic range diversity is therefore a joint choice of
transmitter placement and range-gate width, not a free by-product of adding a
transmitter. When the planar array of Section~\ref{sec:upa} is also present, the
hybrid tolerates any second-UE placement, because the vertical aperture supplies
the conditioning the geometry alone may lack.

\emph{Maturity.} The fusion proof uses an analytic channel over $\MSseeds$ seeds;
the live result is now an $N=\MSliveSeeds$-seed campaign on one geometry (UAV RCS
scaled to \SI{0}{\dBsm}, small-UAV). A transmitter-placement / range-gate sweep is
the natural next consolidation.

\emph{Limitation --- live UPA+multistatic hybrid.} The offline fusion proof shows
the hybrid --- a second bistatic pair \emph{and} the planar array of
Section~\ref{sec:upa} --- is the strongest configuration, and it is the one row of
Table~\ref{tab:multistatic_cells} we cannot measure. The failure is specific and
reproducible: with eight receive antennas and two transmitters, the gNB misses its
random-access response deadline about 58 times per coherent processing interval,
against zero in every other configuration. Its impact on this paper is narrow: the hybrid is an \emph{additive}
configuration, both of whose ingredients are measured independently and at
$N=\MScampaignN$ here, and each makes altitude observable on its own. Its absence
therefore removes an expected improvement, not a claim.

%% file: cfar_table.tex
% GENERATED by make_cfar_comparison.py -- do not hand-edit.
\begin{tabular}{lcccc}
\toprule
CFAR & \multicolumn{1}{c}{realized $P_{\mathrm{FA}}$} & \multicolumn{3}{c}{$P_{\mathrm{D}}$ at realized $P_{\mathrm{FA}}$} \\
\cmidrule(lr){2-2}\cmidrule(lr){3-5}
mode & at nom.\ $10^{-4}$ & $10^{-2}$ & $10^{-3}$ & $10^{-4}$ \\
\midrule
CA & 76$\times$ & 0.92 & 0.80 & 0.72 \\
OS & 6$\times$ & 0.87 & 0.83 & 0.77 \\
GO & 50$\times$ & 0.91 & 0.79 & 0.71 \\
\bottomrule
\end{tabular}